\documentclass{aa}  
\usepackage{graphicx}
\usepackage[varg]{txfonts}
\usepackage{subfig}
\usepackage[labelfont=bf]{caption}
\usepackage{longtable,booktabs}
\usepackage{pdfpages}
\usepackage{import}
\usepackage[maxfloats=256]{morefloats}
\DeclareMathOperator{\sech}{sech}
\newcommand{\kms}{~km~s$^{-1}$}
\newcommand{\msun}{M$_\odot$}

\newcommand{\barolo}{$^{\mathrm{3D}}$\textsc{Barolo} }

\newcommand{\hz}{h_z}

\newcommand{\beq}{\begin{equation}}
\newcommand{\eeq}{\end{equation}}

\newcommand\dd{{\rm d}}
\newcommand{\hseventy}{h_{70\%}}

\newcommand\rhozero{\rho_0}

\newcommand\lambdaRinst{\lambda_{R,{\rm inst}}}
\newcommand\lambdaJ{\lambda_{\rm J}}
\newcommand\Mring{M_{\rm ring}}
\newcommand\Mclump{M_{\rm clump}}
\newcommand\lambdaphi{\lambda_\phi}
\newcommand{\Qthreed}{Q_{\rm 3D}}

\newcommand\rhobar{{\bar{\rho}}}

\graphicspath{{../images/}}

\usepackage[]{hyperref}
\hypersetup{colorlinks=true,linkcolor=red,citecolor=blue,urlcolor=magenta}

\defcitealias{2019Bacchini}{B19a}
\defcitealias{2019Bacchini_b}{B19b}
\defcitealias{2020Bacchini_a}{B20a}
\defcitealias{2020Bacchini_b}{B20b}

\begin{document}

\title{A 3D view on the local gravitational instability of cold gas discs in star-forming galaxies at $0 \lesssim \mathrm{z} \lesssim 5$}
%\subtitle{Testing the 3D stability criterion for thick \& flaring gas discs}

\authorrunning{C. Bacchini et al.}
%\author{Cecilia Bacchini~\inst{}, Carlo Nipoti~\inst{}, Giuliano Iorio~\inst{}, Fernanda Roman-Oliveira~\inst{}, Francesca Rizzo~\inst{}, Pavel E. Mancera Pi\~{n}a~\inst{}, Antonino Marasco~\inst{}, Anita Zanella~\inst{}} % Fernanda, Filippo, Anita, Lelli, Antonino
\author{C. Bacchini~\inst{\ref{inst:dark},\ref{inst:inaf_pd}}, 
		C. Nipoti~\inst{\ref{inst:difa_bo}}, 
		G. Iorio~\inst{\ref{inst:inaf_pd},\ref{inst:difa_pd},\ref{inst:infn_pd}}, 
		F. Roman-Oliveira~\inst{\ref{inst:kapteyn}}, 
		F. Rizzo~\inst{\ref{inst:dawn},\ref{inst:nbi}}, 
		P. E. Mancera Pi\~{n}a~\inst{\ref{inst:leiden}}, 
		A. Marasco~\inst{\ref{inst:inaf_pd}}, 
		A. Zanella~\inst{\ref{inst:inaf_pd}}, and
		F. Lelli~\inst{\ref{inst:inaf_fi}}}

\institute{
DARK, Niels Bohr Institute, University of Copenhagen, Jagtvej 155, 2200 Copenhagen, Denmark \label{inst:dark}\\
\email{cecilia.bacchini@nbi.ku.dk}
\and
INAF - Osservatorio Astronomico di Padova, vicolo dell'Osservatorio 5, 35122 Padova, Italy  \label{inst:inaf_pd}
%\email{cecilia.bacchini@inaf.it}
\and
Dipartimento di Fisica e Astronomia ``Augusto Righi'', Universit\`{a} di Bologna, via Gobetti 93/2, 40129, Bologna, Italy \label{inst:difa_bo}
\and
Dipartimento di Fisica e Astronomia ``Galileo Galilei'', Universit\`{a} di Padova, Vicolo dell’Osservatorio 3, 35122 Padova, Italy  \label{inst:difa_pd}
\and 
INFN - Padova, Via Marzolo 8, 35131 Padova, Italy \label{inst:infn_pd}
\and
Kapteyn Astronomical Institute, University of Groningen, Landleven 12, 9747 AD, Groningen, The Netherlands  \label{inst:kapteyn}
\and
Cosmic Dawn Center (DAWN), University of Copenhagen, Jagtvej 155, 2200 Copenhagen, Denmark  \label{inst:dawn}
\and
Niels Bohr Institute, University of Copenhagen, Jagtvej 128, 2200, Copenhagen, Denmark \label{inst:nbi}
\and
Leiden Observatory, Leiden University, P.O. Box 9513, 2300 RA Leiden, The Netherlands  \label{inst:leiden}
\and
INAF - Arcetri Astrophysical Observatory, Largo Enrico Fermi 5, 50125, Firenze, Italy  \label{inst:inaf_fi}
}

\date{Accepted \today.}

% \abstract{}{}{}{}{} 
% 5 {} token are mandatory
 
\abstract{	
Local gravitational instability (LGI) is considered crucial for regulating star formation and gas turbulence in galaxy discs, especially at high redshift. 
Instability criteria usually assume infinitesimally thin discs or rely on approximations to include the stabilising effect of the gas disc thickness. 
We test a new 3D instability criterion for rotating gas discs that are vertically stratified in an external potential. 
This criterion reads $Q_{\rm 3D}<1$, where $Q_{\rm 3D}$ is the 3D analogue of the Toomre parameter $Q$. 
The advantage of $Q_{\rm 3D}$ is that it allows us to study LGI in and above the galaxy midplane in a rigorous and self-consistent way. 
We apply the criterion to a sample of 44 star-forming galaxies at $0\lesssim\mathrm{z}\lesssim5$ hosting rotating discs of cold gas. 
The sample is representative of galaxies on the main sequence at $\mathrm{z}\approx0$ and includes massive star-forming and starburst galaxies at $1\lesssim\mathrm{z}\lesssim5$. 
For each galaxy, we first apply the Toomre criterion for infinitesimally thin discs, finding ten unstable systems. 
We then obtain maps of $Q_{\rm 3D}$ from a 3D model of the gas disc derived in the combined potential of dark matter, stars and the gas itself. 
According to the 3D criterion, two galaxies with $Q<1$ show no evidence of instability and the unstable regions that are 20\% smaller than those where $Q<1$. 
No unstable disc is found at $0\lesssim\mathrm{z}\lesssim 1$, while $\approx 60$\% of the systems at $2\lesssim\mathrm{z}\lesssim5$ are locally unstable. 
In these latter, a relatively small fraction of the total gas ($\approx 30$\%) is potentially affected by the instability. 
Our results disfavour LGI as the main regulator of star formation and turbulence in moderately star-forming galaxies in the present-day Universe. 
LGI likely becomes important at high redshift, but the input by other mechanisms seems required in a significant portion of the disc. 
We also estimate the expected mass of clumps in the unstable regions, offering testable predictions for observations.
}


   \maketitle

\section{Introduction}\label{sec:intro}
For more than six decades, the local gravitational instability (LGI) of gas discs has been a crucial topic of galaxy formation and evolution studies \citep[e.g.][to cite some early works]{1960Safronov,1961Chandrasekhar,1964LinShu,1964Toomre,1965Goldreich}. 
When perturbed, a gas disc that is prone to LGI is expected to fragment into gas clumps, which can eventually collapse and form new stars \citep[e.g.][]{1989Kennicutt,1997Silk,1998Kennicutt,2001MartinKennicutt}. 
Furthermore, theoretical models predict that the clumps migrate inward, dragging mass towards the galaxy center and increasing the gas turbulence \citep[e.g.][]{2007Bournaud,2009Agertz,2009Dekel_b,2010Krumholz,2012Cacciato,2018Krumholz}. 

Traditionally, the stability of gas discs is investigated using the so-called \cite{1964Toomre} criterion for instability
\begin{equation}\label{eq:q2d}
	Q(R) \equiv \frac{\kappa \sigma}{\pi G \Sigma} <1 \, ,
\end{equation}
where $R$ is the galactocentric distance, $\sigma$ is the gas velocity dispersion, $\Sigma$ is the gas surface density, and $\kappa$ is the epicycle frequency. 
The latter is defined by
\begin{equation}\label{eq:kappa2}
\kappa^2\equiv 4\Omega^2+\frac{\dd \Omega^2}{\dd \ln R},
\end{equation}
where $\Omega \equiv V_\mathrm{rot}/R$ is the angular frequency and $V_\mathrm{rot}$ is the rotation velocity of the gas in the disc. 
Equation~\ref{eq:q2d} shows that the gas pressure and the rotation have a stabilizing effect, while the denser the gas disc is the more it is prone to LGI. 
From Eq.~\ref{eq:q2d}, one can derive the expression for the gas density threshold above which a disc is unstable
\begin{equation}\label{eq:Sigma_th}
	\Sigma_\mathrm{th} = \frac{\sigma \kappa}{\pi G}. 
\end{equation}
It is has been proposed that such threshold could explain the very weak star formation activity observed in the outskirts of spiral galaxies and in dwarfs, where the gas density is very low \citep[e.g.][]{1989Kennicutt,1998Kennicutt}. 

Eqs.~\ref{eq:q2d} and \ref{eq:Sigma_th} are simple and easy to use to study LGI in galaxies, as they involve observable quantities. 
However, Eq.~\ref{eq:q2d} was derived for a rotating gas disc that is infinitesimally thin, which is an unrealistic assumption for galactic gas discs. 
Several works on nearby galaxies showed that gas discs have a non-negligible thickness, which can significantly increase with the galactocentric distance \citep{2010Roychowdhury,2010OBrien_c,2011Yim,2014Yim,2017Peters,2017Marasco,2018Iorio,2019Bacchini,2019Bacchini_b,2020Bacchini_b,2020Patra,2022ManceraPina_b}. 
A few authors have proposed instability criteria for thick discs. 
Such criteria are obtained by modifying the thin-disc dispersion relation, introducing a reduction factor that accounts for the fact that the gravitational potential is weaker if the mass is not confined in a plane 
\citep{1964Toomre,1992Romeo,1994Romeo,2010Bertin,2010Wang,2011Elmegreen,2012Griv,2013Romeo,2015Behrendt}. 
Some authors computed the reduction factor assuming a given scale height for the disc vertical structure \citep{1964Toomre,1984JogSolomon}, while other studies self-consistently calculate this scale height assuming the vertical equilibrium \citep{1970Vandervoort,1992Romeo,1994Romeo}. 
Typically, the critical value for instability is reduced to $Q \approx 0.65-0.70$ \citep[e.g.][]{1994Romeo}, hence thick discs are more stable than thin discs. 

These criteria can be used to study LGI at a given radius (i.e. in 2D), but they do not provide information on the behaviour of the gas as a function of the distance from midplane. 
A few authors have studied specific cases to obtain a 3D instability criterion that allows to study the LGI also above and below the midplane \citep[e.g.][]{1960Safronov,1961Chandrasekhar,1965Goldreich,1975Genkin,1982Bertin,2022Meidt}. 
In particular, \cite{2023Nipoti} has recently proposed a new 3D instability criterion for a rotating gas disc that is vertically stratified in a given gravitational potential. 
\cite{2023Nipoti}'s criterion can be readily used to investigate the gas disc stability in a self-consistent way, provided that the vertical distribution of the gas is known. 

In general, studying the LGI of gas discs requires accurate measurements of the gas distribution and kinematics ($\sigma$ and $V_\mathrm{rot}$ in Eq.~\ref{eq:q2d}). 
Previous studies based on either 2D or 3D criteria often assume a flat rotation curve (thus $\kappa=\sqrt{2}\Omega$) and a fixed value for $\sigma$ constant with $R$. 
However, the rotation curve gradient can be very different from galaxy to galaxy, depending on its mass distribution and on which component dominates the galactic potential at a given $R$ \citep*[see for instance Sect.~4.3.3 in][]{2019TheBook_CFN}. 
Most importantly, $\sigma$ is not constant with the galactocentric radius, but it usually decreases with increasing $R$ \citep[e.g.][]{2002Fraternali,2008Boomsma,2009Tamburro,2015Diteodoro,2016Mogotsi,2017Iorio,2017Marasco,2019Bacchini,2020Bacchini_b,2021Lelli,2023Rizzo}. 
Furthermore, an accurate measurement of the gas kinematics requires a careful modelling of the emission line datacube in order to break the degeneracy between $\sigma$ and $V_\mathrm{rot}$. 
This is particularly important when galaxies are observed at low angular resolution, as the beam smearing effect may artificially broaden the emission line, increasing $\sigma$ and decreasing $V_\mathrm{rot}$ \citep[e.g.][]{1973Warner,1989Begeman,2015Diteodoro}. 
This clearly modifies in a non-trivial way the calculation of the instability parameter. 
Unfortunately, the angular resolution of the observations is usually low for galaxies at high redshift, which are particularly interesting for studying LGI. 
A few observational works have shown that galaxies at high redshifts often host unstable gas discs \citep[e.g][]{2014DeBreuck,2014Genzel,2016Stott,2019Ubler,2022Walter,2023Liu,2024Fujimoto} and a similar result is found for starburst galaxies, which are considered their present-day analogues \citep{2021Girard,2022Fisher,2023Puschnig}.

In this paper, we use the 3D instability criterion by \cite{2023Nipoti} to understand the impact of the vertical structure of the gas disc on its stability against perturbations. 
Although our study has the limitation of neglecting the stellar disc response to the perturbation, which can be important \citep[e.g.][]{1966LinShu,1970Vandervoort_b,1984JogSolomon,2011Romeo}, it contains three significant improvements with respect to previous works. 
First, the vertical structure of the gas disc is derived, by numerical calculation, in a self-consistent way under the assumption of vertical hydrostatic equilibrium  and for a realistic galactic potential including stars, dark matter (DM), and the gas itself. 
Second, we use accurate measurements of the gas kinematics that account for the observational and instrumental effects (i.e. beam smearing, instrumental broadening of emission lines) using state-of-the-art techniques. 
Third, our sample of galaxies is the widest in terms of stellar mass and redshift coverage for which this kind of analysis is carried out in an homogeneous way and based on robust gas kinematics measured from observations\footnote{\cite{2020Romeo_a} and \cite{2023Aditya} analysed the stability of more than 100 galaxies, a sample much larger than ours. However, both authors assumed a fixed and radially constant velocity dispersion for the gas, while we use the radial profile measured from observations.}. 
The paper is structured as follows. 
Section~\ref{sec:method} explains how we model a galaxy and its mass components, the instability criterion by \cite{2023Nipoti}, and the method used to derive the vertical structure of the gas disc. 
Section~\ref{sec:sample} describes the galaxy sample and the data used for the instability analysis. 
We present and discuss our results in Sections~\ref{sec:results} and~\ref{sec:discussion}, respectively. 
Section~\ref{sec:conclusions} summarises the paper and the main conclusions. 

%%%%%%%%%%%%%%%%%%%%%%%%%%%%%%%%%%%%%%%%%%%%%%%%%%%%%%%%%%%%%%%%%%%%%%%%%%%%%%%%%%%%%%%%%%%%%%%%%%%%%%%%%%%%%%%%%%%%%%%%%

\section{Method}\label{sec:method}
In this section, we explain the procedure to obtain a self-consistent model of a gas disc in the gravitational potential of a star-forming galaxy. 
This gas disc model can then be used to apply the 3D instability criterion by \cite{2023Nipoti}.

\subsection{Galaxy model}\label{sec:method_model}
We consider the typical case of a rotating gas disc in the gravitational potential of a galaxy.  
This whole system is symmetric with respect to the rotation axis and the midplane.  
Therefore, the total gravitational potential of the galaxy is axisymmetric: $\Phi_\mathrm{tot}=\Phi_\mathrm{tot}(R,z)$, where $R$ and $z$ are the cylindrical coordinates.  
The unperturbed gas disc is assumed to stationary rotate with $V_\mathrm{rot} = V_\mathrm{rot}(R)$ and has no radial or vertical motions. 
We further assume that the gas disc is in vertical hydrostatic equilibrium in the galactic potential. 
We also take the gas pressure to be $P=\rho \sigma^2$ with $\rho$ the gas volume density. 
For simplicity, we assume that the velocity dispersion is isotropic and independent of $z$, thus $\sigma = \sigma(R)$.  
The gas vertical distribution at a given radius is then \citep[e.g.][]{1995Olling}
\begin{equation}\label{eq:rho_Rz_HE}
	\rho(R,z) = \rho(R,0) \exp{\left[ - \frac{\Phi_\mathrm{tot}(R,z)-\Phi_\mathrm{tot}(R,0)}{\sigma^2(R)} \right]} \, ,
\end{equation}
where $\rho(R,0)$ and $\Phi_\mathrm{tot}(R,0)$ are, respectively, the gas volume density and the total gravitational potential evaluated in the midplane. 
In particular, $\Phi_\mathrm{tot}$ can be calculated via numerical integration assuming a model of the mass distribution of the galaxy. 

The main mass components of a star-forming galaxy are the dark matter (DM) halo, the stellar disc (or discs), the stellar bulge (when present), and the cold gas mostly consisting of atomic gas and molecular gas, which are usually distributed in discs. 
We rely on models of the mass distribution available in the literature (see Sect.~\ref{sec:sample} for details). 
The functional forms used for each mass component are described in the following sections.

\subsubsection{Dark matter halo}\label{sec:dm_halo_model}
We model the DM density distribution using either a pseudo-isothermal halo \citep{1985Vanalbada} or a Navarro-Frenk-White (NFW) halo \citep{1996NavarroFrenkWhite} for the galaxies with $V_\mathrm{rot} \gtrsim 50$~\kms~\footnote{We anticipate that our conclusions remain unchanged if the NFW halo is adopted for all galaxies with $V_\mathrm{rot} \gtrsim 50$~\kms, as all the systems modelled with an ISO halo have $Q>1$ (see Sect.~\ref{sec:results_q2d}). }
A core-NFW (cNFW) profile \citep{2016Read_b,2016Read_c,2017Read} is chosen for the dwarf galaxies with $V_\mathrm{rot} \lesssim 50$~\kms.  
For simplicity, the DM halo is assumed spherical for all galaxies.  

The pseudo-isothermal density profile is
\begin{equation}\label{eq:rho_isohalo}
	\rho_\mathrm{DM}(r) = \rho_\mathrm{DM,0} \left( 1 + \frac{r^2}{r_\mathrm{c}^2} \right)^{-1} \, ,
\end{equation}
where $r=\sqrt{R^2+z^2}$, $\rho_{\mathrm{DM},0}$ is the central volume density and $r_\mathrm{c}$ the core radius. 

The NFW profile is
\begin{equation}\label{eq:rho_nfwhalo}
	\rho_\mathrm{DM}(r) = \rho_\mathrm{DM,0} \left(\frac{r}{r_\mathrm{s}} \right)^{-1} \left( 1 + \frac{r}{r_\mathrm{s}} \right)^{-2}\, ,
\end{equation}
where $r_\mathrm{s}$ is the scale radius. 

The cNFW profile was obtained by \cite{2016Read_b,2016Read_c} simulating the evolution of dwarf galaxies with different halo masses and having initially an NFW profile. 
In those simulations, the cusped halo gradually develops a core because the stellar feedback ``heats'' the DM by injecting energy and momentum into the gas, provoking a significant modification of the innermost gravitational potential \citep{2005Read,2014Pontzen}. 
The DM profile of the simulated dwarfs is well described by
\begin{equation}\label{eq:rho_cnfwhalo}
	\rho_\mathrm{DM}(r) = f^\eta \rho_\mathrm{NFW}(r) + \frac{\eta f^{\eta-1} \left( 1 - f^2 \right)}{ 4 \pi r^2 r_\mathrm{c} } M_\mathrm{NFW}(r) \, ,
\end{equation}
where $\rho_\mathrm{NFW}$ and $M_\mathrm{NFW}$ are, respectively, the NFW density profile (Eq.~\ref{eq:rho_nfwhalo}) and mass \citep{1996NavarroFrenkWhite}, $r_\mathrm{c}$ is the core radius, and $f = \tanh \left( r/r_\mathrm{c} \right)$ is a function than regulates the shape and the extent of the core. 

\subsubsection{Stellar components}\label{sec:stars_model}
The mass distribution of stellar discs is modelled by an exponential radial profile with scale length $R_\star$, and either a $\sech^2$ or an exponential vertical profile \citep{1981VanderKruit_a}:
\begin{equation}\label{eq:rho_disc}
	\rho_\star (R,z) = \rho_{\star,0} \exp \left( - \frac{R}{R_\star} \right) \xi(z) \, ,
\end{equation}
where $\rho_{\star,0}=\Sigma_{\star,0}/(2 z_\star)$ is the central density, and $\xi$ describes the vertical distribution, which is $\xi (z)= \sech^2 (z/z_\star)$ or $\xi (z) = \exp (- z/z_\star)$ with $z_\star$ being the scale height.

The bulges and the stellar distributions of some galaxies at high redshift (see Sect.~\ref{sec:sample} for details) are modelled using a spherical distribution that gives, when projected, a \cite{1963Sersic} surface brightness profile . 
The density profile of such spherical distribution is \citep{1997Prugniel,2005Terzic}
\begin{equation}\label{eq:rho_sersic}
	\rho_\mathrm{b}(r) =  \rho_\mathrm{b,0} \left( \frac{r}{R_\mathrm{e}} \right)^{-p_n} \exp \left[ - b_n \left( \frac{r}{R_\mathrm{e}}\right) ^{1/n} \right] \,
\end{equation}
where $R_\mathrm{e}$ is the effective radius, $n$ is the S\'{e}rsic index, $b_n \approx 2n - 1/3 + 4/(405n)$, and $p_n \approx 1 - 0.6097/n + 0.05563/n^2 $ for $0.6 < n < 10$ and $10^{-2} \leq R/R_\mathrm{e} \leq 10^3$ \citep{1999Ciotti,1999LimaNeto,2000Marquez}.
In particular, $n=1$ and $n=4$ correspond to an exponential profile and a \cite{1948DeVaucoulers} profile, respectively. 

The only exception is the MW bulge, which is modelled using the exponentially truncated power-law profile by \cite{2017McMillan}
\begin{equation}\label{eq:mw_bulge}
	\rho_\mathrm{b}(r) =  \rho_\mathrm{b,0} \left(1 +\frac{m}{r_0} \right)^\alpha \exp \left[ -\left( \frac{m}{r_\mathrm{cut}} \right)^{2} \right]
\end{equation}
where $m=\sqrt{R^2+(z/q)^2}$ with $q=0.5$, $\rho_\mathrm{b,0} = 9.8 \times 10^{10}$~$M_\odot\mathrm{kpc}^{-3}$, $\alpha=-1.8$, $r_0 = 0.075$~kpc, and $r_\mathrm{cut}=2.1$~kpc.

\subsubsection{Gas surface density}\label{sec:gas_model}
A flexible function is necessary to model the variety of gas distributions observed in galaxies. 
We thus adopt a combination of a polynomial and an exponential function:
\begin{equation}\label{eq:poly}
	\Sigma (R) = \Sigma_0 \left( 1 + \sum_{i=1}^{N=4} C_i R^i \right) \exp \left( -\frac{R}{R_\Sigma} \right) \,,
\end{equation}
where $\Sigma_0$ is the central surface density, $C_i$ are the polynomial coefficients, and $R_\Sigma$ is the scale radius. 
These parameters are obtained by fitting Eq.~\ref{eq:poly} to the radial profiles of the gas surface density measured from observations (see Sect.~\ref{sec:sample}). 
In Sect.~\ref{sec:method_3Dgas_distribution}, we describe the method use to derive the vertical distribution of the gas disc. 

\subsection{3D instability criterion}\label{sec:method_instab}
\cite{2023Nipoti} has shown that a criterion sufficient for the LGI of a gaseous disc is
\begin{equation}\label{eq:q3d_nipoti23}
	Q_\mathrm{3D} = \frac{\sqrt{\kappa^2 + \nu^2} + \sigma h_z^{-1}}{\sqrt{4\pi G \rho}} < 1\, ,
\end{equation}
where $\kappa$ is the epicycle frequency (Eq.~\ref{eq:kappa2}), $h_z$ is the gas disc thickness defined as the height of an infinitesimal vertical strip centred at $z=0$ and containing $\approx 70$\% of the mass per unit surface, and $\nu$ is a frequency related to vertical gradients of the gas pressure and density
\begin{equation}\label{eq:nu2}
	\nu^2
	\equiv \frac{\partial \rho}{\partial z} \frac{\partial P}{\partial z} \frac{1}{\rho^2}
	=\left( \frac{\sigma}{\rho} \frac{\partial \rho}{\partial z}\right)^2
	\, ,
\end{equation}
with the last equality holding under the assumptions $\sigma=\sigma(R)$ and $P=\rho \sigma^2$. 
We note that $\rho=\rho(R,z)$, $\kappa = \kappa(R)$ (as we assumed cylindrical rotation), and $h_z = h_z (R)$ by definition, thus $Q_\mathrm{3D}(R,z)$.  
In Eq.~\ref{eq:q3d_nipoti23},  we have substituted the gas sound speed $c_\mathrm{s} =  \sqrt{\partial P/\partial \rho}$ in \cite{2023Nipoti} with $\sigma$, implicitly assuming that the perturbations occur at constant $\sigma$. 
We stress that Eq.~\ref{eq:q3d_nipoti23} is a sufficient criterion for instability, hence it does not guarantee that the disc is stable where $\Qthreed>1$\footnote{Eq.~12 in \cite{2023Nipoti} provides a sufficient criterion for stability, however this is never verified in our case as $\partial \sigma / \partial z=0$ and $\partial P = \sigma^2 \partial \rho$.}. 
In particular, this criterion was derived assuming axisymmetric perturbations, but it cannot inform us about the stability of the disc against non-axisymmetric perturbations (see Sect.~\ref{sec:discussion_limitations} for discussion). 

For our gas disc in vertical hydrostatic equilibrium, Eq.~\ref{eq:q3d_nipoti23} can be evaluated once $V_\mathrm{rot}$, $\sigma$, and $\Phi_\mathrm{tot}$ are known.  
The first two ingredients, $V_\mathrm{rot}$ and $\sigma$, can be measured from emission line observations of gas tracers (see Sect.~\ref{sec:sample} for details), while $\Phi_\mathrm{tot}$ can be calculated via numerical integration of the mass models described in Sect.~\ref{sec:method_model}. 
However, to obtain a self-consistent 3D model of the gas disc, the self-gravity of the gas disc must be taken into account. 
We explain our methodology in the following section. 

\subsection{Derivation of the 3D gas distribution}\label{sec:method_3Dgas_distribution}
To calculate $\Qthreed$ using Eq.~\ref{eq:q3d_nipoti23}, we need the gas volume density as a function of $R$ and $z$, and the radial profile of the gas disc thickness. 
Under the assumptions described in Sect.~\ref{sec:method_model}, the vertical distribution of our gas disc is regulated by the hydrostatic equilibrium between the gravitational potential of the galaxy $\Phi_\mathrm{tot}$ and the gas pressure. 
Crucially, $\Phi_\mathrm{tot}$ includes not only the stellar and DM potentials, but also the gas potential $\Phi$, which depends on the 3D distribution of the gas itself. 

\textsc{Galpynamics}\footnote{\url{https://gitlab.com/iogiul/galpynamics}} \citep{2018Iorio} is an ideal tool to obtain $\rho(R,z)$ taking into account the whole potential. 
This is a Python module for galaxy dynamics that includes an iterative algorithm to account for the gas self-gravity \citep[see also][]{2008Abramova,2011Banerjee}. 
Given a set of mass components, \textsc{Galpynamics} can be used to calculate, via numerical integration, the gravitational potential of a galaxy, the 3D density distribution of a gas disc in vertical hydrostatic equilibrium in such potential, and the radial profile of gas scale height. 
The gas in the disc has a user-defined velocity dispersion, which we model using the exponential profile
\begin{equation}\label{eq:exp_vdisp}
	\sigma(R) = \sigma_{0} \exp \left( -\frac{R}{R_\sigma} \right) \, ,
\end{equation}
with $\sigma_{0}$ being the gas velocity dispersion at the galaxy centre and $R_\sigma$ a scale radius. 
These parameters are obtained by fitting Eq.~\ref{eq:exp_vdisp} to the radial profile of the gas velocity dispersion measured from observations (see Sect.~\ref{sec:sample}). 
The observed velocity dispersion is used in the rare cases when Eq.~\ref{eq:exp_vdisp} cannot describe the observed profile (e.g. non-monotonic trends). 

For a galaxy model including a DM halo, a stellar component, and a gas disc, \textsc{Galpynamics} applies the following iterative procedure. 
In a preliminary stage, the software calculates the external and fixed potential of DM and stars (i.e $\Phi_\mathrm{ext}=\Phi_\mathrm{DM}+\Phi_\star$) on a user-defined grid. 
Initially, the gas disc is assumed infinitesimally thin to estimate its gravitational potential $\Phi$ by numerical integration. 
The total galactic potential is then set to $\Phi_\mathrm{tot} = \Phi_\mathrm{ext}+\Phi$. 
A first estimate of the gas 3D distribution is obtained through Eq.~\ref{eq:rho_Rz_HE}.  
An initial guess of the gas scale height ($z_\mathrm{gas}$) is also derived by fitting the gas vertical distribution at each $R$ using either the Gaussian profile $\exp [ - z^2/(2 z_\mathrm{gas}^2 )]$ or the function $\sech^2(z/z_\mathrm{gas})$. 
The profile that is most suitable to fit the gas vertical profile is chosen as the one that minimises the residuals between the best-fit curve and $\rho(R,z)$ calculated numerically through Eq.~\ref{eq:rho_Rz_HE}. 
The resulting scale height is adopted to re-evaluate $\Phi$, which is then used to update $\Phi_\mathrm{tot}$. 
This procedure is iterated until two successive computations of the scale height differ by less than a tolerance factor chosen by the user. 
When convergence is reached, the gas volume density is calculated on the user-defined grid and the gas disc thickness is obtained as $h_z=2 z_\mathrm{gas}$. 
Thus, $h_z$ contains $\approx 68$\% and $\approx 76$\% of the mass per unit surface for the Gaussian profile and the $\sech^2$ distribution respectively, which is within the fiducial range $\approx 60-80$\% recommended by \cite{2023Nipoti}. 

In the case of present-day galaxies hosting an atomic gas disc and a molecular gas disc (see Sect.~\ref{sec:sample}), we follow the same approach as in \cite{2019Bacchini}. 
We first calculate $\rho(R,z)$ and $h_z$ for the disc of atomic gas and we then do the same for the molecular gas disc but including also the potential of the atomic gas disc using the model obtained in the previous step. 
Therefore, the distribution of the atomic gas disc is not influenced by the molecular gas. 
This choice does not significantly affect the results, as the molecular gas disc is subdominant with respect to the other mass components, at least for the present-day galaxies studied here \citep[for a thorough study, see][]{2022ManceraPina_b}. 

It is worth noting that the procedure used here to derive the 3D distribution of galactic gas discs is a standard methodology that has been applied to several nearby galaxies \citep[e.g.][but also \citealt{2011Banerjee} for a similar approach]{2018Iorio,2019Bacchini,2019Bacchini_b,2020Bacchini,2020Bacchini_b,2022ManceraPina_b}. 
As explained in the following section, our sample includes not only present-day galaxies, but also systems at high redshift for which this procedure to derive the 3D gas distribution has never been applied before. 
We extend the application of this methodology to high-z galaxies relying on the fact that the properties of their gas discs are quite similar to those of galaxies in the present-day Universe \citep[e.g. the gas kinematics is rotation-dominated, see][]{2020Rizzo,2021Rizzo,2021Lelli,2023RomanOliveira}.

%%%%%%%%%%%%%%%%%%%%%%%%%%%%%%%%%%%%%%%%%%%%%%%%%%%%%%%%%%%%%%%%%%%%%%%%%%%%%%%%%%%%%%%%%%%%%%%%%%%%%%%%%%%%%%%%%%%%%%%%%%%%%%%%%%%%%%%%%%%%%%%%%%%%%%%%%%%%%%
\section{Sample and data description}\label{sec:sample}
We aim at analysing the gas disc instability using accurate measurements of the gas kinematics that are corrected for the limited resolution of the observations. 
We carefully inspected the literature to select galaxies with 1) robust measurements of the gas kinematics corrected for beam smearing and instrumental broadening, and 2) mass models derived through the decomposition of the galaxy rotation curve in the contribution of single mass components. 
We also required that the gas disc is covered by at least three resolution elements in order to have enough points to estimate the rotation curve gradient using the central difference method.
In particular, for all the galaxies in our sample, the kinematics of the cold gas was derived using state-of-the-art 3D techniques based on the tilted-ring modelling procedure \citep[e.g.][]{1973Warner,1974Rogstad,1989Begeman}. 
In practice, the gas disc is modelled as a series of circular rings described by geometrical (e.g. inclination, position angle, kinematic centre) and kinematic (e.g. $V_\mathrm{rot}$, $\sigma$, systemic velocity, radial velocity~\footnote{Gas radial motions are negligible with respect to rotation for the galaxies in our sample.}) parameters, which are allowed to vary between the rings. 
This model is used to simulate mock observations that are convolved with the telescope beam and then fitted, channel by channel, to the observed datacube until the procedure finds the set of parameters that best reproduces the data. 
This approach allows us to take into account the beam smearing effect and obtain robust measurements of the gas kinematics \citep[e.g.][]{2015Diteodoro}. 
The rotation curve was corrected for the asymmetric drift when necessary  \citep[typically when $V_\mathrm{rot}/\sigma \lesssim 5$; see e.g.][]{2017Iorio,2023Lelli} and used to derive the parameters for the mass models described in Sect.~\ref{sec:method_model}. 
In Appendix~\ref{app:mass_models}, Table~\ref{tab:mass_models} reports the model parameters for the stellar components and the DM halo for the whole sample. 

\begin{figure}
	\includegraphics[width=1\columnwidth]{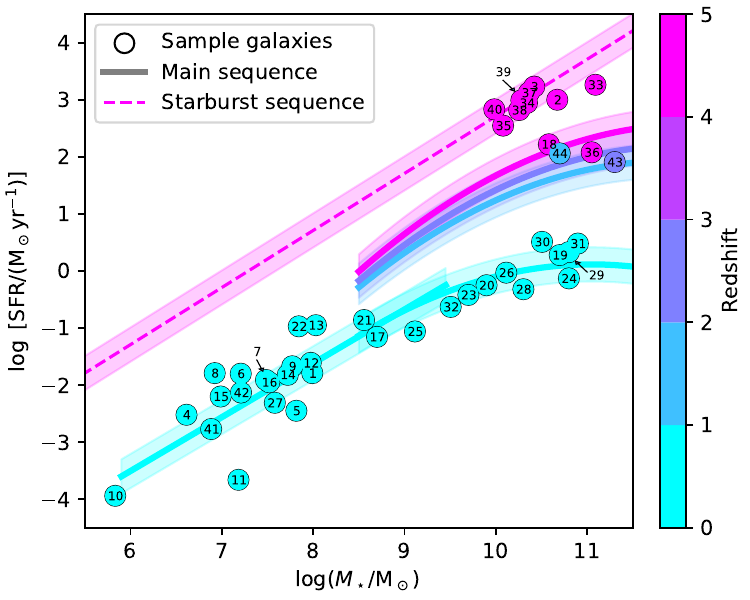}
	\caption{Stellar mass versus SFR for our sample of galaxies (circles). 
		The numbers indicate the galaxy ID reported in Table~\ref{tab:sample}. 
		The solid curves show the star-forming main sequence at different redshifts for galaxies with $8.5 \lesssim \log (M_\star/\mathrm{M}_\odot) \lesssim 11.5$ \citep{2023Popesso}. 
		The solid line is the main sequence for galaxies at $\mathrm{z}\approx 0$ with $6 \lesssim \log (M_\star/\mathrm{M}_\odot) \lesssim 9.5$ \citep{2012Berg,2022Berg}. 
		The dashed line is the starburst sequence (SBS) at $4 \lesssim \mathrm{z} \lesssim 5$ from \cite{2022Rinaldi}. 
		The shaded areas show the $1\sigma$ scatter.}
	\label{fig:Mstar_SFR}
\end{figure}

In the following, we describe the observations and the gas tracers used for each sub-sample of galaxies analysed here. 
We anticipate that the final sample consists of 44 star-forming galaxies at $0 \lesssim \mathrm{z}\lesssim 5$ hosting rotating discs of cold gas.  
This sample covers a wide range of stellar masses ($5.8 \lesssim \log (M_\star/\mathrm{M}_\odot) \lesssim 11.3$) and star formation rates (SFRs; $-3.9 \lesssim \log [\mathrm{SFR}/(\mathrm{M}_\odot \mathrm{yr}^{-1})] \lesssim 3.3$). 
Figure~\ref{fig:Mstar_SFR} shows our sample in the $M_\star$-SFR plane. 
Each galaxy is identified by an ID number, which is reported in Table~\ref{tab:sample} together with the main properties of the galaxy. 
The sub-sample at $\mathrm{z} \approx 0$ is representative of galaxies on the star-forming main sequence (SFMS) in the present-day Universe, from dwarfs to spirals. 
The sub-sample at high redshift is biased towards high stellar masses, but it contains both galaxies on the SFMS and starbursts. 

We note that the gas distribution and kinematics were measured from observations with spatial and spectral (i.e. velocity) resolutions that are not homogeneous across the whole sample. 
Moreover, we use different tracers of the cold gas. 
For both local and high-z galaxies, we use carbon monoxide (CO) to probe the molecular gas, which is typically dense and prevalently distributed in clouds \citep[e.g.][]{2013Bolatto}. 
In addition, we use tracers of diffuse gas phases: the 21-cm emission line of atomic hydrogen (HI) for galaxies at $\mathrm{z}\approx 0$ and the 158~$\mu$m emission line of the ionised carbon ([CII]) for high-z systems. 
This latter is one of the brightest fine-structure line in high-z star-forming galaxies \citep{2018Lagache} and traces multiple gas phases, mostly molecular and neutral gas \citep{2013Pineda,2013Carilli,2015Gullberg,2018Zanella,2021Tarantino,2021RamosPadilla,2021Heintz,2022Vizgan_a,2022Vizgan_c,2022Wolfire}. 
Given the limited redshift-coverage of HI observations and the small spatial extent of CO emission, [CII] is currently considered the best tracer of the cold gas kinematics in high-z galaxies ($\mathrm{z} \gtrsim 3.5$) over large areas of their discs \citep[e.g.][]{2020Neeleman,2020Rizzo,2021Rizzo,2021Fraternali,2021Lelli,2021Zanella,2023RomanOliveira}. 

\renewcommand{\arraystretch}{1.}
\setlength{\tabcolsep}{2.pt}
\begin{table}
	\centering
	\caption{Properties of the galaxies in the sample. 
		The top and bottom parts of the table are for the low and high redshift samples, respectively. 
		The columns report: ID number in Fig.~\ref{fig:Mstar_SFR}, galaxy name, morphological type (for nearby galaxies) or classification based on position on either the SFMS or the SBS (for high-z galaxies), distance (for nearby galaxies) or redshift (for high-z systems), disc inclination in the sky plane, stellar mass, SFR, and cold gas tracer(s).}
	\label{tab:sample}
	\begin{tabular}{cccccccc}
		\hline\hline
		ID 	& Galaxy 	& Type 	& D 	& $i$	& $\log M_\star$	& $\log$SFR 							& Gas tracer \\
		%		ID & Galaxy 	& Type 	& $\frac{\mathrm{D}}{\mathrm{Mpc}}$ & $\log \left( \frac{M_\star}{\mathrm{M}_\odot}\right)$	&  $\log \left( \frac{\mathrm{SFR}}{\mathrm{M}_\odot/\mathrm{yr}}\right)$ \\
			&        	&      	& Mpc	& deg	& $\mathrm{M}_\odot$&$\mathrm{M}_\odot \mathrm{yr}^{-1}$    &             \\
		\hline
		1	&AGC114905	&UDG	&78.27	& 31.5	& 7.95				&-1.79									& HI \\
		4	&CVIDWA		&dIrr	&3.6	& 49.2	& 6.61				&-2.52									& HI  \\
		5	&DDO101		&dIrr	&6.4	& 52.4	& 7.82				&-2.45									& HI \\
		6	&DDO126		&dIrr	&4.9	& 62.2	& 7.21				&-1.80									& HI \\
		7	&DDO133		&dIrr	&3.5	& 38.9	& 7.48				&-1.91									& HI  \\
		8	&DDO154		&dIrr	&3.7	& 67.9	& 6.92				&-1.79									& HI  \\
		9	&DDO168		&dIrr	&4.3	& 62.0	& 7.77				&-1.67									& HI \\
		10	&DDO210		&dIrr	&0.9	& 63.2	& 5.83				&-3.94									& HI \\
		11	&DDO216		&dIrr	&1.1	& 70.0	& 7.18				&-3.66									& HI  \\
		12	&DDO47		&dIrr	&5.3	& 37.4	& 7.97				&-1.61									& HI  \\
		13	&DDO50		&dIrr	&3.4	& 33.1	& 8.03				&-0.95									& HI \\
		14	&DDO52		&dIrr	&10.3	& 55.1	& 7.72				&-1.82									& HI  \\
		15	&DDO53		&dIrr	&3.6	& 37.0	& 6.99				&-2.20									& HI \\
		16	&DDO87		&dIrr	&7.4	& 42.7	& 7.52				&-1.95									& HI  \\
		17	&IC2574		&Sm		&3.9	& 66.4	& 8.70				&-1.15									& HI  \\
		19	&MW			&SBbc	&0.0	& - 	& 10.70				&0.28									& HI \\
		20	&NGC0925	&SABd	&9.2	& 58.0	& 9.90				&-0.25 									& HI, CO \\
		21	&NGC1569	&dIrr	&3.4	& 67.0	& 8.56				&-0.86 									& HI \\
		22	&NGC2366	&dIrr	&3.4	& 65.1	& 7.84				&-0.97									& HI \\
		23	&NGC2403	&Scd	&3.2	& 61.0	& 9.70				&-0.42									& HI, CO  \\
		24	&NGC2841	&Sb		&14.1	& 73.7	& 10.80				&-0.13									& HI  \\
		25	&NGC2976	&Sc		&3.6	& 61.0	& 9.11				&-1.06									& HI, CO \\
		26	&NGC3198	&Sc		&13.8	& 71.5	& 10.11				&-0.03									& HI, CO \\
		27	&NGC3741	&Sm		&3.2	& 68.0	& 7.58				&-2.31									& HI \\
		28	&NGC4736	&SABa	&4.7	& 41.4	& 10.30				&-0.32									& HI, CO \\
		29	&NGC5055	&Sbc	&9.9	& 55.0	& 10.80				&0.33									& HI, CO \\
		30	&NGC6946	&Scd	&5.5	& 33.0	& 10.51				&0.51									& HI, CO \\
		31	&NGC7331	&Sb		&14.7	& 75.8	& 10.90				&0.48									& HI, CO  \\
		32	&NGC7793	&Sd		&3.6	& 47.0	& 9.51				&-0.63									& HI  \\
		41	&UGC8508	&dIrr	&2.6	& 67.6	& 6.88				&-2.77									& HI  \\
		42	&WLM		&dIrr	&1.0	& 74.0	& 7.21				&-2.13									& HI  \\
		\hline
		\hline
		ID 	& Galaxy 	& Type 	& $z$	& $i$	& $\log M_\star$ 	& $\log$SFR 							& Gas tracer \\
			&        	&      	& 		& deg & $\mathrm{M}_\odot$&$\mathrm{M}_\odot \mathrm{yr}^{-1}$    &             \\
		\hline
		2 &ALESS073.1	&SBS	&4.75	& 22.1	& 10.67				&3.00									& [CII] \\		
		3 &BRI1335-0417	&SBS	&4.41	& 42.0	& 10.42				&3.23									& [CII] \\
		18&J81740		&SFMS	&4.26	& 43.0	& 10.58				&2.22									& [CII] \\
		33&SGP38326-1	&SBS	&4.42	& 42.0	& 11.09				&3.26									& [CII] \\
		34&SGP38326-2	&SBS	&4.43	& 41.0	& 10.34				&2.95									& [CII] \\
		35&SPT0113-46	&SFMS	&4.23	& 70.0	& 11.05				&2.08									& [CII]  \\
		36&SPT0345-47	&SBS	&4.30	& 53.0	& 10.36				&3.12									& [CII]  \\
		37&SPT0418-47	&SBS	&4.23	& 54.0	& 10.08				&2.55									& [CII]  \\
		38&SPT0441-46	&SBS	&4.48	& 57.0	& 10.26				&2.82									& [CII]  \\
		39&SPT2132-58	&SBS	&4.57	& 52.0	& 10.28				&2.99									& [CII]  \\
		40&SPT2146-55	&SBS	&4.77	& 47.0	& 9.98				&2.83									& [CII]  \\
		43&zC400569		&SFMS	&2.24	& 54.6	& 11.30				&1.91									& CO  \\
		44&zC488879		&SFMS	&1.47	& 71.5	& 10.70				&2.06									& CO  \\
		\hline
	\end{tabular}
\end{table}

\subsection{Galaxies at $\mathrm{z}\approx 0$}
We collected the data for 31 galaxies in the present-day Universe consisting of the following sub-samples. 

\paragraph{Eleven spiral galaxies.}
The kinematics of the atomic gas and the molecular gas were derived by \cite{2019Bacchini} and \cite{2020Bacchini} using 21-cm emission line datacubes from The HI Nearby Galaxy Survey \citep[THINGS,][]{2008Walter} and the CO(2-1) emission line datacubes from the HERA CO-Line Extragalactic Survey \citep[HERACLES,][]{2005Leroy}, respectively. 
The 21-cm datacubes have channel separation 2.6~\kms~$\leq \Delta v \leq$~5.2~\kms and were smoothed to a common spatial resolution of $\approx 400$~pc \citep[see][]{2019Bacchini}. 
The CO datacubes have $\Delta v = 5.2$~\kms and angular resolution of 13~\arcsec. 
The gas kinematics was analysed using \barolo\footnote{\url{http://editeodoro.github.io/Bbarolo/}}, a software designed to perform a 3D tilted-ring modeling of emission line datacubes of galaxies \citep{2015Diteodoro}. 
The atomic gas surface densities are from \cite{2008Leroy} and \cite{2010Bigiel}, while the molecular gas surface density is from \cite{2016Frank}.  
This is derived using the CO-H$_2$ conversion factor ($\alpha_\mathrm{CO}$) measured by \cite{2013Sandstrom}, who took account of the dust-to-gas ratio and the metallicity gradient to accurately estimate $\alpha_\mathrm{CO}$. 
The mass models (Table~\ref{tab:mass_models}) were taken from \cite{2008deBlok} for all the galaxies except NGC~7793, for which we use the model revised by \cite{2019Bacchini}. 
The mass components consist of an exponential stellar disc, a bulge (if present), an NFW or pseudo-isothermal DM halo, an atomic gas disc, and a molecular gas disc (for the galaxies where CO emission is detected). 
The stellar distribution is constrained by the 3.6~$\mu$m emission, which traces the old stellar populations containing the bulk of the stellar mass \citep[e.g.][]{2014McGaugh,2019Schombert}. 
This latter is calculated using the mass-to-light ratio ($M/L$) obtained by fitting the HI rotation curve for each galaxy \citep[see][]{2008deBlok}. 
The SFRs in Table~\ref{tab:sample} were derived by \cite{2008Leroy} combining far-ultraviolet (FUV) emission and 2.4~$\mu$m emission, tracing unobscured and dust obscured star formation, respectively. 
	
\paragraph{The Milky Way (MW).} 
The atomic gas and molecular gas distribution and kinematics were derived by \cite{2017Marasco} by modelling HI and CO emission-line datacubes from the Leiden-Argentine-Bonn (LAB) survey \citep{2005Kalberla} and the CO(1-0) survey \citep{2001Dame}, respectively. 
The molecular gas surface density is obtained using $\alpha_\mathrm{CO}=4.3$~\msun(K \kms pc$^2$)$^{-1}$ recommended by \cite{2013Bolatto} . 
The LAB data have angular resolution of $\approx 0.6$\textdegree and $\Delta v \approx 2$~\kms, while the CO data have angular resolution of $\approx 0.15$\textdegree and $\Delta v \approx 1.3$~\kms. 
\cite{2017Marasco} developed a methodology based on the same assumptions as the tilted-ring technique to reproduce these observations. 
The mass models for the stellar and DM components of the MW were taken from \cite{2017McMillan} and consist of two exponential stellar discs with different thickness (thin and thick disc; see Table~\ref{tab:mass_models}), the bulge (Eq.~\ref{eq:mw_bulge}), and an NFW DM halo. 
The stellar distribution is characterized by \cite{2011McMillan} combining kinematics tracers with near-infrared (NIR) and optical data \citep[see][]{2002Bissantz,2008Juric}. 
The SFR in Table~\ref{tab:sample} was obtained by \cite{2019Bacchini_b} averaging literature values based on different star formation tracers. 

\paragraph{Eighteen dwarf irregular galaxies.}
This sub-sample mostly consists in 17 galaxies that are part of the Local Irregulars That Trace Luminosity Extremes, The HI Nearby Galaxy Survey \citep[LITTLE THINGS;][]{2012Hunter}. 
The distribution and kinematics of the atomic gas in the LITTLE THINGS sample was derived by \cite{2017Iorio} using \barolo, while the mass models in Table~\ref{tab:mass_models} were taken from \cite{2017Read}. 
The asymmetric drift correction is applied to the whole sample \citep[see][]{2017Iorio}. 
The eighteenth galaxy is NGC~3741: the atomic gas distribution and kinematics was derived by \cite{2022Annibali} using \barolo, while the mass model in Table~\ref{tab:mass_models} were taken from \cite{2017Allaert}. 
The 21-cm datacubes of the LITTLE THINGS sample and NGC~3741 have channel separation 1.3~\kms~$\leq \Delta v \leq$~2.6~\kms and spatial resolution $\approx 100-400$~pc.
The mass models include an exponential stellar disc, a cNFW DM halo, and the atomic gas disc. 
The stellar surface density was derived by fitting the spectral energy distribution (SED) for the LITTLE THINGS sample \citep[see][]{2012Zhang} and by converting the 3.6~$\mu$m surface brightness profile using the $M/L$ obtained by fitting the HI rotation curve for NGC~3741 \citep{2017Allaert}. 
The SFRs were derived from FUV images by \cite{2004Hunter} and \cite{2020Bacchini} for the LITTLE THINGS sample, and by \cite{2013Karachentsev} for NGC~3741. 

\paragraph{An HI-rich ultra-diffuse galaxy, AGC114509.}
The atomic gas distribution and kinematics were derived by \cite{2024ManceraPina} using \barolo. 
With respect to their previous work \citep{2022ManceraPina}, the authors used improved measurements of the galaxy inclination and stellar mass distribution obtained with ultra-deep optical observations. 
We use the data from \cite{2024ManceraPina} for our fiducial model, but we also perform the analysis using the data from \cite{2022ManceraPina} for comparisons with previous works (see Sect.~\ref{sec:discussion_literature}).
The HI datacube has spatial resolution of $\approx 3$~kpc and $\Delta v \approx 3.4$~\kms. 
We adopt the mass model by \cite{2024ManceraPina}. 
This includes the atomic gas disc, which is the dominant baryonic component, and a stellar disc modelled using Eq.~\ref{eq:poly}. 
The stellar surface density is obtained from the optical emission assuming $M/L$-color relations calibrated for UDGs \citep[][]{2020Du}. 
This mass model also includes a cNFW halo with $\eta = 1$ (see Table~\ref{tab:mass_models}). 
We anticipate that AGC114509 is particularly interesting for our scope, as its low DM content is expected to make this system particularly prone to LGI \citep[see discussions in][]{2022ManceraPina,2022Sellwood,2024ManceraPina}.
The SFR in Table~\ref{tab:sample} was derived by \cite{2020Durbala} from near-ultraviolet (NUV) emission.

\subsection{Galaxies at $1 \lesssim \mathrm{z} \lesssim 5$.}
We collected the data for 13 galaxies at $\mathrm{z}\gtrsim 1$, consisting in the following three sub-samples. 

\paragraph{Two galaxies at cosmic noon.}
These are the main-sequence star-forming galaxies zC-400569 at $\mathrm{z}\approx 2.24$ and zC-488879 $\mathrm{z}\approx 1.47$. 
The molecular gas kinematics and the mass models in Table~\ref{tab:mass_models} were derived by \cite{2023Lelli} using \barolo on CO emission line datacubes, specifically CO(3-2) and CO(4-3) lines for zC-400569, and CO(2-1) and CO(3-2) lines for zC-488879. 
The CO surface brightness was converted into molecular gas surface density using values of $\alpha_\mathrm{CO}$ obtained by fitting the CO rotation curve \citep{2023Lelli}. 
The velocity resolution of the datacubes is about 30~\kms and the spatial resolution of these observations is 3-4~kpc. 
As discussed in \cite{2023Lelli}, the measurements of the gas velocity dispersion may be uncertain due to the limited resolution and sensitivity of the observations. 
To be conservative, we assume the fiducial upper limits of 15\kms recommended by \cite{2023Lelli}. 
The mass models include an exponential stellar disc, a \cite{1948DeVaucoulers} bulge, the molecular gas disc, and a NFW DM halo. 
\cite{2023Lelli} constrained the stellar distribution using the rest-frame optical emission, which was converted into mass using the $M/L$ obtained by fitting the CO rotation curve. 
\cite{2019Liu_D} obtained the SFRs in Table~\ref{tab:sample} by fitting the galaxy SED.

\paragraph{Six gravitationally lensed galaxies at $4 \lesssim z \lesssim 5$.} 
These are dusty star-forming galaxies (DSFG) with SFRs typical of starburst or main-sequence galaxies. 
The gas distribution and kinematics were measured from [CII] emission-line datacubes by \cite{2020Rizzo} and \cite{2021Rizzo}, who also derived the mass models (Table~\ref{tab:mass_models}) and the SFRs (Table~\ref{tab:sample}). 
The spatial resolution of the observations for these lensed systems ranges from $\approx 200-300$~pc to about 1~kpc, and the datacubes have $\Delta v \approx 30$~\kms. 
\cite{2020Rizzo,2021Rizzo} used a Bayesian method that reconstructs simultaneously the mass distribution of the lens and the kinematics of the source \citep[see][]{2009Vegetti,2018Rizzo}. 
Their approach includes a modified version of \barolo in which the radial profiles of $V_\mathrm{rot}$ and $\sigma$ were modelled by functional forms described by a set of free parameters, rather than being determined in a non-parametric form like in the classical tilted-ring modelling. 
The mass models of these galaxies include an exponential disc for the gas, an NFW halo for the DM, and a spherical S\'{e}rsic distribution for the stellar component. 
This choice for the stellar distribution was motivated by the lack of spatially resolved observations of the rest-frame optical or near-infrared (NIR) emission for these sources, which hampered disentangling the bulge and disc distribution. 
Both the stellar mass and the [CII]-to-gas mass conversion factor ($\alpha_\mathrm{[CII]}$) to convert [CII] surface brightness into gas surface density were derived by fitting the [CII] rotation curve. 

\paragraph{Five galaxies at $4 \lesssim \mathrm{z} \lesssim 5$.} 
Among these, we have the starburst galaxy ALESS073.1, for which the gas kinematics was derived by \cite{2021Lelli} using \barolo on [CII] emission line datacubes. 
The data have spatial resolution and velocity resolution of about 700~pc and 28~\kms, respectively.  
\cite{2021Lelli} also derived the mass models, which include an exponential stellar disc, a \cite{1948DeVaucoulers} bulge, a gas disc, and a NFW DM halo (see Table~\ref{tab:mass_models}). 
This galaxy is an interesting case of study as it seems potentially subject to LGI \citep{2021Lelli}. 
The other four galaxies in this sub-sample were taken from \cite{2023RomanOliveira} and consist of a star-forming galaxy (J81740) and three starbursts (BRI1335-0417, SGP38326-1, and  SGP38326-2). 
The gas kinematics was derived by \cite{2023RomanOliveira} using \barolo on [CII] emission line observations. 
These data have spatial resolution of about 1-2~kpc and velocity resolution of 15~\kms for BRI1335-0417 and 26~\kms for the other three galaxies. 
The mass models were taken from \cite{2024RomanOliveira} and include an exponential disc for the gas, an NFW halo for the DM, and a spherical S\'{e}rsic distribution for the stellar component (see Table~\ref{tab:mass_models}). 
We note that, spatially resolved observations of the rest-frame optical/NIR emission are not available for this sub-sample, as in the case of the lensed galaxies described in the previous paragraph.  
For this reason, \cite{2024RomanOliveira} assumed a S\'{e}rsic distribution to model the stellar component, while \cite{2021Lelli} used the rest-frame NUV emission and dust continuum to constrain the bulge and disc distribution, respectively. 
The [CII] surface brightness profile was used to constrain the gas radial distribution, while the gas mass is obtained from the CO luminosity multiplied by a normalisation factor derived by fitting the rotation curve \citep[this normalisation factor includes both $\alpha_\mathrm{CO}$ and CO line ratios, see ][]{2021Lelli,2024RomanOliveira}. 
The SFRs in Table~\ref{tab:sample} were derived from warm dust continuum emission \citep[see][]{2009Coppin,2016Oteo,2020Neeleman,2023Tsukui}.

Clearly, the lack of spatially resolved observations of the stellar component and the loose constraints on the value of $\alpha_\mathrm{CO}$ and $\alpha_\mathrm{[CII]}$ make our analysis on the high-z sample more uncertain than the results for the galaxies in the present-day Universe. 
In Sect.~\ref{sec:discussion_limitations}, these issues are discussed in detail. 
We anticipate that our analysis represents an improvement with respect to previous studies of LGI in galaxies at $\mathrm{z}>0$, which typically rely on some assumption to fix the stellar and gas masses (see also Sect.~\ref{ap:Qformulaccia} for further discussion). 
We also note that the stellar and gas masses used here have been independently confirmed by other works \citep[see Sect.~\ref{sec:discussion_limitations} and discussions in][]{2021Rizzo}. 

%%%%%%%%%%%%%%%%%%%%%%%%%%%%%%%%%%%%%%%%%%%%%%%%%%%%%%%%%%%%%%%%%%%%%%%%%%%%%%%%%%%%%%%%%%%%%%%%%%%%%%%%%%%%%%%%%%%%%%%%%%%%%%%%%%%%%%%%%%%%%
\section{Results}\label{sec:results}
This section presents the main results of this work: the radial profiles of the gas disc scale heights (Sect.~\ref{sec:results_hgas}) and of $Q(R)$ (Sect.~\ref{sec:results_q2d}), and the maps of $\Qthreed(R,z)$ (Sect.~\ref{sec:results_q3d}). 

\subsection{Gas scale heights in local and high-z galaxies}\label{sec:results_hgas}
\begin{figure*}
	\includegraphics[width=2\columnwidth]{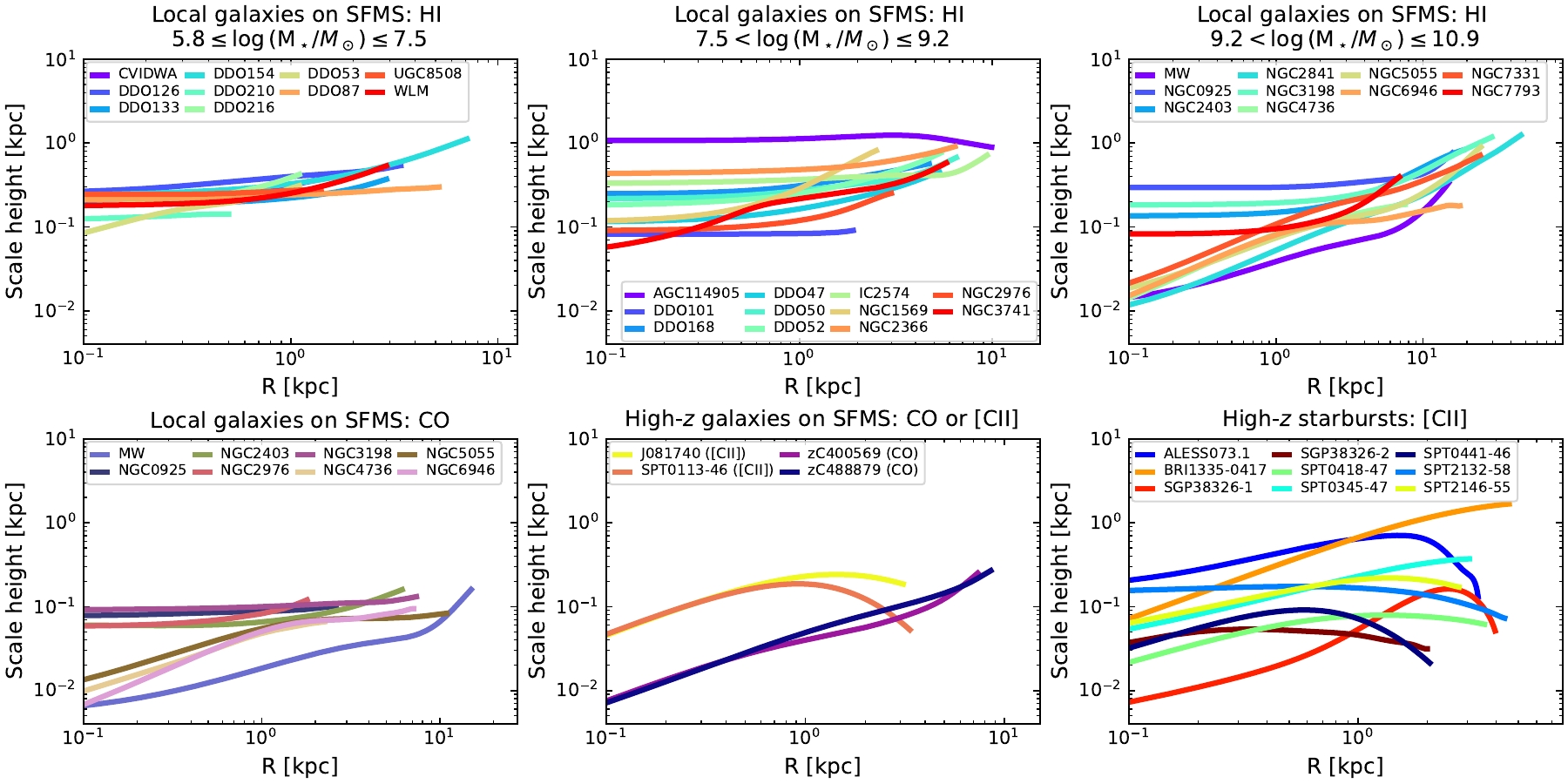}
	\caption{Radial profiles of the gas disc scale heights in the sample of galaxies. 
		The top panels are for the atomic gas discs in present-day Universe divided in bins of increasing stellar mass from left to right. 
		The bottom left panel is for the molecular gas discs at $\mathrm{z} \approx 0$. 
		The central and right panels in the bottom row are for gas discs, traced by either CO or [CII] emission lines, in high-z galaxies on the SFMS and the SBS, respectively.}
	\label{fig:h_all}
\end{figure*}
Figure~\ref{fig:h_all} shows the scale height profiles $z_\mathrm{gas}$ of the gas discs for galaxy in the sample (see Figs.~\ref{fig:h_Q_Q3D} for the scale height profiles plotted in linear scale). 
The top panels are for the atomic gas discs in the present-day Universe, while the bottom panels are for the molecular gas discs in nearby spirals (left) and for cold gas discs of SFMS galaxies (center) and starbursts (right) at high redshift. 
We note that the scale heights of nearby spirals (including the MW) and of some dwarfs were already displayed in previous works \citep{2019Bacchini,2019Bacchini_b,2020Bacchini_b,2022ManceraPina_b}, but this is the first time that the radial profiles of the gas disc scale heights are shown for the high-z sample. 
The scale heights of the atomic gas discs in present-day galaxies (top panels in Fig.~\ref{fig:h_all}) tend to be a few hundreds parsecs in the inner regions and then increase up to about 1~kpc in the outskirts, with bulge hosts having very thin discs in their innermost regions because the stellar mass is centrally concentrated. 
The same happens for the molecular gas discs at $\mathrm{z} \approx 0$ (bottom right panel in Fig.~\ref{fig:h_all}), which tend to be thinner than the atomic gas discs due to the lower velocity dispersion of the molecular gas \citep{2020Bacchini_b}. 
As pointed out by \cite{2022ManceraPina_b}, despite the radial profiles of the atomic and molecular gas scale heights share some similarities, there is little evidence for the "universal" profile proposed by \cite{2019Patra,2020Patra,2020Patra_b}. 
In these works, the author postulated the existence of a universal linear profile for the gas scale height when normalised to a characteristic scale. 
However, the precise shape of $z_\mathrm{gas}(R)$ depends not only on the relative contribution of the stellar disc and the DM halo at a given radius, but also on the bulge contribution and, most importantly, on the gas velocity dispersion \citep[for futher discussion, see][]{2022ManceraPina_b}. 
All these factors can be significantly different from galaxy to galaxy, creating a variety of profiles that is difficult to describe with a simple functional form.

The galaxies at cosmic noon (zC400569 and zC488879) host flaring molecular gas discs very similar to those of present-day galaxies (see bottom central panel in Fig.~\ref{fig:h_all}). 
We note that, since the CO velocity dispersion in these systems is an upper limit \citep[which may be itself uncertain, see][]{2023Lelli}, the molecular gas scale height may be overestimated. 
Nevertheless, the scale heights of zC400569 and zC488879 in Fig.~\ref{fig:h_all} are very similar to those of molecular gas discs in present-day galaxies, which were obtained using $\sigma$ measured from high resolution observations. 
In the case of gas discs observed via [CII] emission (bottom central and right panels in Fig.~\ref{fig:h_all}), the scale heights tend to be similar, on average, to those of present-day galaxies, which is consistent with the idea that the [CII] emission can trace both the atomic and molecular gas \citep[e.g.][]{2013Carilli,2018Zanella,2022Vizgan_c}. 
We do not find striking differences between the scale heights of main-sequence and starburst galaxies at high redshift. 
In the latter, the strong supernova feedback is expected to enhance the gas turbulence with respect to moderately star-forming galaxies, increasing the gas disc thickness. 
The lack of a dependence of the gas scale heights on the SFR is not straightforward to interpret, as the gas scale height depends not only on the gas velocity dispersion, but also on the galaxy potential. 
We note though that BRI1135-0417 is one of the galaxies with the highest SFR (see Table~\ref{tab:sample}) and hosts the thickest gas disc ($\approx 2$~kpc). 
Overall, these results indicate that the scale height of gas discs at massive high-z galaxies is of the order of 100~pc. 

For the majority of the gas discs, the scale height profiles increase monotonically for increasing galactocentric distance, indicating a flaring. 
However, the scale heights of the UDG (AGC114905) and nine high-z galaxies increase with $R$ in the central parts up to a peak value and decreases outward. 
The non-monotonic behaviour in the outer parts can be understood from Eq.~\ref{eq:rho_Rz_HE}: the scale height drops when $\sigma^2$ decreases more rapidly with $R$ than the difference between the gravitational potential in the midplane and at $|z|>0$. 
In practice, the gas pressure is not sufficient to support the flaring and the radial profile of the gas scale height starts decreasing. 
Typically, the gas scale height does not increase monotonically outwards in the galaxies in which the contribution of the DM halo to the galactic potential is subdominant with respect to the baryonic component \citep[see \S~4.6.2 in][]{2019TheBook_CFN}. 

\cite{2023RomanOliveira} performed some experiments to obtain a tentative estimate of the gas disc scale height in J81740, SGP38326-1, SGP38326-2, and BRI1135-0417 using the Python routine \textsc{Cannubi}\footnote{\url{https://www.filippofraternali.com/cannubi}}. 
This was designed to estimate geometrical parameters of a rotating disc performing an iterative 3D modelling of the observations using \barolo. 
\cite{2023RomanOliveira} report scale heights of the order of 1~kpc, while we obtain $z_\mathrm{gas} \lesssim 0.2$~kpc using \textsc{Galpynamics}. 
The only exception is BRI1135-0417: we find $z_\mathrm{gas} \approx 1-1.5$~kpc, which is close to the values reported by \cite{2023RomanOliveira}. 
The origin of the discrepancy between the scale heights for J81740, SGP38326-1, and SGP38326-2 is not easy to understand, as the methods used are not fully comparable: \textsc{Cannubi} assumes that the thickness is constant within the disc, while \textsc{Galpynamics} derives a radial profile using a dynamical modelling based on the vertical hydrostatic equilibrium. 
As pointed out by \cite{2023RomanOliveira}, further tests are needed to understand how to break the degeneracy between thickness and inclination of the observed discs, in particular when the spatial scale resolved by the data is much larger than the gas scale height (as in the case of these systems). 
We therefore decided to use the profile of $z_\mathrm{gas}$ obtained with \textsc{Galpynamics} in our analysis and leave further investigations for future works. 

\subsection{2D instability analysis}\label{sec:results_q2d}
\begin{figure*}
	\includegraphics[width=2\columnwidth]{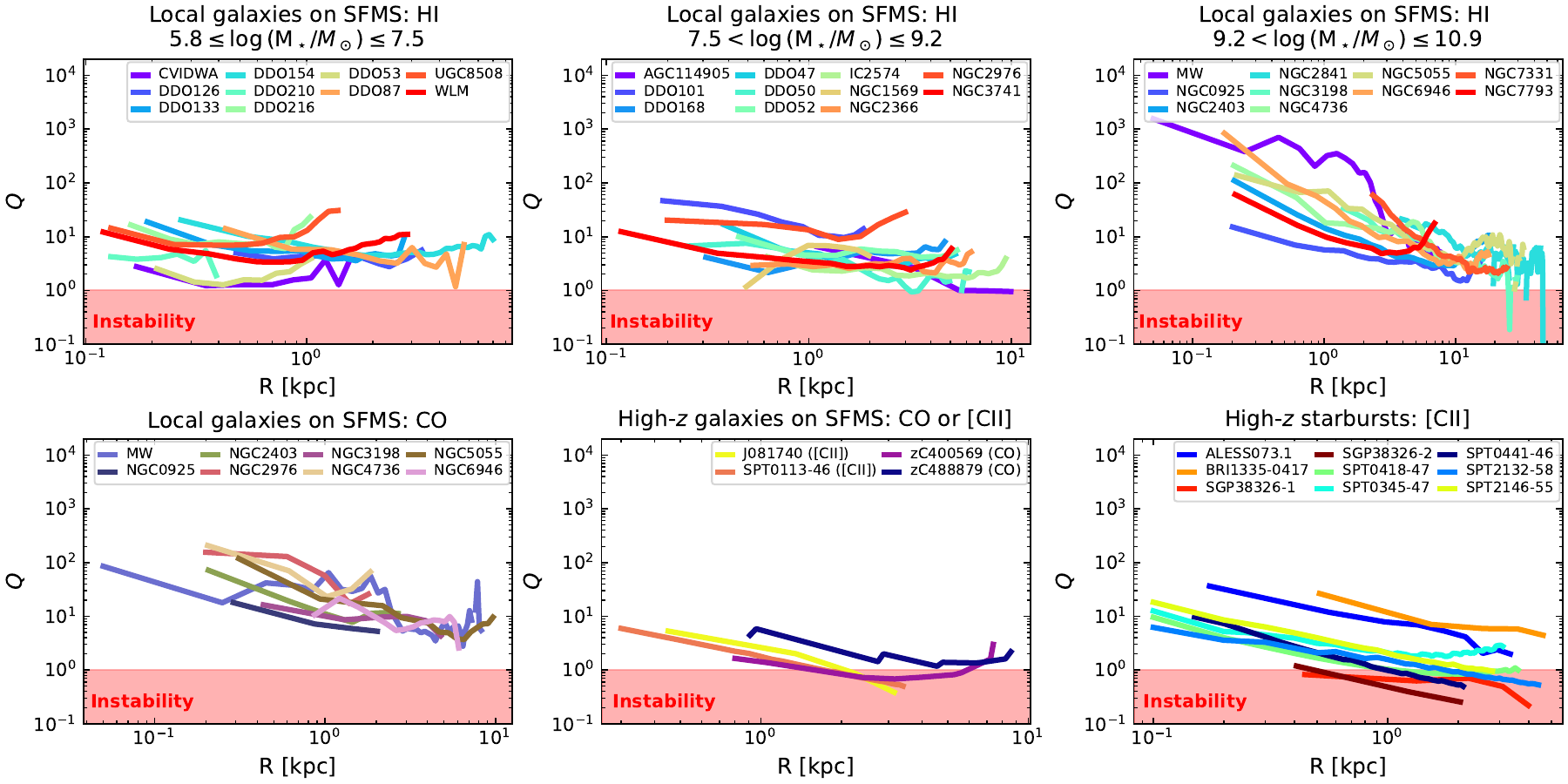}
	\caption{Same as Fig.~\ref{fig:h_all}, but for the radial profiles of the 2D instability parameter $Q$ (Eq.~\ref{eq:q2d}). }
\label{fig:q2d_all}
\end{figure*}
We derive the radial profiles of $Q$ (Eq.~\ref{eq:q2d}) using $V_\mathrm{rot}$, $\sigma$, and $\Sigma$~\footnote{The atomic gas surface density of galaxies at $\mathrm{z}\approx 0$ includes a multiplicative factor of 1.36 to account for helium.} measured from observations, searching for gas discs with $Q<1$ in some radial interval. 
The resulting profiles are shown in Fig.~\ref{fig:q2d_all} (see also Figs.~\ref{fig:h_Q_Q3D} for $Q(R)$ shown in linear scale): the top panels are for atomic gas discs in present-day galaxies, while the bottom panels are for molecular gas discs in local spirals (left) and for high-z galaxies on the SFMS (center) and the SBS (right). 
If we classify as potentially unstable only galaxies with $Q < 1$ in a region larger than the resolution element of the observations, we find that 12 galaxies host potentially unstable gas discs according to the 2D stability criterion.
However, the unstable regions in DDO~50 and NGC~3198 (see Figs.~\ref{fig:h_Q_Q3D}) are most likely due to fluctuations in the observed profiles rather than a real feature of the disc. 
Indeed, the uncertainties on observables ($\sigma$ in particular) are large in these cases (see also Sect.~\ref{sec:discussion_limitations}).  
We thus conclude that, at $\mathrm{z}\approx 0$, only the atomic gas disc of the UDG AGC114905 can be classified as locally unstable across a significantly large region disc. 
Unsurprisingly, the atomic gas disc of AGC114905 is also unstable if we use the data from \cite{2022ManceraPina}, who measure lower $\sigma$ with respect to \cite{2024ManceraPina}. 
For AGC114905, the radial extent of the unstable region where $Q \lesssim 1$ is $L_Q \approx 4$~kpc (see Table~\ref{tab:gas_inst}). 
In the high-z sample, we find that nine out of 13 galaxies (i.e. $\approx 70$\%) have $Q < 1$ beyond a certain galactocentric distance, while only ALESS073.1, BRI1335-0417, SPT0345-47, and zC488879 have no unstable region. 
Typically, the radial extent of the unstable regions ranges from $\sim 1$~kpc to a few kpc when the 2D instability criterion is adopted. 

We estimate the fraction of the unstable gas as 
\begin{equation}\label{eq:finst}
	f_\mathrm{inst} \equiv \frac{M_\mathrm{inst}}{M_\mathrm{gas}} \, ,
\end{equation}
where $M_\mathrm{gas}$ is the total mass of the gaseous disc and $M_\mathrm{inst}$ is the mass of the unstable gas obtained by integrating the gas surface density in the region where $Q(R)<1$. 
We find a broad range of mass fractions ($0.1 \lesssim f_\mathrm{inst} \lesssim 1.0$) with median and $1\sigma$ uncertainty of $\langle f_\mathrm{inst} \rangle = 0.6^{+0.2}_{-0.3}$. 
Table~\ref{tab:gas_inst} reports the total gas mass, the properties of the unstable regions (mass and radial extent), and the fraction of unstable gas for the galaxies with $Q<1$. 
\setlength{\tabcolsep}{3pt}
\begin{table}
\centering
\caption{Properties of galaxies with unstable regions. }
\label{tab:gas_inst}
\begin{tabular}{l|c|cc|cc|cc}
\hline\hline
Galaxy		& $\log M_\mathrm{gas}$		& \multicolumn{2}{c|}{$\log M_\mathrm{inst}$}	& \multicolumn{2}{c|}{$f_\mathrm{inst}$}& \multicolumn{2}{c}{$L$}	\\
%Galaxy		& $\log \frac{M_\mathrm{gas}}{M_\odot}$		& \multicolumn{2}{c|}{$\log \frac{M_\mathrm{inst}}{M_\odot}$}		& \multicolumn{2}{c|}{$f_\mathrm{inst}$}& \multicolumn{2}{c}{$L$/kpc}	\\
			& $\mathrm{M}_\odot$		& \multicolumn{2}{c|}{$\mathrm{M}_\odot$}		& \multicolumn{2}{c|}{}					& \multicolumn{2}{c}{kpc}	\\
			& 							& 2D		& 3D								& 2D		& 3D						& 2D	& 3D				\\
\hline
AGC114905	& 9.1						& 8.8		& 0.0								& 0.5		& 0.0						& 3.7	& 0.0				\\
J81740		& 10.6						& 10.2		& 9.8								& 0.4		& 0.2						& 1.0	& 0.9				\\
SGP38326-1	& 11.4						& 11.4		& 11.1								& 1.0		& 0.5						& 3.5	& 2.7				\\
SGP38326-2	& 11.1						& 11.0		& 10.6								& 0.9		& 0.3						& 1.4	& 0.9				\\
SPT0418-47	& 10.1						& 9.9		& 9.1								& 0.6		& 0.1						& 2.3	& 1.7				\\
SPT0113-46	& 10.6						& 10.3		& 10.0								& 0.5		& 0.2						& 1.6	& 1.5				\\
SPT0441-46	& 10.2						& 9.8		& 9.4								& 0.4		& 0.2						& 1.1	& 0.9				\\
SPT2132-58	& 10.3						& 10.1		& 9.8								& 0.7		& 0.4						& 3.2	& 2.3				\\
SPT2146-55	& 10.1						& 8.7		& 0.0								& 0.1		& 0.0						& 0.2	& 0.0				\\
zC400569	& 10.4						& 10.2		& 9.8								& 0.7 		& 0.3						& 4.0	& 3.6				\\
\hline
\end{tabular}
\tablefoot{From left to right: Galaxy name, total gas mass, gas mass in the unstable regions based on $Q$ and $\Qthreed$, corresponding fraction of the total gas, and radial extent of the unstable region in 2D and 3D. }

\end{table}

%AGC114905	& 0.14						& 0.08		& 0.0								& 0.5		& 0.0						& 3.8	& 0.0				\\
%J81740		& 4.0						& 1.1		& 0.5								& 0.3		& 0.1						& 1.0	& 0.9				\\
%SGP38326-1	& 25.1						& 25.1		& 12.0								& 1.0		& 0.5						& 3.5	& 2.7				\\
%SGP38326-2	& 12.6						& 8.0		& 4.0								& 0.6		& 0.3						& 1.4	& 1.0				\\
%SPT0418-47	& 1.3						& 1.4		& 0.3								& 1.0		& 0.2						& 2.3	& 1.7				\\
%SPT0113-46	& 4.3						& 2.9		& 1.3								& 0.7		& 0.3						& 1.7	& 1.5				\\
%SPT0441-46	& 1.4						& 1.0		& 0.4								& 0.7		& 0.3						& 1.1	& 0.7				\\
%SPT2132-58	& 1.9						& 1.9		& 0.7								& 1.0		& 0.4						& 3.3	& 1.9				\\
%SPT2146-55	& 1.2						& 0.1		& 0.0								& 0.1		& 0.0						& 0.2	& 0.0				\\
%zC400569	& 2.3						& 1.3		& 0.0								& 0.6 		& 0.0						& 4.5	& 0.0				\\

\subsection{3D instability analysis}\label{sec:results_q3d}
\begin{figure*}
	\includegraphics[width=2\columnwidth]{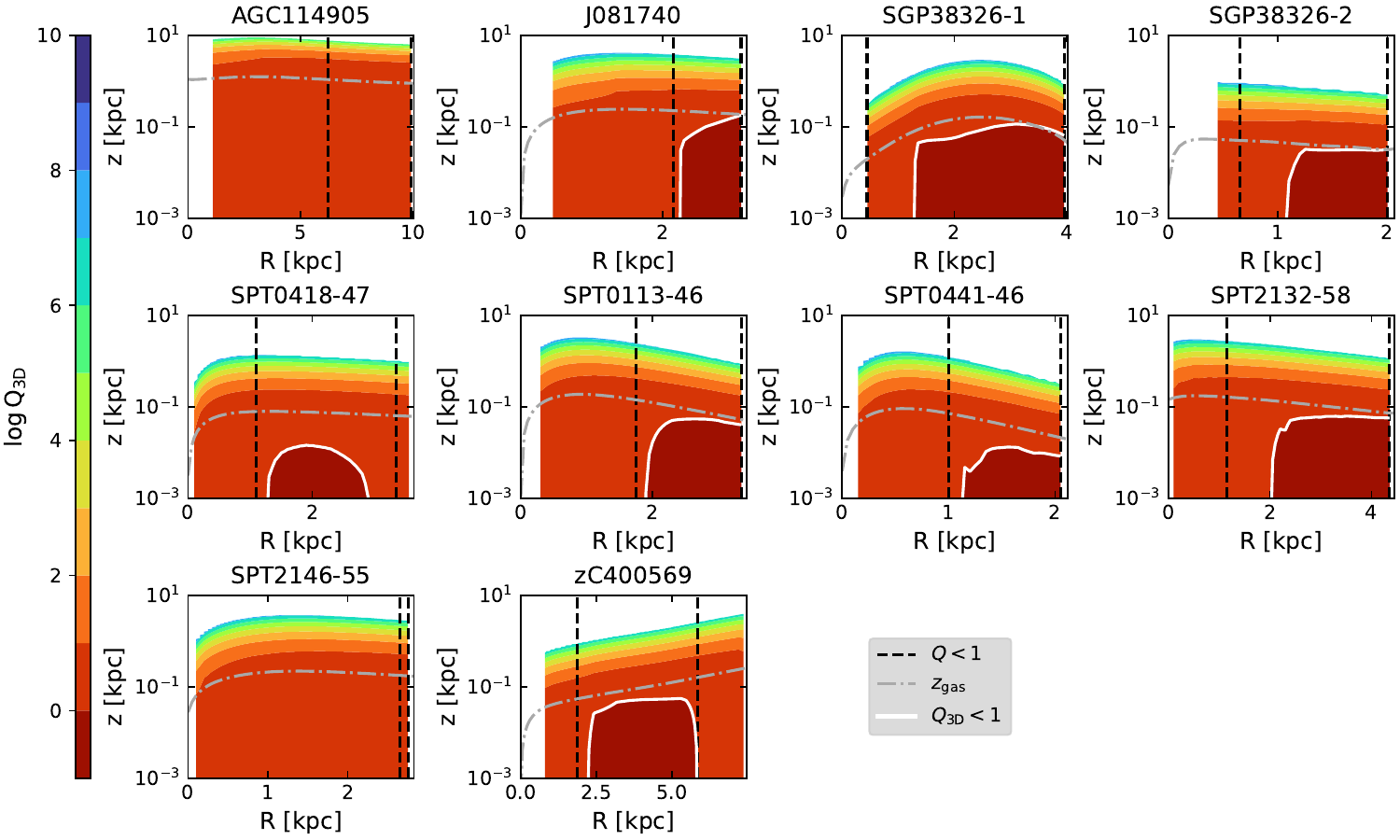}
	\caption{Maps of $\Qthreed(R,z)$ for the galaxies in our sample having gravitationally unstable regions (delimited by the black dashed lines) according to the $Q$ criterion. 
		The white contour encompasses the region that is locally unstable when $\Qthreed$ criterion is adopted. 
		The gray dash-dotted curve represents the gas disc scale height $z_\mathrm{gas}$. }
	\label{fig:q3d_all}
\end{figure*}
We then apply the 3D instability criterion by \cite{2023Nipoti} using Eq.~\ref{eq:q3d_nipoti23}, aiming to understand whether there is any gas disc that is unstable (or partially unstable) when its vertical structure is taken into account. 
For each galaxy, we obtain $\rho(R,z)$ using the procedure described in Sect.~\ref{sec:method_3Dgas_distribution} and we calculate Eqs.~\ref{eq:kappa2} and~\ref{eq:nu2} using the central difference method. 
The resulting maps of $\Qthreed(R,z)$ are shown in Fig.~\ref{fig:q3d_all} for the galaxies with $Q<1$ (see Figs.~\ref{fig:h_Q_Q3D} for the whole sample). 
Interestingly, we find that two galaxies with $Q<1$ in some radial interval, namely AGC114905 and SPT2146-55, have $\Qthreed>1$ everywhere throughout the disc. 
Therefore, no galaxy in our sample at $\mathrm{z} \approx 0 $ host unstable gas discs, considering both the atomic gas and the molecular gas. 
We note that, using the data from \cite{2022ManceraPina}, we would find that AGC114905 is locally unstable, as the stabilising effect of the gas vertical structure is less effective since $\sigma$ is slightly lower than in \cite{2024ManceraPina}. 
Only eight galaxies out of a sample of 44 ($\approx 18$\%) host locally unstable gas discs according to the 3D criterion. 
These are all at redshift $\mathrm{2} \lesssim \mathrm{z} \lesssim 5$ and account for $\approx 60$\% of the whole sample of galaxies at high redshift. 
The gas tracer is [CII] for all the locally unstable systems, except zC400569, which is traced by CO. 

Figure~\ref{fig:q3d_all} shows that the vertical extent of the unstable region is always smaller than the gas scale height (grey dot-dashed curve). 
This confirms that the disc is more prone to LGI near the midplane \citep[][]{2023Nipoti}. 
Figure~\ref{fig:q3d_all} also shows that the radial extent $L_{\Qthreed}$ of an unstable region according to $\Qthreed$ is smaller than $L_\mathrm{Q}$. 
Table~\ref{tab:gas_inst} compares the values of $L_{\Qthreed}$ and $L_Q$. 
If we consider only the galaxies that are unstable based on the 3D criterion, the median value of $L_{\Qthreed}/L_Q$ is $\approx 0.8 \pm 0.1$ (the uncertainty corresponds to $1\sigma$). 
Hence, the instability region is $\approx 20$\% smaller when the vertical stratification of the gas disc is taken into account using $\Qthreed$ rather than $Q$. 

The mass of the gas that is located in the unstable regions ($M_\mathrm{inst}$) can be calculated by integrating $\rho(R,z)$ in $R$ and $z$ where $\Qthreed(R,z)<1$, which is the region encompassed by the white contour in Fig.~\ref{fig:q3d_all}. 
The resulting values can then be used in Eq.~\ref{eq:finst} to calculate the fraction of unstable gas $f_\mathrm{inst}$. 
The results are reported in Table~\ref{tab:gas_inst}: we find $0.1 \lesssim f_\mathrm{inst} \lesssim 0.5$ with median and $1\sigma$ error $\langle f_\mathrm{inst} \rangle = 0.3 \pm 0.1$. 
Hence, $f_\mathrm{inst}$ based on the 3D criterion is reduced by a factor of 2 on overage and covers a narrower range of values compared to the estimates resulting from the 2D analysis. 
These results also indicate that a  small fraction of the total gas in the disc is potentially involved in the disc fragmentation. 

\subsection{Expected masses and number of clumps}\label{sec:results_mclump}
The expected outcome of the instability is the formation of gas clumps following the disc fragmentation. 
Though our instability analysis is based on axisymmetric perturbations, we attempt here to estimate the characteristic mass and number of the clumps expected in the galaxies with unstable regions.
If we consider an unstable gas disc, axisymmetric perturbations lead to the formation of rings. 
\cite{2023Nipoti} shows that the most unstable modes have wavelength $\lambdaRinst\approx 2\pi\hz$, which corresponds to the radial separation between two nearby rings. 
For the galaxies with $\Qthreed<1$, we estimate that the size of the unstable regions ($L_{\Qthreed})$ corresponds to about 1-3$\times \lambdaRinst$, meaning that, for strictly axisymmetric perturbations, between one and three rings might form in the unstable regions. 
At radius $R$ where $\Qthreed<1$, a ring has typical mass
\begin{equation}\label{eq:Mring}
	\Mring\approx 2\pi R \lambdaRinst(R) \Sigma_\mathrm{inst}(R)=4\pi^2 R\hz(R) \Sigma_\mathrm{inst}(R), 
\end{equation}
where $\Sigma_\mathrm{inst}$ is the surface density of the gas in the unstable region. 
More realistically, the unstable ring will fragment into clumps. 
We assume that, azimuthally, the ring will break in portions of azimuthal size $\lambdaphi$ \citep[which must not be confused with the clump final size; see][]{2015Behrendt}, so that 
\begin{equation}\label{eq:Mclump}
	\Mclump\approx  2\pi \hz(R) \lambdaphi \Sigma_\mathrm{inst}(R).
\end{equation}
The azimuthal wavelength $\lambdaphi$ is not determined in \citet{2023Nipoti}'s model, where only axisymmetric perturbations are considered. 
However, we can do physically motivated assumptions on $\lambdaphi$ \citep[see also][]{2001Fuchs}. 
One possibility is to assume that $\lambdaphi=\lambdaRinst\approx 2\pi\hz$, just based on the assumption that the proto-clumps are not elongated in the equatorial plane. 
Another possibility is to assume that the azimuthal size of the unstable region is determined by the Jeans criterion, so $\lambdaphi\approx \lambdaJ$, where
\beq
\lambdaJ = \sigma \sqrt{\frac{\pi}{G\rhobar}}
\eeq
is the Jeans wavelength and
\begin{equation}
	\rhobar(R)\equiv\frac{1}{\Sigma(R)}\int_{-\infty}^\infty\rho^2(R,z)\dd z
\end{equation}
is the average density at $R$. 
To relate $\lambdaJ$ to $\hz$, we specialize to the case of a vertical isothermal distribution, for which $\rhobar(R)=(2/3)\rhozero(R)$ and $\hz(R)=\hseventy(R)\simeq 1.73 b(R)$ with $b(R)=\sigma(R)/\sqrt{2\pi G\rhozero(R)}$. 
In this case, we obtain $\lambdaJ \simeq \pi \hz$ and can thus write $\lambdaphi=f\lambdaRinst\approx f 2\pi\hz$. 
Eq.~\ref{eq:Mclump} becomes
\begin{equation}\label{eq:clump_mass}
	\begin{split}
		\Mclump & \approx  4f\pi^2\hz^2(R)\Sigma_\mathrm{inst}(R) \\
		& \simeq 4 \times 10^6 f \left( \frac{\hz}{100~\mathrm{pc}}\right)^2  \left( \frac{\Sigma_\mathrm{inst}}{10~M_\odot \mathrm{pc}^{-2}}\right) M_\odot,
	\end{split}
\end{equation}
with $0.5\lesssim f \lesssim 1$ to bracket the expected values of $\Mclump$. 

It is useful to define, as a function of radius $R$, the effective azimuthal wavenumber associated to $\lambdaphi$, which also gives a rough estimate of the number of clumps expected from a perturbation centred at $R$ \citep[see also][]{2010Wang}. 
Dividing the ring mass (Eq.~\ref{eq:Mring}) by the clump mass (Eq.~\ref{eq:clump_mass}), we obtain
\begin{equation}\label{eq:clump_number}
	m(R) = \frac{2\pi R}{\lambdaphi}=\frac{1}{f}\frac{R}{\hz}	
	\simeq \frac{10}{f} \left( \frac{R}{1~\mathrm{kpc}} \right) \left( \frac{\hz}{100~\mathrm{pc}} \right)^{-1}.
\end{equation}

\begin{figure}
	\includegraphics[width=1\columnwidth]{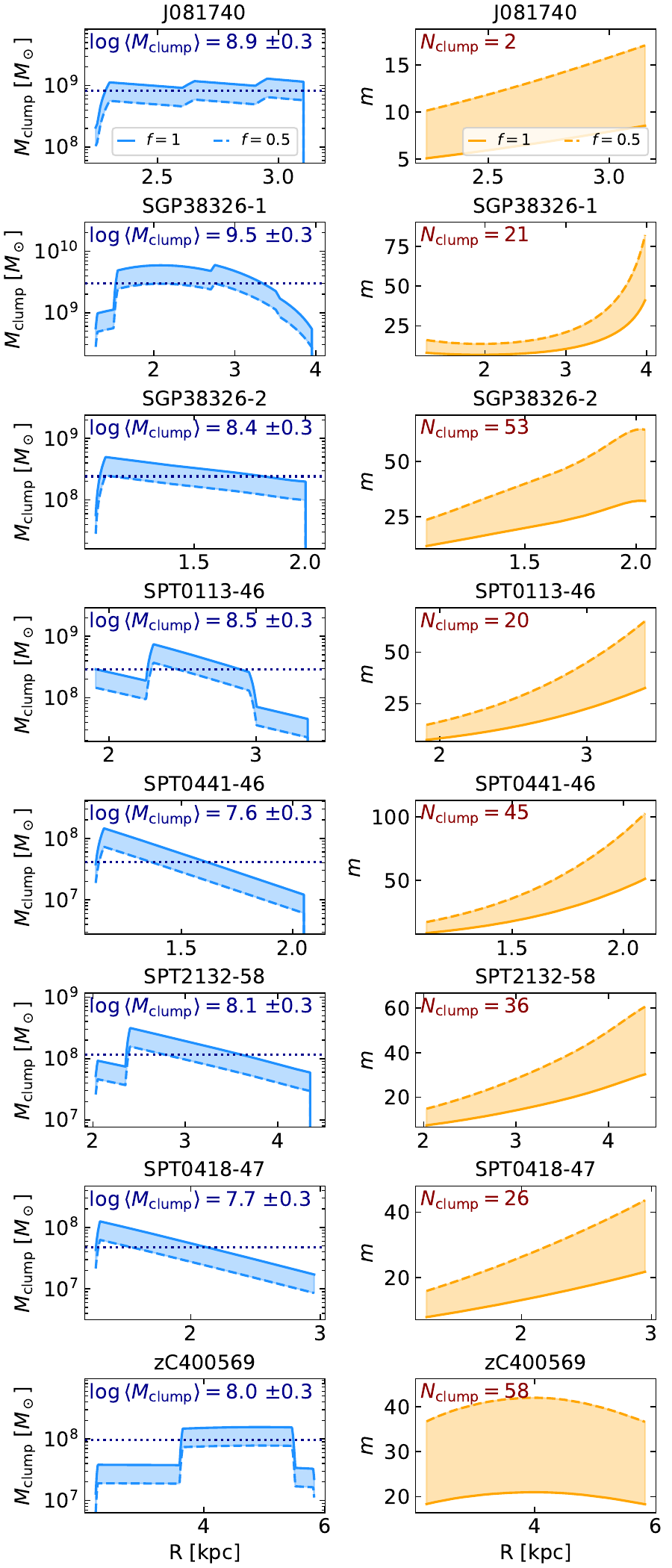}
	\caption{Radial trends of the expected mass of clumps (left) and the most unstable effective azimuthal wavenumber (right) for the galaxies with $\Qthreed<1$. 
		The shaded areas show the range of values for $0.5 \leq f \leq 1$ (see Sect.~\ref{sec:results_mclump} for details). 
		In each panel, we also report the expected clump mass $\langle M_\mathrm{clump} \rangle$  (blue dotted line) with the uncertainty given by $0.5 \leq f \leq 1$ and the corresponding expected number of clumps $N_\mathrm{clump}$.}
	\label{fig:clump_props}
\end{figure}
Figure~\ref{fig:clump_props} shows the clump mass (Eq.~\ref{eq:clump_mass}) and the most unstable effective azimuthal wavenumber (Eq.~\ref{eq:clump_number}) in the regions of our galaxies where $\Qthreed<1$. 
To obtain $\Sigma_\mathrm{inst}$ in Eq.~\ref{eq:clump_mass}, we integrated the gas volume density along $z$ only in the regions where $\Qthreed(R,z)<1$ (see Fig.~\ref{fig:q3d_all}). 
The clump masses range from a few $10^6~M_\odot$ up to a few $10^{9}~M_\odot$, while $m(R)$ varies from five up to about 100.
With the exception of J081740 and zC400569, the clumps tend to become less massive (although not always monotonically) and $m$ higher with increasing distance from the galaxy centre. 
Each panel in Fig.~\ref{fig:clump_props} also reports the expected average clump mass
\begin{equation}\label{eq:mclump_average}
	\langle M_\mathrm{clump} \rangle \equiv \frac{2\pi}{M_\mathrm{inst}} \int_{R_\mathrm{min}}^{R_\mathrm{max}} M_\mathrm{clump}(R) \Sigma_\mathrm{inst}(R) R \mathrm{d}R \, ,
\end{equation}
where $R_\mathrm{min}$ and $R_\mathrm{max}$ are the minimum and maximum $R$ where $\Qthreed<1$, and the expected clump number $N_\mathrm{clump} = M_\mathrm{inst}/\langle M_\mathrm{clump} \rangle$. 
We find that the average clump masses are $7.6 \lesssim \log \left( \langle M_\mathrm{clump} \rangle / \mathrm{M}_\odot\right) \lesssim 9.5$, while the expected clump number is approximately 20-60. 
We note that $N_\mathrm{clump} =2$ in the case of J081740, but this number should be taken with caution as the size of the radial extent of the unstable region $L_{\Qthreed}$ is smaller than $\lambdaRinst$. 
This is probably a consequence of the spatial resolution of the observations for this system, which might affect the precision of our estimate of $L_{\Qthreed}$. 

%%%%%%%%%%%%%%%%%%%%%%%%%%%%%%%%%%%%%%%%%%%%%%%%%%%%%%%%%%%%%%%%%%%%%%%%%%%%%%%%%%%%%%%%%%%%%%%%%%
\section{Discussion}\label{sec:discussion}
In this section, we compare our results with the existing literature on LGI. 
We then discuss limitations and systematic uncertainties of our approach, together with future improvements. 
We also comment on potential implications for the origin of gas turbulence and the regulation of star formation in galactic discs, and compare the observed properties of cold gas clumps in galaxies with those predicted by our model for unstable discs. 

\subsection{Comparison with previous works based on 2D criteria}\label{sec:discussion_literature}
The literature on LGI of galactic discs is vast and spans more than six decades. 
We therefore limit our discussions to recent works investigating the stability of gas discs only, under the simplifying assumption that the stellar discs is insensitive to the perturbation (see Sect.~\ref{sec:discussion_limitations}). 
We also select only the studies of galaxy samples having some overlap with ours.

\begin{itemize}
\item We find that cold gas discs (considering both atomic and molecular gas) in the present-day Universe have $Q>1$. 
This result is in agreement with the existing literature investigating LGI in nearby spiral and dwarf galaxies \citep[e.g.][]{1997Combes,1998Hunter,2001MartinKennicutt,2003Boissier,2008Leroy,2008Begum_a,2011Romeo,2012Elson,2013Romeo,2015Elmegreen,2015ElmegreenHunter,2017Namumba,2018Watts,2022Meidt,2022Lizee,2023Aditya}.
We highlight though that our study includes a significant improvement in the calculation of $Q$, as we use the observed radial profiles of $\sigma$ rather than choosing a radially constant value like in the majority of these previous works. 
In the literature, some studies have taken into account the radial gradient of the velocity dispersion and used $\sigma(R)$ measured from observations  \citep[e.g.][]{2015Romeo,2016Romeo,2017Romeo}. 
However, these measurements are typically affected by the beam smearing effect, which can significantly influence the results of the instability analysis (see Sect.~\ref{sec:intro}). 
Another important improvement with respect to these prior studies consists in using measurements of the gas kinematics obtained with state-of-the-art techniques that efficiently minimise the beam smearing effect (see Sect.~\ref{sec:sample}). 

\item Recently, \cite{2023Aditya} has investigated the stability of nearby galaxies in the \textit{Spitzer} Photometry and Accurate Rotation Curves
(SPARC) database \citep{2016Lelli} and of the high-z DSFGs from \cite{2020Rizzo}. 
\cite{2023Aditya} finds that the nearby gas discs have $Q>1$, which is confirmed by our analysis. 
However, they find only two unstable gas discs (SPT0113-46 and SPT2132-58) among the DSFGs, while we find five. 
The origin of this difference might be the assumption of a radially constant value for the gas velocity dispersion in the instability analysis performed by  \cite{2023Aditya}. 

\item The UDG AGC114509 is a problematic case that deserves ad hoc discussion. 
By definition, UDGs have significantly larger effective radii than the typical dwarf galaxies with similar stellar mass  \cite[e.g.][]{2015Mihos,2015VanDokkum,2019ManceraPina}. 
\cite{2022ManceraPina} show that the HI rotation curve of AGC114509 can be explained almost entirely by the contribution of baryons alone, leaving little or no room for a DM halo. 
The gravitational potential of galaxies with low DM fraction is dominated by baryons with an important contribution from the gas disc. 
Therefore, these systems are expected to be unstable, which originally prompted the hypothesis for galactic DM halos \citep{1973Ostriker}. 
\cite{2022Sellwood} run N-body simulations of AGC114905 based on the measurements by \cite{2022ManceraPina} finding that the atomic gas disc cannot rotate regularly, as it would be rapidly disrupted by the gravitational instability. 
If we use the data from \cite{2022ManceraPina}, our method gives consistent results with the simulations by \cite{2022Sellwood}, as the atomic gas disc is unstable according to both the 2D and the 3D instability criteria. 
On the other hand, we showed that $\Qthreed>1$ across the whole disc when using the latest measurements by \cite{2024ManceraPina}, which corroborates the point by \cite{2022Sellwood} that higher $\sigma$ may assist in avoiding the instability, even in the case of an extremely shallow DM halo potential, subdominant within the disc region. 
Nonetheless, we note that $\Qthreed$ is just slightly above unity in the regions close to the midplane and at $R \gtrsim 8.5$~kpc so, considering the uncertainties on the measurements, we cannot completely exclude that the some regions of the disc are unstable. 
HI observations with higher resolution and sensitivity appear necessary to reach more robust conclusions.

\item \cite{2014DeBreuk} analysed the stability of ALESS~73.1 finding $Q <1$ across the whole galaxy, while we obtain $Q>1$. 
This difference is due to the fact that \cite{2021Lelli} uses an improved estimate of the galaxy inclination and new observations with higher resolution. 
Most importantly, they used \barolo to model the gas kinematics taking into account the beam smearing. 
Our result is consistent with the conclusions by \cite{2018Gullberg}, who analyse the dust continuum emission of ALESS~73.1 and found no evidence for gas clumps. 

\item \cite{2014Genzel} analysed the stability of the molecular gas disc in zC400569 finding $Q<1$ across the whole disc except the very innermost regions ($R \lesssim 1$~kpc), while we obtain $Q<1$ for $R \gtrsim 2.5$~kpc. 
This discrepancy is partially due to the different technique used by \cite{2023Lelli} to model the gas kinematics, which allowed to properly account for the beam smearing effect \citep[see discussions in][]{2023Lelli}. 
However, a one-to-one comparison is not fully appropriate. 
In fact, \cite{2014Genzel} calculated $Q$ using the velocity dispersion measured from H$\alpha$ emission line datacubes (probing the ionised gas phase rather than the molecular gas) and the gas surface density extrapolated from the SFR surface density using the Kennicutt-Schmidt relation \citep{1959Schmidt,1989Kennicutt,1998Kennicutt} rather than the observed $\Sigma$. 
Observations of the H$\alpha$ emission of zC400569 revealed the presence of a chain of clumps \citep{2017Genzel,2018FosterSchreiber}, which has been ascribed to LGI \citep[e.g.][]{2009Elmegreen}. 
On the other hand, \cite{2023Lelli} suggest that the clumps visible in the optical images are low mass companions based on the lack of a CO emission associated with the H$\alpha$ clumps. 
Our results are consistent with the scenario in which non-axisymmetric perturbations induced by these companions may have triggered the instability in the CO disc. 

\item \cite{1992Romeo,1994Romeo} proposed that the 2D criterion $Q < 2/3$ can be used to analyse LGI taking into account the stabilising effect of the gas disc thickness \citep[see also][]{2002Kim,2010Bertin,2010Wang,2013Romeo,2015Behrendt}. 
As pointed out by \cite{2023Nipoti}, this value of the critical $Q$ for the instability is consistent with those found using the 3D approach. 
However, the criterion $Q < 2/3$ has two limitations with respect to $\Qthreed<1$: it cannot take into account the detailed vertical structure of the gas disc and it cannot be used to study how the instability properties vary with the distance from the midplane. 
To quantity the impact of these limitations, we repeated the analysis adopting $Q < 2/3$ and compared the results with those based on $\Qthreed<1$ (see figures in Appendix~\ref{ap:q_all_sample}). 
We find no disc with $Q < 2/3$ at $z \approx 0$, which is in agreement with the result based on $\Qthreed<1$. 

Contrary to the 3D analysis, zC400569, SPT2146-55 and SPT0418-47 have $Q \gtrsim 2/3$ everywhere. 
Therefore, we find six unstable discs  based on $Q < 2/3$ and these are all at $\mathrm{z} \approx 4-5 $. 
We estimate that, on average, the regions where $Q < 2/3$ are $\approx 50$\% smaller than the regions where $Q < 1$. 
This indicates that the 2D criterion for thick discs tends to underestimate the extent of the unstable regions compared to $\Qthreed<1$, which gives radial extents of the unstable regions about 20\% smaller than $Q < 1$. 
However, the fraction of the gas affected by the instability for $Q < 2/3$ is $\approx 20$\% on average, which is consistent within the uncertainties with the result for $\Qthreed<1$. 
This can be explained by the fact that 2D criteria implicitly assume that, at each $R$ within the instability region, the whole gas layer (i.e. at any $z$) is affected by the instability, while only the gas at heights $|z| \lesssim z_\mathrm{gas}$ is unstable based on the 3D analysis. 
Overall, these results indicate that the 2D instability criterion $Q < 2/3$ is a fairly good approximation for thick gas discs, although it misses some of the information provided by the 3D criterion.
\end{itemize}

\subsection{Comparison between \cite{1964Toomre}'s $Q$ and the approximation $Q_\mathrm{approx} = a \sigma /(V_\mathrm{rot} f_\mathrm{gas})$}\label{ap:Qformulaccia}
In recent years, several authors have adopted an approximated version of the \cite{1964Toomre}'s instability parameter $Q$ (Eq.~\ref{eq:q2d}) to study LGI in gas discs of high-z star-forming galaxies and their dynamical evolution \citep[e.g.][]{2011Genzel,2014Genzel,2015Wisnioski,2016Stott,2016Simons,2016Mieda,2018Johnson,2018Girard,2019Girard}. 
The approximation reads
\begin{equation}\label{eq:formulaccia}
	Q_\mathrm{approx} = \frac{a \sigma }{V_\mathrm{rot} f_\mathrm{gas}}\, ,
\end{equation}
where $f_\mathrm{gas}$ is the gas mass fraction in the disc and $a$ is an approximation for $\kappa/\Omega$, with $a=1$ for a Keplerian rotation curve, $a=\sqrt{2}$ for a flat rotation curve, $a=\sqrt{3}$ for a disc with uniform density, and $a=2$ for a solid body-body rotation \citep[e.g.][]{2013Glazebrook}.  
Frequently, Eq.~\ref{eq:formulaccia} is calculated by using $\sigma$ averaged across the whole gas disc, the maximum $V_\mathrm{rot}$, and the gas fraction obtained as the ratio between the cold gas (typically, molecular gas) mass and the total baryonic mass (or the stellar mass) of a galaxy \citep[see for instance the assumptions in][]{2014Green,2015Wisnioski,2016Stott,2018Girard,2019Girard,2019Ubler}. 
Some authors have applied Eq.~\ref{eq:formulaccia} (as well as Eq.~\ref{eq:q2d} or similar formulas) pixel by pixel on maps of the gas distribution and kinematics \citep[e.g.][]{2011Genzel,2014Genzel,2014DeBreuck,2023Liu,2024Fujimoto}, even thought these instability criteria are derived assuming that the system and the perturbations are axisymmetric. 
Therefore, such criteria should not be used for a pixel-by-pixel analysis.
\begin{figure*}
	\centering
	\includegraphics[width=1.3\columnwidth]{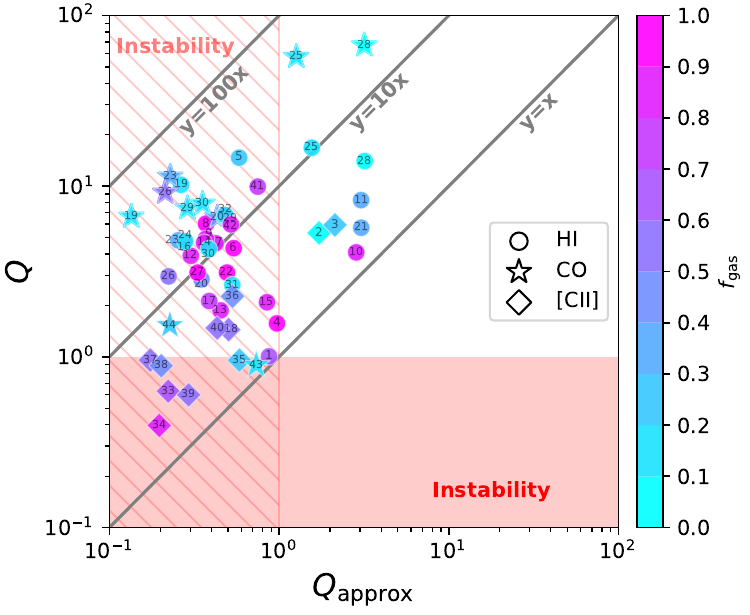}
	\caption{Comparison between $Q_\mathrm{approx}$ (Eq.~\ref{eq:formulaccia}) and the median value of $Q(R)$ (Eq.~\ref{eq:q2d}). 
		The different symbols indicate the gas tracer, with circles for HI, stars for CO, and diamonds for [CII]. 
		The symbols are coloured according to the mass fraction of cold gas with respect to baryons and the numbers indicate the galaxy ID (see Table~\ref{tab:sample}).
		The areas in shaded red and hatched show the instability regions for $Q<1$ and $Q_\mathrm{approx}<1$, respectively. 
		The grey solid lines represent the different ratios between the values on $y$ and $x$ axes. 
		This figure shows that Eq.~\ref{eq:formulaccia} strongly underestimate $Q$. 
		}
	\label{fig:QToomre_vs_Qformulaccia}
\end{figure*}

We test the accuracy of Eq.~\ref{eq:formulaccia} by comparing $Q_\mathrm{approx}$ with $Q$. 
For each galaxy, we calculated $Q_\mathrm{approx}$ using the median value of $\sigma$ across the whole gas disc and the maximum $V_\mathrm{rot}$. 
We also assumed $f_\mathrm{gas}=M_\mathrm{gas}/(M_\mathrm{gas} + M_\star)$ with $M_\mathrm{gas}$ being the sum of the atomic gas mass and the molecular gas mass (when available) for the present-day galaxies or, for high-z systems, the gas mass estimated from CO or [CII] emission. 
We took $a=2$ for galaxies with $\log (M_\star/M_\odot) < 9$, as dwarf galaxies typically have slowly rising rotation curves, and $a=\sqrt{2}$ for galaxies with $\log (M_\star/M_\odot) \geq 9$, as their rotation curve is essentially flat. 
We also checked that our overall conclusions do not change if we assume $f_\mathrm{gas}=M_\mathrm{gas}/M_\star$ or use the median $V_\mathrm{rot}$. 
For each galaxy, we also derived the median value of $Q(R)$ (Eq.~\ref{eq:q2d}; see Fig.~\ref{fig:q2d_all}) across the whole gas disc. 
Figure~\ref{fig:QToomre_vs_Qformulaccia} clearly shows that Eq.~\ref{eq:formulaccia} dramatically underestimates the instability parameter. 
Except for two fortuitous cases (AGC114905 and zC400569), $Q_\mathrm{approx}$ is at least a factor two smaller than the median $Q$, but this difference can be more than one order of magnitude. 
Based on $Q_\mathrm{approx}<1$, there are 37 unstable galaxies out of 44 ($\approx 84$\%). 
This greatly exceeds the number of unstable systems found in Sect.~\ref{sec:results_q2d}, which is ten out of 44 ($\approx 23$\%), and also the number of systems with median $Q$ below unity in Fig.~\ref{fig:QToomre_vs_Qformulaccia}, which is only seven ($\approx 16$\%). 
We therefore conclude that $Q_\mathrm{approx}$ is not an accurate tool for studying LGI in galactic gas discs and gives biased results towards a high fraction of unstable discs. 
Figure~\ref{fig:QToomre_vs_Qformulaccia} also shows that there is no apparent correlation between the median values of $Q$ and the gas fraction, and that galaxies with high baryon fraction do not seem to be systematically closer the 1:1 line, further supporting that $Q_\mathrm{approx}$ is not a good approximation for $Q$	 regardless of the gas fraction. 
We note that the median $Q(R)$ for molecular gas discs tend to be higher than the values for atomic gas discs as the molecular gas is typically located in the inner parts of the disc, where $Q(R)$ is the highest.

\subsection{Limitations and uncertainties}\label{sec:discussion_limitations}
The instability criterion by \cite{2023Nipoti} allows us to take into account the vertical stratification of the gas discs in a self-consistent way, a significant improvement with respect to other criteria, which either assume a razor-thin disc or modify thin-disc criteria with some approximations to account for the disc thickness. 
However, it is worth discussing some critical points of our approach. 

\subsubsection{Limitations of the analytical approach}\label{sec:discussion_limitations_analytical}
We address here the limitations due to the assumptions in the analytical derivation of $\Qthreed$. 

\begin{itemize}
	
	\item \cite{2023Nipoti} assumes that only the gas disc reacts to the perturbation, while the stellar disc and the DM halo are insensitive to it. 
	It is expected that introducing a perturbation of the external potential fosters the instability, as pointed out by a few works investigating the effect of perturbations acting on more than one mass component \citep[e.g.][]{1992Romeo,2003Boissier,2011Elmegreen,2011Romeo,2013Romeo,2016Romeo,2017Romeo,2023Aditya,2023Puschnig}. 
	In particular, the instability of a razor-thin disc can occur up to $Q\approx 2-3$ when the stellar disc response is considered \citep{2011Romeo,2013Romeo,2016Romeo,2017Romeo,2018MarchukSotnikova,2018Marchuk,2023Aditya}.
	As already noted by \cite{2017Romeo}, unstable stellar discs can drive the instability of the gas disc. 
	The LGI of 3D two-component discs is studied in \cite{2024Nipoti}, who show that, for each component, $\Qthreed <1$ remains a sufficient condition for instability also in the presence of a second responsive component. 
	We note though that a few discs have $Q >2-3$ everywhere or $Q <2-3$ only in some regions (see figures in Appendix~\ref{ap:q_all_sample}). 
	This suggests that, as a rule, the LGI is not expected to occur in these galaxies, with crucial implications for star formation and turbulence regulation (see Sects.~\ref{sec:discussion_turb} and~\ref{sec:discussion_SF}). 
	The case of high-$\mathrm{z}$ systems, which host massive and compact stellar components, is the most uncertain, as detailed studies are hampered by the lack of spatially resolved observations probing the stellar distribution and kinematics. 
	This issue will be potentially overcome thanks to James Webb Space Telescope (JWST) observations, which may permit measurements of the stellar distribution and kinematics, and accurate studies of their stability. 

	\item 
	$\Qthreed$ can be used to study only the instability for axisymmetric perturbations, but it is does not guarantee that the gas disc is stable against non-axisymmetric perturbations. 
	For infinitesimally thin discs, these grow for higher values of $Q$ than the axisymmetric perturbations, meaning that the gas disc is more prone to instability \citep{1964LinShu,1978Lovelace,2008BinneyTremaine,2012Griv,2016KratterLodato,2016Inoue}. 
	Similarly, we cannot exclude that 3D discs with $\Qthreed>1$, might be unstable for more general perturbations. 
	Studies based on numerical simulations showed that even discs with $Q \approx 2-3$ can be unstable because of non-axisymmetric or non-linear perturbations \cite[e.g.][]{2012Griv,2016Inoue}. 
	As mentioned in the previous paragraph, a few galaxies have $Q >2-3$ over the entire disc or over significant radial range. 
	This suggests that these galaxies are not expected to have widespread LGI, even when more general perturbations are considered.
	However, this hypothesis needs to tested in the case of a responsive stellar disc to reach a robust conclusion. 
	
	\item
	The critical value for instability is set to 1, for both $Q$ and $\Qthreed$. 
	However, a few factors that can increase this value above unity (at least for 2D criteria) besides those discussed above. 
	The critical $Q$ can be increased by $\approx50$\% or more in the presence of magnetic fields \citep[e.g.][]{2003Kim} or by a factor 2-3 for rapid dissipation of turbulence \cite[][]{2011Elmegreen}. 
	To first approximation, one may apply the same reasoning as in the previous paragraphs: a few gas discs have (either locally or globally) $Q >2-3$, which should disfavour widespread instability even when magnetic fields and turbulence dissipation are considered. 
	Given that the neglected processes (e.g. responsive stellar disc, non-axisymmetric perturbations, magnetic fields) favour the instability, our results on the unstable galaxies can considered robust in this respect. 
	
	\item 
	\cite{2010Romeo} analysed the instability of a clumpy turbulent disc and showed that gas turbulence can excite instabilities on physical scales smaller than those typically probed by observations, as the velocity and density fluctuations in a turbulent medium depend on the spatial scale following the so-called \cite{1981Larson}'s laws. 
	Hence, 2D instability criteria can be unreliable in such systems \citep[][]{2010Romeo,2014Romeo,2015Agertz,2021Renaud}, a consideration that may also apply to 3D criteria for gas discs.
	Our results for the atomic gas discs should be robust in this respect, as HI discs belong to a regime in which instability criteria (at least, in 2D) do not depend on the physical scale probed by observations \citep{2014Romeo}. 
	On the other hand, CO and [CII] emissions typically trace the molecular gas, which is more clumpy and turbulent than HI. 
	Hence, depending on the slope of Larson's law for molecular clouds, these discs may be unstable on scales smaller than those probed by our observations ($\lesssim z_\mathrm{gas}$, except galaxies from \citealt{2023Lelli} and \citealt{2023RomanOliveira}) even when $Q, \Qthreed >1$ \citep{2014Romeo}. 
	Empirical determinations of Larson's laws in the MW and nearby galaxies indicate that present-day discs likely belong to the regime of small scale instabilities \citep{2014Romeo}, although observational uncertainties are non-negligible \citep[see discussion in][Sect. II.D]{2004MacLowKlessen}. 
	At $\mathrm{z}>0$, the properties of spatially resolved GMCs have been measured for only two lensed galaxies, but the low statistics does not allow to fully constrain slope of Larson's laws \citep{2019DessaugesZavadsky,2023DessaugesZavadsky} and understand whether small scale instabilities could play an important role.

\end{itemize}

\subsubsection{Observational limitations}\label{sec:discussion_limitations_observational}
We now discuss potential issues due to observational limitations. 
We anticipate that these are not expected to affect the results for the low-z sample, but maybe important for high-z galaxies. 

\begin{itemize}
	
	\item
	The spatial resolution influences how accurately we can measure the extent of the unstable regions and the fraction of unstable gas. 
	We expect that this issue does not affect our sample at $\mathrm{z}\approx 0$ (see Sect.~\ref{sec:sample}), as the analysis was performed using observations with high spatial resolution \citep[e.g., $\approx 400$~pc for the sub-sample from][]{2019Bacchini,2020Bacchini}. 
	However, this may be important for the non-lensed high-z sample, whose discs are covered by relatively few resolution elements. 
	Nonetheless, the spatial resolution seems to have a relatively small effect in most cases, as our results for poorly resolved galaxies are similar to those for lensed systems (see e.g. $f_\mathrm{inst}$ and $L$ values in Table~\ref{tab:gas_inst}). 
	
	\item
	Clearly, $\alpha_\mathrm{CO}$ and $\alpha_\mathrm{[CII]}$ may represent significant source of uncertainty on $M_\mathrm{gas}$ and $\Sigma$, affecting the outcome of the stability analysis (both in 2D and in 3D) and the estimate of clump masses. 
	Using dynamically constrained $\alpha_\mathrm{CO}$ and $\alpha_\mathrm{[CII]}$ is arguably a better choice than extrapolating the gas surface density using some empirical scaling relation or assuming a given value. 
	Moreover, $\alpha_\mathrm{CO}$ and $\alpha_\mathrm{[CII]}$ obtained by \cite{2020Rizzo,2021Rizzo} and \cite{2023Lelli} with the dynamical modelling of the galaxy rotation curves are consistent with the expected values for high-z galaxies \cite[e.g.][but see also \citealt{2018Zanella,2020Madden}]{2021Sommovigo,2022Sommovigo,2022Dunne,2022Vizgan_a}. 
	Another source of uncertainty on the gas mass distribution is the potential presence of CO-dark gas \citep{2010Wolfire}, which may influence the instability analysis \citep[e.g.][]{2015Romeo}. 
	At high-$\mathrm{z}$, [CII] emission can trace CO dark gas \citep{2022WolfireValliniChevance}, so its contribution is in principle included in the dynamically constrained $\alpha_\mathrm{[CII]}$. 
	The mass fraction of CO-dark gas is approximately 30\% in spiral galaxies at $\mathrm{z}\approx 0$ \citep{2022WolfireValliniChevance}, so it is unlikely that its mass contribution would make present-day CO discs unstable.  
	
	\item
	More in general, a potential source of uncertainty is represented by the degeneracies in the mass modelling for the galaxies at $\mathrm{z > 0}$, which were broken by imposing physically motivated priors in the rotation curve fitting \citep[see discussions in][]{2020Rizzo,2021Rizzo,2021Lelli,2023Lelli,2024RomanOliveira}. 
	Recently, the stellar mass of SPT0418-47 estimated by \cite{2020Rizzo} has been confirmed using JWST observations \citep{2023Cathey}, suggesting that the masses obtained via dynamical modelling are fairly robust. 
	Upcoming JWST data will likely improve the situation by delivering spatially resolved observations of the stellar components. 
	
	\item
	A potential source of uncertainty on the mass model and the instability analysis is also the inclination, which is key for determining the amplitude of the rotation curve \citep{1987Begeman,1991Begeman}. 
	This issue can be important for nearly face-on systems (i.e. $i \lesssim 30 \deg$; see Table~\ref{tab:sample}), such as AGC114905 and ALESS073.1, and for poorly resolved gas discs without spatially resolved observations of the stellar counterpart, like those from \cite{2023RomanOliveira,2024RomanOliveira}. 
	We note tough that the inclination of AGC114905 is well constrained by high-quality observations of the stellar component \citep[see][]{2024ManceraPina} and that the the gas disc of ALESS073.1 is resolved by several resolution elements, so the inclination is reasonably robust \citep[see][]{2021Lelli}. 
	\cite{2023RomanOliveira} performed several tests on mock observations, finding that \textsc{Cannubi} can recover the correct disc geometry for $30 \deg \lesssim i \lesssim 70 \deg$ if the discs is relatively thin ($z_\mathrm{gas} \lesssim 1$~kpc), which is the case of J081740, SGP38326-1, and SGP38326-2. 
	BRI1335-0417 has $z_\mathrm{gas} \gtrsim 1$~kpc, but the SNR and the spatial resolution of the data is good enough to recover the disc	 geometry \citep{2023RomanOliveira}. 
	
	\item
	Our methodology to obtain a 3D model of the gas disc and derive $\Qthreed$ does not take into account the uncertainties on the input parameters, which are the observable quantities ($V_\mathrm{rot}$, $\sigma$, and $\Sigma$) and the mass models. 
	This task would require several realisations of the gas disc model obtained by varying the input parameters within their uncertainties. 
	For simplicity, we just explore the effect of varying $\sigma$ within its uncertainty. 
	This is expected to be the dominant source of error for most of the galaxies in our sample, as the vertical distribution of the gas disc is strongly influenced by the gas velocity dispersion \citep[see][]{2019Bacchini}. 
	We repeated the whole procedure described in Sects.~\ref{sec:method} and~\ref{sec:results} using two "extreme" gas disc models: a "thinner" disc model obtained with $\sigma-\Delta \sigma$ and a "thicker" disc derived using $\sigma+\Delta \sigma$, where $\Delta \sigma$ is the uncertainty on the observed $\sigma$. 
	Thus, the stabilising effect of the gas vertical stratification is essentially minimised and maximised in the low-$\sigma$ and high-$\sigma$ disc model, respectively~\footnote{Exploring models with $\Sigma+\Delta \Sigma$ ("heavier" disc) and $\Sigma-\Delta \Sigma$ ("lighter" disc), where $\Delta \Sigma$ is the uncertainty on $\Sigma$, would likely have similar outcome as the low-$\sigma$ and high-$\sigma$ cases, respectively. }. 
	Based on the 3D instability criterion, we find that: 
	i) three galaxies with no unstable region in the fiducial model (AGC114905, SPT02146-56, and zc488879) are locally unstable for the low-$\sigma$ modelling; 
	ii) four galaxies that are locally unstable based on the fiducial model (J081740, SPT0418-47, SPT0441-46, and zc400569) have no unstable region in the high-$\sigma$ model;
	iii) for the rest of the sample, the galaxies with $\Qthreed>1$ everywhere in the fiducial model do not have unstable regions in the low-$\sigma$ model either, and the galaxies with $\Qthreed<1$ in the fiducial model have unstable regions also in the high-$\sigma$ model.
	This test confirms that essentially none of the galaxies at $\mathrm{z} \approx 0$ host unstable gas discs, while about half of the system at high-z are locally unstable. 
	We emphasize that this test considers extreme cases and does not take into account other uncertainties than may affect the conclusions of the instability analysis, such as those on the gravitational potential and the gas surface density. 
	However, this exercise is still useful to underline the primary importance of using robust measurements of the gas kinematics and of increasing the galaxy sample to include more UDGs and high-z galaxies. 
\end{itemize}

\subsection{Possible implications for the origin of gas turbulence}\label{sec:discussion_turb}
Gas discs in galaxies are notoriously turbulent, but there is much debate on which mechanism can inject enough energy into the interstellar medium to maintain such turbulence \cite[e.g.][]{2004MacLowKlessen,2004ElmegreenScalo}. 
The primary candidates are supernova (SN) explosions, but some authors showed that these are insufficient to sustain the gas turbulence in nearby star-forming galaxies, especially in their outskirts \citep{2009Tamburro,2013Stilp,2019Utomo}. 
Therefore, alternative mechanisms have been proposed \citep[for a review, see][]{2004MacLowKlessen}, including LGI. 
In practice, LGI can convert gravitational energy into turbulent energy via torques, which make the clump migrate towards the galaxy center and drive radial gas flows \citep[e.g.][]{2002Wada,2007Bournaud,2009Agertz,2009Dekel_b,2010Krumholz,2012Cacciato,2018Krumholz}. 

Our findings suggest that, in the present-day Universe, LGI does not significantly contribute in driving turbulence (and thus radial flows) in the gas disc of star-forming galaxies, both low mass dwarfs and massive spirals. 
This supports the results by \cite{2021DiTeodoro} showing that large-scale radial flows of atomic gas are essentially negligible in nearby spiral galaxies, as expected for stable discs. 
Our findings are also consistent with the conclusions of \cite{2020Bacchini}, who show that the SN feedback alone can sustain the cold gas turbulence in 10 nearby galaxies in our sample, provided that the reduced dissipation of the gas turbulence due to the gas disc thickness is taken into account. 
Recently, \cite{2021Hunter} and \cite{2022Elmegreen} have investigated the origin of turbulence in 11 spiral galaxies and the dwarf DDO~154, which are in both \cite{2020Bacchini}'s sample and ours. 
They argue that, locally, stellar feedback does not significantly increase the HI turbulence. 
This conclusion is based on the lack of a correlation between the azimuthal variations (i.e. the difference between the value of the azimuthally averaged profile at given $R$ and the value in a pixel at the same $R$) of $\sigma$ measured from 2nd moment maps of the HI datacube and the star formation rate surface density measured from FUV images. 
Therefore, they propose that the atomic gas turbulence is driven by instabilities \citep{2022Elmegreen,2023Elmegreen,2024Hunter}. 
We stress that LGI is not a viable option to explain the HI turbulence in spiral and dwarf galaxies at $\mathrm{z} \approx 0$, as both $Q$ and $\Qthreed$ are well above unity in the majority of these systems. 
Moreover, pixel-by-pixel measurements of $\sigma$ from 2nd moment maps should be interpreted with caution, as they may be overestimated because of beam smearing and spurious wings of the line profile, especially for data with low S/N. 
Besides these considerations, starburst dwarf galaxies show systematically more complex gas kinematics than moderately star-forming dwarfs \citep[e.g.][]{2014Lelli,2023Marasco}, suggesting that, locally, stellar feedback does stir the gas inside the potential well and promote complex non-circular motions.

Taken at face value, our results suggest that the situation is likely different at high redshift, as LGI may contribute to driving turbulence in $\approx 60$~\% of the systems. 
The fact that the unstable regions are typically at the disc edges suggests that the contribution of some other mechanism is needed beside LGI. 
\cite{2020Rizzo,2021Rizzo} show that the median velocity dispersion of the gas in the disc of SPT galaxies is consistent with the predictions of analytical models of turbulence driven by stellar feedback, while models including gravitational processes predict median values of $\sigma$ that are much higher. 
The only exception is SPT0113-46, where LGI may contribute to driving turbulence together with feedback. 
This galaxy is indeed unstable according to the $\Qthreed$ criterion. 
Similarly, \cite{2024RomanOliveira} show that the gas turbulence in J81740, SGP38326-1, and SGP38326-2 appears to be primarily driven by stellar feedback rather than LGI. 
Overall, these results potentially suggest that LGI and stellar feedback dominate in driving turbulence in different regions of the disc. 
Spatially resolved observations in the rest-frame optical/FUV would be very useful to understand whether there is a transition between these two regimes.

\subsection{Possible implications for star formation regulation}\label{sec:discussion_SF}
Understanding the physics regulating star formation is one of the main objectives of galaxy evolution studies. 
Various theories of star formation have been put forward until today, but we limit the following discuss to two families of models that are most relevant for our study: the pressure-regulated feedback-modulated (PRFM) theory \citep{2009Koyama,2010Ostriker,2011Kim,2013Kim,2022Ostriker} and the models based on LGI \citep[e.g.][]{1989Kennicutt,1998Kennicutt,2001MartinKennicutt}. 

In the PRFM theory, the SFR in the galactic disc self-regulates so that the gas pressure, which depends on the energy and momentum injected by stellar feedback into the surrounding gas, will balance the gravitational restoring force.
In this framework, the SFR surface density is expected to correlate with the pressure needed to reach this equilibrium, a prediction that has been verified empirically by several authors \citep[e.g.][but see also \citealp{2023Williams}]{2019Fisher,2022Ostriker,2022KadoFong,2023Sun,2024Ellison,2024Zhai}. 

The models based on LGI typically rely on the idea that an unstable gas discs will fragment into clumps and form GMCs, which will potentially collapse and form new stars \citep[e.g.][]{2022Meidt}. 
Such models usually imply the existence of a density threshold for star formation (Eq.~\ref{eq:Sigma_th}). 
\cite{2001MartinKennicutt} showed that this scenario could explain the radial distribution of HII regions in nearby star-forming discs. 
However, \cite{2008Leroy} later found that the star formation efficiency does not correlate with $Q$ in local spirals. 
On the other hand, LGI-based models seem to be suitable to describe extreme star-forming systems, such as starbursts at low and high redshifts \citep[e.g.][]{2016Romeo,2018Tadaki,2019Litke,2022Fisher}. 
In practice, the hypothesis is that star formation self-regulates as stellar feedback maintains the gas velocity dispersion at the level needed for marginal LGI \citep[e.g.][]{1997Silk,2013FaucherGiguere,2018Krumholz,2023Elmegreen}.   

The fact that there is no evidence of unstable regions in the nearby gas discs disfavours the hypothesis that star formation is primarily regulated by LGI and the density threshold (Eq.~\ref{eq:Sigma_th}), at least for galaxies in the present-day Universe that are on the SFMS. 
This supports the recent results by \cite{2019Bacchini,2019Bacchini_b,2020Bacchini_b}, who investigate the origin of the break in the Schmidt-Kennicutt law, the empirical correlation between the cold gas and SFR surface densities \citep{1959Schmidt,1989Kennicutt,1998Kennicutt}. 
They show that the break is not due to a drop in the star formation efficiency where $\Sigma < \Sigma_\mathrm{th}$, but it is rather due to the projection effects caused by the gas disc thickness. 
\cite{2019Bacchini,2019Bacchini_b,2020Bacchini_b} converted the observed surface densities into the de-projected volume densities using the gas disc scale height, which was derived using the same methodology used here. 
This approach was applied to 23 spiral and dwarf galaxies in our low-$\mathrm{z}$ sample, finding that they follow a very tight power-law correlation between the gas and SFR volume densities. 
This volumetric star formation (VSF) law holds from the low-density to the high-density regions of galaxies, without any clear indication of a break. 
Overall, the VSF law and the results of this work suggest that star formation in nearby galaxies is regulated by the gas volume density, which depends on the balance between gravity and the gas pressure due to thermal motions and SN-driven turbulence \citep{2020Bacchini}. 
This clearly supports the PRFM theory rather than regulation by LGI. 
It is worth noting that the scatter of the VSF law ($\approx 0.1$~dex) is smaller than the scatter of the SFR-pressure correlation ($\approx 0.2-0.3$~dex), possibly suggesting that the VSF law is more fundamental than the SFR-pressure correlation. 

Taking our results at face value, about 60\% of the galaxies in the high-z sample are locally unstable. 
Among these, we find SPT0418-47, SGP38326-1, SGP38326-2, and zc400569, which are classified as potentially interacting systems because of the presence of companion galaxies \citep[see][]{2023RomanOliveira,2023Spilker,2023Lelli,2023Cathey}. 
In these cases, the interaction may have triggered LGI and star formation. 
The presence of unstable regions does not seem to correlate with the SFR, as only five out of nine starburst galaxies have $\Qthreed<1$. 
These findings seem to disfavour LGI as the sole regulator of star formation in high-z systems, although it cannot be completely ruled out (see discussion in Sect.~\ref{sec:discussion_limitations}). 
As noted in Sect.~\ref{sec:discussion_turb}, the fact that the unstable regions are confined at the disc edges may indicate a transition between two different regimes of star formation. 

We note that our low-z sample can be considered representative of the population of galaxies on the SFMS in the present-day Universe (see Sect.~\ref{sec:sample}), but it does not contain any massive starburst.  
The sample at high redshift is relatively small and biased towards massive systems, but it contains both SFMS galaxies and starbursts. 
Unfortunately, since the high-z sample is mostly at redshift $4 \lesssim \mathrm{z} \lesssim 5$, we cannot drive any conclusion about the cosmic evolution of star formation regulation and turbulence driving. 
Such task that would require a large sample covering a wide redshift range; future efforts will be devoted to this goal. 

When considering possible relationships between LGI and star formation, it is important to keep in mind that both $Q$ and $\Qthreed$ can just tell us which parts of the gas disc may fragment into clumps and be potentially converted into stars. 
It is not guaranteed that all this gas will be used for star formation, a complex phenomenon that involves different physical processes at different spatial scales that are beyond the scope of this study (see Sect.~\ref{sec:discussion_limitations_analytical}). 

\subsection{Comparing clump masses with observations and simulations}\label{sec:discussion_clumps}
Rest-frame optical/UV observations of high-z galaxies revealed that they show an irregular and clumpy morphology \citep[e.g.][]{1996Abraham,1998Brinchmann,2004Conselice,2005Elmegreen,2011Genzel,2014Conselice,2015Zanella,2015Guo,2016Shibuya,2017DessaugesZavadsky,2019Zanella,2022Mestric,2023Claeyssens,2023HuertasCompany,2024Kalita,2024Messa,2024GimenezArteaga}. 
Investigating the origin and fate of such clumps is crucial to understand whether the build up stellar mass mainly proceeds via mergers or secular processes such as LGI. 
Massive clumps observed in rest-frame optical/UV images of high-z galaxies may simply be multiple star-forming regions, which are unresolved and blurred in observations at low spatial resolution \citep[e.g.][]{2017Tamburello}. 
Nonetheless, some authors argue that stellar clumps have \textit{ex situ} origin, being the remnants of merger events with satellite galaxies \citep[e.g.][]{2009Puech,2016Shibuya,2024Nakazato}. 
Alternatively, clumps may form \textit{in situ} from the fragmentation of the gas disc due to LGI \citep[e.g.][but see also \citealt{2021Zanella}]{2008Bournaud,2008Genzel,2015Guo,2015Zanella,2019Zanella}.  
There is a lively debate also about the fate of clumps. 
A few authors argue that clumps are long-lived structures, that can migrate inward and contribute to proto-bulge formation in high-z systems  \citep{2009Dekel_b,2010Ceverino,2015Zolotov}. 
However, other works showed that stellar feedback can rapidly destroy such clumps \citep[][]{2012Genel,2015Tamburello,2017Oklopvic}. 

Interferometric observations of dust continuum and CO or [CII] emission lines seem to indicate the presence of massive clumps of cold gas in high-z galaxies \citep[e.g.][]{2012Hodge,2015Swinbank,2017DessaugesZavadsky,2017Canameras,2018Canameras,2019Hodge,2021Ushio,2021Calura,2022Spilker,2023DessaugesZavadsky,2023Liu}, possibly suggesting that clumps form \textit{in situ} and can survive stellar feedback. 
Unfortunately, clump detection using interferometric data can be very uncertain. 
In fact, some authors showed that correlated noise structures can be misinterpreted as clumps, even for spatially resolved observations with moderate S/N \citep{2016Hodge,2018Gullberg,2019Rujopakarn,2020Ivison}. 
Moreover, clump properties are dramatically affected by the limited resolution of the observations, which can artificially merge different clumps into more massive and bigger objects, and by their sensitivity, which can cut out the population of faint clumps \citep[e.g.][]{2017Tamburello,2017DessaugesZavadsky,2017Fisher_a,2018Cava,2019Rujopakarn}. 
These issues can be mitigated by the effect of magnification in gravitationally lensed galaxies, which are among the best systems for measuring clump properties in distant galaxies \citep[e.g.][]{2019DessaugesZavadsky,2023DessaugesZavadsky,2024Zanella}. 

In order to understand whether the observed clumps are consistent or not with being a result of LGI, we compare the mass of cold gas clumps expected in unstable regions of our galaxies with measurements from observations. 
Unfortunately, interferometric observations of high-z galaxies typically do not have sufficient spatial resolution and S/N to unambiguously identify gas clumps \citep[for instance, see][]{2024RomanOliveira}, in particular in the unstable regions located at the disc edge where the S/N is typically low. 
Therefore, we decided to rely on literature measurements of clump masses in lensed systems. 
\cite{2019DessaugesZavadsky,2023DessaugesZavadsky} detect 17 and 14 CO clumps in two lensed star-forming galaxies at $\mathrm{z} \approx 1.04$ and measure clump masses of $8 \times 10^6 \lesssim \Mclump/\mathrm{M}_\odot\lesssim 1\times 10^9$ and $8 \times 10^5 \lesssim \Mclump/\mathrm{M}_\odot \lesssim 8\times 10^7~\mathrm{M}_\odot$, respectively. 
Furthermore, \cite{2012Hodge} identified five CO clumps with $M_\mathrm{clump} \approx 4-6 \times 10^9 \mathrm{M}_\odot$ in a sub-millimetre galaxy at $\mathrm{z} \approx 4.05$. 
Clump detections based on [CII] emission are limited to \cite{2022Spilker}, who find at least a dozen of clumps with $\Mclump \sim 10^{10}~\mathrm{M}_\odot$ in a lensed system at $\mathrm{z} \approx 6.9$, and \cite{2024Zanella}, who find three clumps with $M_\mathrm{clump} \approx 2.2-3.5 \times 10^8 \mathrm{M}_\odot$ in a lensed galaxy at $\mathrm{z} \approx 3.4$. 
The clump masses in Fig.~\ref{fig:clump_props} are overall within the range of these measurements at high redshift. 
The expected clump numbers in Fig.~\ref{fig:clump_props} tend to be higher than the number of observed clumps. 
This may be due to the limited resolution and sensitivity of the observations, which cannot resolve multiple clumps and/or detect only the brightest ones. 

Starbursts in the present-day Universe are typically rich of dense and turbulent gas, thus they are considered analogues of high-z galaxies \citep{2017Fisher_a}. 
Nearby starbursts host GMCs with typical masses of $\Mclump \sim 10^7-10^9~\mathrm{M}_\odot$ \citep[e.g.][]{2014ZaragozaCardiel,2015Leroy,2020Mok}, which are consistent with the clump masses in Fig.~\ref{fig:clump_props} and those measured in high-z systems \citep{2019DessaugesZavadsky,2023DessaugesZavadsky}. 
In interacting starbursts, \cite{2014ZaragozaCardiel} and \cite{2020Mok} find more that one hundred GMCs, which is higher than the expected clumps number in Fig.~\ref{fig:clump_props}. 
We stress tough that $N_\mathrm{clump}$ is an indicative value obtained by assuming that all clumps have mass equal to $\langle M_\mathrm{clump} \rangle$, as our methodology cannot predict the clump mass distribution. 

Also numerical simulations of galaxies at high redshift find that gas discs fragment into clumps with masses $\sim 10^7-10^9~\mathrm{M}_\odot$ \citep[e.g.][]{2009Agertz,2014Mandelker,2015Tamburello,2016Behrendt,2017Mandelker}. 
In general, the clump masses in Fig.~\ref{fig:clump_props} decreases with increasing galactocentric distances. 
This is consistent with the clumps properties in numerical simulations \citep[e.g.][]{2014Mandelker,2017Mandelker} and the observed mass of stellar clumps, which show a negative radial gradient \citep[e.g.][]{2018Guo,2019Zanella,2024Kalita}. 
In conclusion, our results are congruent with the scenario in which GMCs and, potentially, stellar clumps form \textit{in situ} because of LGI, at least at $z\gtrsim 2$. 
We note though that this conclusion relies on the implicit assumption that the observed clumps are not remnants of merging satellites, a scenario that cannot be ruled out with our approach. 
In fact, numerical simulations showed that there is some overlap between the typical masses of \textit{in situ} and \textit{ex situ} clumps, even though the former tend to be less massive but more numerous than the latter \citep[e.g.][]{2014Mandelker,2019Zanella}. 

The unstable regions are typically found at the edge of the disc of our galaxies (see Fig.~\ref{fig:q3d_all}), suggesting that clumps formed \textit{in situ} via LGI should be preferentially found at large galactocentric distance. 
However, this is not necessarily the case if clumps migrate towards the galaxy centre, provided that they are not immediately destroyed by stellar feedback. 
Given the potential issues affecting clump detections from interferometric observations, the radial distribution of gas clumps in high-z galaxies is essentially unconstrained, hampering a sound comparison. 
Since also \textit{ex situ} clumps are expected to be preferentially found at the disc edge \citep[e.g.][]{2014Mandelker,2017Mandelker}, high-resolution spectroscopic observations of the clump stellar component are also required in order to disentangle between \textit{in situ} clumps and accreted structures \citep[see also][]{2019Zanella}.

%%%%%%%%%%%%%%%%%%%%%%%%%%%%%%%%%%%%%%%%%%%%%%%%%%%%%%%%%%%%%%%%%%%%%%%%%%%%%%%%%%%%%%%%%%%%%%%%%%%%%%
\section{Summary and conclusions}\label{sec:conclusions}
We investigate the local gravitational instability (LGI) of gaseous discs in a sample of 44 star-forming galaxies at $0 \lesssim \mathrm{z} \lesssim 5$. 
Our methodology uses the novel 3D instability criterion $\Qthreed<1$ (Eq.~\ref{eq:q3d_nipoti23}) by \cite{2023Nipoti}, which allows us to self-consistently account for the stabilising effect of the gas vertical stratification in the galactic potential. 
The results obtained with the 3D criterion are compared with those based on the classical 2D instability criterion $Q<1$ (Eq.~\ref{eq:q2d}) by \cite{1964Toomre}, which assumes an infinitesimally thin gas disc. 
Based on \cite{1964Toomre}'s criterion, the only unstable system at $\mathrm{z}\approx 0$ is a DM-poor ultra-diffuse galaxy, while nine high-z systems out of 13 ($\approx 70$\%) are unstable. 
Using \cite{2023Nipoti}'s 3D criterion, we find that the ultra-diffuse galaxy and one high-z system have $Q<1$ but $\Qthreed>1$, indicating no instability. 
Hence, eight high-z galaxies ($\approx 60$\%) are locally unstable based on $\Qthreed<1$. 
The unstable regions are typically found at the disc edge and their radial extent is on average $\approx 20$\% smaller compared to the 2D analysis. 
We estimate that at most $\approx 50$\% of the total gas in the disc is in the unstable region, with median value of $\approx 30$\%. 
This is about half of the gas that would be affected by the instability according to the 2D analysis. 
The typical mass of gas clumps that are expected to form in the unstable regions is $7.6 \lesssim \log \left( \langle M_\mathrm{clump} \rangle / \mathrm{M}_\odot\right) \lesssim 9.5$. 
This is congruent with the mass of gas clumps observed in high-z galaxies and GMCs in low-z starbursts, which are considered their present-day analogues. 
This result is consistent with the hypothesis that the clumpy morphology of galaxies at high redshift may be the result of LGI rather than mergers, although the latter cannot be completely ruled out and may contribute in triggering LGI. 

The main conclusions of this work are the following. 
\begin{enumerate}
\item The classical expression for $Q$ is a rather simple and useful tool, but it is fundamentally limited by the unrealistic assumption of an infinitesimally thin gas disc, which is more prone to instability than a thick disc. 
The vertical structure of gas discs in galaxies has a significant stabilising effect that must be taken into account. 
$\Qthreed$ is very helpful for this task, as it allows a self-consistent treatment of the gas vertical structure in the instability analysis. 
 
\item The 3D analysis of LGI indicates that less discs are unstable and less gas is affected by the instability compared to the classical 2D approach. 
In fact, the fraction of unstable gas with respect to the total gas in the disc is lower by a factor of 2 when the 3D approach is adopted. 
The unstable regions, when present, are $\approx 20$\% less extended and located at larger galactocentric distances when $\Qthreed$ is used instead of $Q$.

\item Based on the $\Qthreed$ criterion, we do not find unstable regions in the cold gas discs (either atomic or molecular) in nearby dwarf and spiral galaxies, corroborating previous studies. 
This suggests that LGI does not play a major role in regulating star formation and gas turbulence in the present-day Universe, at least for galaxies on the star-forming main sequence. 
Our results are consistent with other empirical studies supporting that star formation self-regulates via stellar feedback in present-day galaxies. 

\item About 60\% of the galaxies in the high-z sample host gas discs that are locally unstable according to $\Qthreed$. 
Typically, the unstable regions are about 1-3~kpc in size and located at the disc edge, with about $30$\% of the total gas being involved by the instability. 
This suggest that LGI can contribute to driving turbulence and promote the star formation activity at high redshift, but the input by other mechanisms seems required in a significant portion of the disc. 
\end{enumerate}
The approach tested in this paper is relatively easy to implement, but requires careful modelling of gas kinematics and mass distributions in galaxies. 
These pieces of information are already available for many galaxies in the present-day Universe and, hopefully, will be accessible for an increasing number of distant systems thanks to new observational facilities.
This is crucial to reduce the uncertainties on this kind of studies, which are significantly larger for high-z galaxies than those in the present-day Universe.  
Follow-up studies will aim at expanding the galaxy sample by including low-z starbursts and more systems at high redshift. 
This work also highlights the potential importance of the gas vertical structure in preventing LGI in gas discs, especially in baryon-dominated regions. 

\section*{Acknowledgements}	
We thank the referee, Alessandro Romeo, for thoroughly reading the paper and for providing a prompt and constructive report. 
We would like to thank Filippo Fraternali for useful discussions on this work. 
CB acknowledges support from the European Research Council (ERC) under the European Union’s Horizon 2020 research and innovation program (grant agreement No. 833824, GASP project), and from the Carlsberg Foundation Fellowship Programme by Carlsbergfondet. 
The research activities described in this paper have been co-funded by the European Union – NextGenerationEU within PRIN 2022 project n.20229YBSAN - Globular clusters in cosmological simulations and in lensed fields: from their birth to the present epoch. 
PEMP acknowledges support from the Dutch Research Council (NWO) through the Veni grant VI.Veni.222.364.

%
%-------------------------------------------------------------------

\bibliographystyle{aa}
\bibliography{paty.bib}

%-------------------------------------------------------------------
\onecolumn
\begin{appendix}

\section{Mass models}\label{app:mass_models}
Table~\ref{tab:mass_models} reports, for each galaxy in the sample, the parameters to derive the mass distribution of the stellar components and the DM halo using the models described in Sects.~\ref{sec:method_model} and~\ref{sec:sample}. 
\begin{table*}[h]
\centering
\caption{Parametric mass models for the DM and stellar components in our sample. 
$\ddag$~The stellar disc is modelled using Eq.~\ref{eq:poly} with the best-fit parameters given by \cite{2024ManceraPina}
% $C_1=-1.6$, $C_2=0.09$, $C_3 \approx 0$, and $C_4 \approx 0$. 
$\dagger$~The top and bottom rows in the third block are for the MW thin and thick stellar discs, respectively.}
\label{tab:mass_models}
\begin{tabular}{c|l|ccccc|cccc|ccc}
\hline\hline
ID & Galaxy   & \multicolumn{5}{c|}{Dark matter halo}                & \multicolumn{4}{c|}{Stellar disc} 				  & \multicolumn{3}{c}{Sersic component}	\\
\hline

&          & Type &$\rho_\mathrm{DM,0}$& $r_\mathrm{s}$& $r_\mathrm{c}$& $\eta$&$\Sigma_{\star,0}$&$R_\star$& $\xi$   & $z_\star$& $\rho_\mathrm{b,0}$   & $r_\mathrm{e}$ &  $n$ \\
&          &      & $\mathrm{M}_\odot$kpc$^{-3}$& kpc			  & kpc		      &       &$\mathrm{M}_\odot$kpc$^{-2}$& kpc    &         & kpc      &$\mathrm{M}_\odot$kpc$^{-3}$ & kpc            &      \\
\hline
1  & AGC114905$\ddag$& cNFW & 6.5$\times 10^{4}$& 38.77 		  & 0.17			  & 1    & 3.7$\times 10^{6}$& 1.3   &$\sech^2$& 0.20     & -                  & -              & -		\\

2  & ALESS073.1& NFW & 2.0$\times 10^{7}$& 17.3			  & -     		  & - 	  & 7.8$\times 10^{9}$& 0.70   &$\exp$   & 0.30     & 4.0$\times 10^{13}$& 0.30            & 4		\\

3  & BRI1335-0417& NFW & 2.4$\times 10^{7}$& 8.97 	      & - 			  & - 	  & -                 & -      & -       & -        & 1.4$\times 10^{13}$& 2.18 & 6.5\\

4  & CVIDWA   & cNFW & 4.1$\times 10^{7}$& 1.15           & 2.0           & 0.96  & 1.4$\times 10^{6}$& 0.68   &$\sech^2$& 0.14     & -                  & -              & -	\\

5  & DDO101   & cNFW & 9.0$\times 10^{7}$& 2.66           & 1.7           & 0.99  & 3.1$\times 10^{7}$& 0.58   &$\sech^2$& 0.12     & -                  & -              & -	\\

6  & DDO126   & cNFW & 3.7$\times 10^{7}$& 180            & 2.4           & 0.95  & 3.8$\times 10^{6}$& 0.82   &$\sech^2$& 0.16     & -                  & -              & -	\\

7  & DDO133   & cNFW & 6.7$\times 10^{7}$& 2.01           & 2.4           & 0.98  & 7.5$\times 10^{6}$& 0.80   &$\sech^2$& 0.16     & -                  & -              & -	\\

8  & DDO154  & cNFW & 1.5$\times 10^{7}$ & 3.38           & 1.6           & 0.81  & 4.6$\times 10^{6}$& 0.54   &$\sech^2$& 0.11     & -                  & -              & -	\\

9  & DDO168  & cNFW & 2.2$\times 10^{7}$ & 3.39           & 2.4           & 0.88  & 1.4$\times 10^{7}$& 0.82   &$\sech^2$& 0.16     & -                  & -              & -	\\

10  & DDO210  & cNFW & 4.1$\times 10^{7}$ & 0.85           & 0.6           & 0.96  & 2.2$\times 10^{6}$& 0.22   &$\sech^2$& 0.04     & -                  & -              & -	\\

11 & DDO216  & cNFW & 5.2$\times 10^{7}$ & 0.77           & 1.5           & 0.97  & 9.0$\times 10^{6}$& 0.52   &$\sech^2$& 0.10     & -                  & -              & -	\\

12 & DDO47   & cNFW & 3.7$\times 10^{7}$ & 3.55           & 2.1           & 0.95  & 3.1$\times 10^{7}$& 0.70   &$\sech^2$& 0.14     & -                  & -              & -	\\

13 & DDO50   & cNFW & 7.0$\times 10^{7}$ & 1.16           & 2.6           & 0.99  & 2.2$\times 10^{7}$& 0.89   &$\sech^2$& 0.18     & -                  & -              & -	\\

14 & DDO52   & cNFW & 2.4$\times 10^{7}$ & 2.73           & 2.8           & 0.89  & 9.5$\times 10^{7}$& 0.94   &$\sech^2$& 0.19     & -                  & -              & -	\\

15 & DDO53   & cNFW & 4.2$\times 10^{7}$ & 1.50           & 2.6           & 0.96  & 3.0$\times 10^{6}$& 0.89   &$\sech^2$& 0.18     & -                  & -              & -	\\

16 & DDO87   & cNFW & 2.5$\times 10^{7}$ & 2.63           & 3.3           & 0.90  & 4.1$\times 10^{6}$& 1.13   &$\sech^2$& 0.23     & -                  & -              & -	\\

17 & IC2574   & ISO  & 5.0$\times 10^{6}$& - 			  & 6.2           & -     & 1.5$\times 10^{7}$& 2.85   &$\sech^2$& 0.57     & -                  & -              & -	\\

18 & MW$\dagger$& NFW  & 8.5$\times 10^{6}$& 19.6         & -             & -     & 8.9$\times 10^{8}$ & 2.50  &$\exp$   & 0.30     & -                  & -              & -	\\
&          &      &					 &				  &				  &		  & 1.8$\times 10^{8}$ & 3.02  &$\exp$   & 0.90     & -                  & -              & -	\\
19 & J81740   & NFW  & 2.2$\times 10^{7}$& 21.29		  & - 			  & - 	  & -				  & - 	   & - 		 & - 		& 1.8$\times 10^{13}$& 2.76 & 6.8\\
20 & NGC0925  & cNFW & 6.5$\times 10^{6}$& -              & 8.9           & -     & 6.9$\times 10^{7}$& 4.10   &$\sech^2$& 0.82     & -                  & -              & -	\\
21 & NGC1569  & cNFW & 3.3$\times 10^{7}$& 2.10           & 1.3           & 0.94  & 2.8$\times 10^{8}$& 0.45   &$\sech^2$& 0.09     & -                  & -              & -	\\
22 & NGC2366  & cNFW & 2.4$\times 10^{7}$& 3.44           & 4.5           & 0.89  & 4.7$\times 10^{6}$& 1.54   &$\sech^2$& 0.31     & -                  & -              & -	\\
23 & NGC2403  & ISO  & 1.4$\times 10^{8}$& -		      &1.5            & -     & 1.8$\times 10^{8}$& 1.81   &$\sech^2$& 0.36     & -                  & -              & -	\\
24 & NGC2841  & NFW  & 5.6$\times 10^{7}$& 10.39          & -             & -     & 6.8$\times 10^{8}$& 4.20   &$\sech^2$& 0.84     & 1.4$\times 10^{10}$& 0.66           & 1	\\
25 & NGC2976  & ISO  & 4.3$\times 10^{7}$& -              & 2.6           & -     & 2.5$\times 10^{8}$& 0.90   &$\sech^2$& 0.18     & -                  & -              & -	\\
26 & NGC3198  & ISO  & 4.5$\times 10^{7}$& -              & 2.8           & -     & 3.0$\times 10^{8}$& 3.06   &$\sech^2$& 0.61     & -                  &  -              & - 	\\
27 & NGC3741  & cNFW & 2.2$\times 10^{7}$& 3.08           & 4.1           & 0.88  & 1.4$\times 10^{8}$& 0.16   &$\sech^2$& 0.03     & -                  & -              & -	\\
28 & NGC4736  &  NFW & 2.9$\times 10^{9}$& 0.56           & -             & -     & 5.3$\times 10^{8}$& 1.99   &$\sech^2$& 0.34     & 6.1$\times 10^{10}$& 0.23           & 1	\\
29 & NGC5055  & ISO  & 1.1$\times 10^{7}$& - 			  & 7.2           & -     & 1.2$\times 10^{9}$& 3.2    &$\sech^2$& 0.64     & 4.5$\times 10^{10}$& 0.32           & 1	\\
30 & NGC6946  & ISO  & 3.1$\times 10^{7}$& - 	          & 4.8           & -     & 7.5$\times 10^{8}$& 2.97   &$\sech^2$& 0.59     & 2.4$\times 10^{11}$& 0.14           & 1	\\
31 & NGC7331  &  NFW & 4.7$\times 10^{6}$& 27.48          & -             & -     & 1.2$\times 10^{9}$& 3.3    &$\sech^2$& 0.66     & 1.2$\times 10^{11}$& 0.29           & 1   \\
32 & NGC7793  & ISO  & 9.4$\times 10^{7}$& -              & 2.0           & -     & 4.2$\times 10^{8}$& 1.3    &$\sech^2$& 0.26     & -                  & -              & -   \\
33 & SGP38326-1 & NFW & 2.4$\times 10^{7}$& 43.15 		  & - 			  & - 	  & - 				  & - 	   & - 		 & - 	    & 8.5$\times 10^{12}$& 2.59 & 5.7\\
34 & SGP38326-2 & NFW & 2.4$\times 10^{7}$& 47.28		  & -             & -     & -                 & -      & -       & -        & 2.4$\times 10^{12}$& 1.84 & 5.4\\
35 & SPT0418-47& NFW & 1.7$\times 10^{7}$& 23.1           & -             & -     & -                 & -      & -       & -        & 1.9$\times 10^{12}$& 0.22           & 2.2	\\
36 & SPT0113-46& NFW & 1.7$\times 10^{7}$& 20.5           & -             & -     & -                 & -      & -       & -        & 4.6$\times 10^{13}$& 1.40           & 6.0	\\
37 & SPT0345-47& NFW & 1.8$\times 10^{7}$& 11.2           & -             & -     & -                 & -      & -       & -        & 4.2$\times 10^{11}$& 0.30           & 1.5	\\
38 & SPT0441-46& NFW & 2.0$\times 10^{7}$& 35.7           & -             & -     & -                 & -      & -       & -        & 1.6$\times 10^{13}$& 0.11           & 2.0	\\
39 & SPT2132-58& NFW & 2.3$\times 10^{7}$& 5.6            & -             & -     & -                 & -      & -       & -        & 1.4$\times 10^{9}$ & 1.40           & 0.9 \\
40 & SPT2146-55& NFW & 2.1$\times 10^{7}$& 9.0            & -             & -     & -                 & -      & -       & -        & 9.9$\times 10^{10}$& 0.39           & 1.6	\\
41 & UGC8508  & cNFW & 3.9$\times 10^{7}$& 1.81           & 0.9           & 0.95  & 1.3$\times 10^{7}$& 0.31   &$\sech^2$& 0.06     & -                  & -              & - 	\\
42 & WLM      & cNFW & 2.3$\times 10^{7}$& 2.46 		  & 2.2			  & 0.89  & 4.6$\times 10^{6}$& 0.75   &$\sech^2$& 0.15     & -                  & -              & -	\\
43 & zC400569  & NFW & 8.5$\times 10^{6}$& 29.1           & -             & -     &	2.4$\times 10^{8}$& 3.82   &$\exp$   & 0.46     & 1.1$\times 10^{12}$&1.01			  & 4	\\
44 & zC488879  & NFW & 3.1$\times 10^{6}$& 80.8           & -             & -     & 1.3$\times 10^{8}$& 4.55   &$\exp$   & 0.51     & 2.2$\times 10^{11}$& 3.04           & 4	\\
\hline
\end{tabular}
\end{table*}

\newpage

\section{Gas scale heights, $Q(R)$, and $\Qthreed(R,z)$ for the whole sample.}\label{ap:q_all_sample}
In Figs.~\ref{fig:h_Q_Q3D}, the left panels display the radial profiles of the gas scale height (solid black curve) with the corresponding uncertainty (grey band). 
The central panels show the radial profiles of $Q$ (solid black curve) with the corresponding uncertainty obtained with the error propagation rules (grey band). 
The red dashed area in the bottom part of the panel indicates the instability regime ($Q<1$) and the vertical black dashed lines bracket the unstable regions. 
The blue dotted line indicates $Q=2/3$, which is the critical value for thick discs (see \citealt{1992Romeo,1994Romeo} and Sect.~\ref{sec:discussion_limitations_analytical}). 
The right panels show the map of $\Qthreed$ coloured according to its value in logarithmic scale. 
The unstable regions where $\Qthreed<1$ are, if present, coloured in dark-red and encompassed by a white solid contour. 
The vertical black dashed lines are the same as in the central panels and indicate where $Q<1$. 
The gray dot-dashed curve shows the scale height of the gas disc (same as the left panel). 
We note that the gaps in $Q(R)$ and $\Qthreed(R,z)$ indicate the regions where $\kappa^2<0$ because of local fluctuations in the observed data.

\begin{figure*}[htp] 
\includegraphics[width=1\columnwidth]{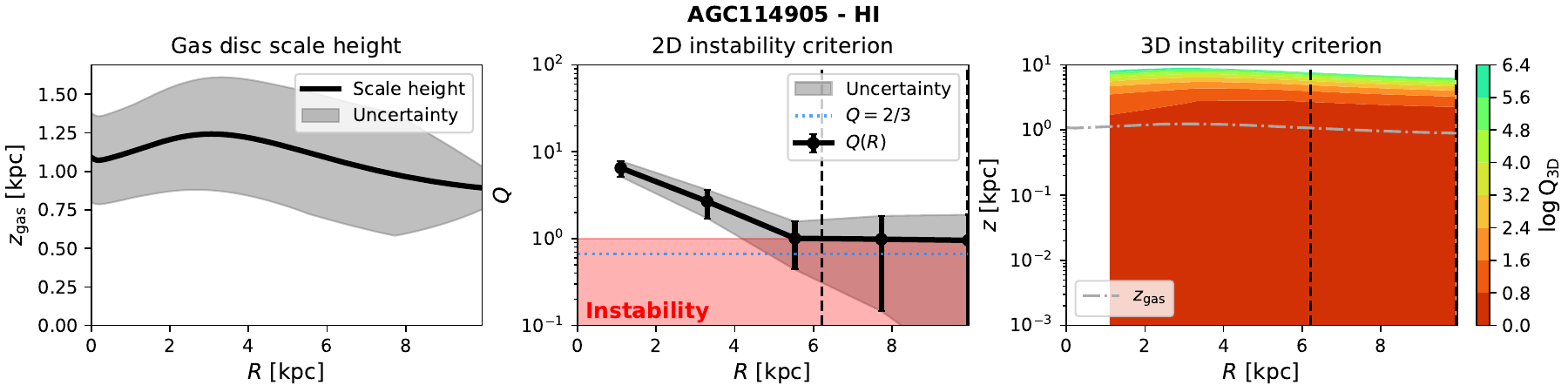} 
\includegraphics[width=1\columnwidth]{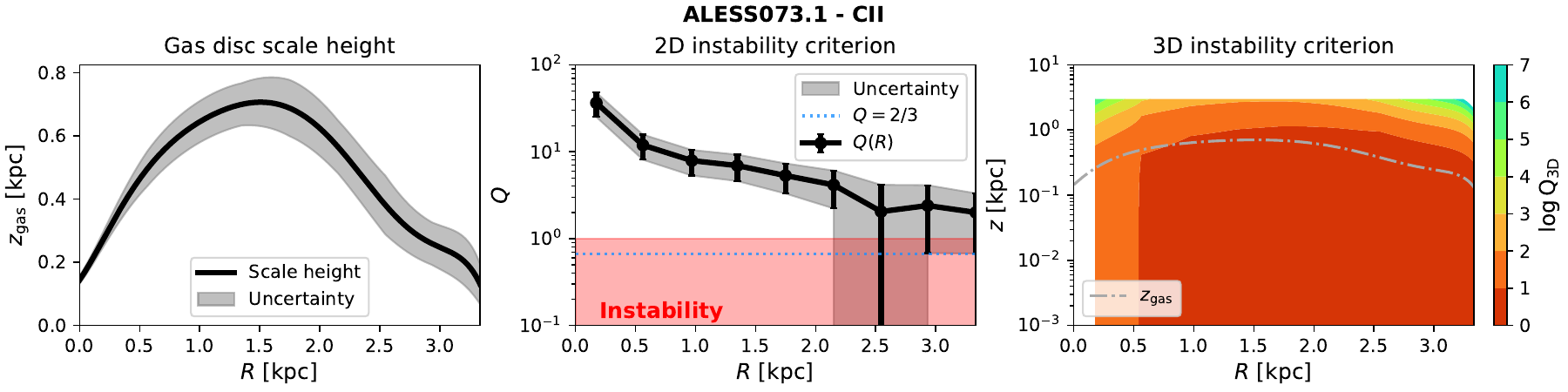} 
%\caption{Radial profiles of the gas disc scale heights in the sample of galaxies. 
%The top panels are for the atomic gas discs in present-day Universe divided in bins of increasing stellar mass from left to right. 
%	The bottom left panel is for the molecular gas discs at $\mathrm{z} \approx 0$. 
%	The central and right panels in the bottom row are for gas discs, traced by either CO or [CII] emission lines, in high-z galaxies on the SFMS and the SBS, respectively.}
\includegraphics[width=1\columnwidth]{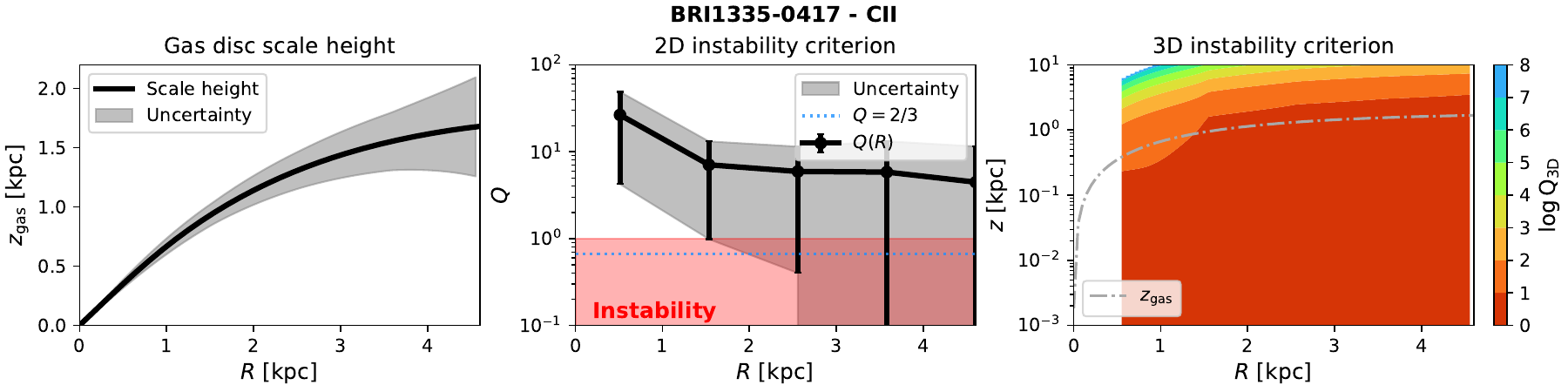} 
\includegraphics[width=1\columnwidth]{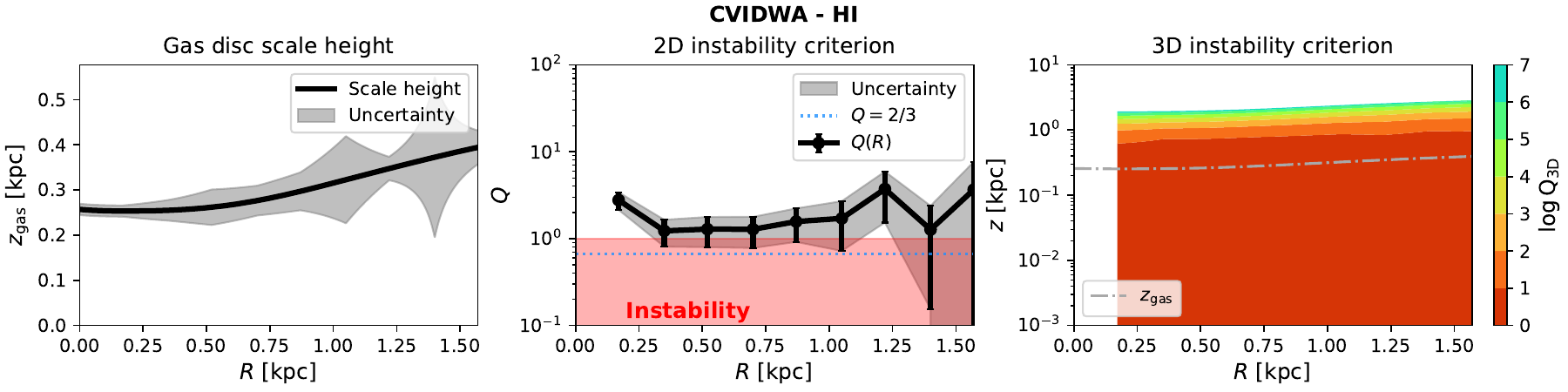} 
\caption{}
\label{fig:h_Q_Q3D}
\end{figure*}

\begin{figure*}[htp] \ContinuedFloat 
 \includegraphics[width=1\columnwidth]{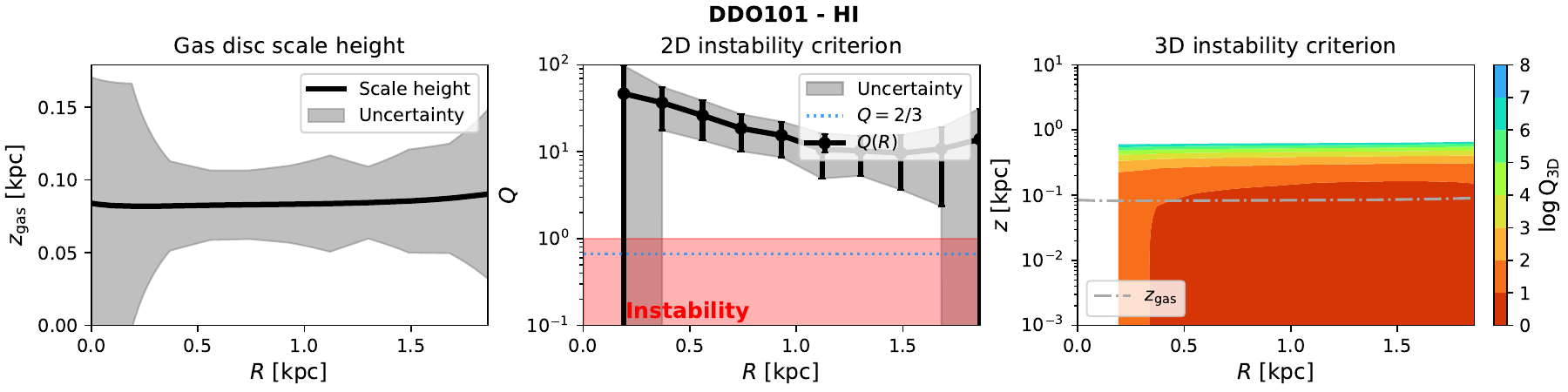} 
 \includegraphics[width=1\columnwidth]{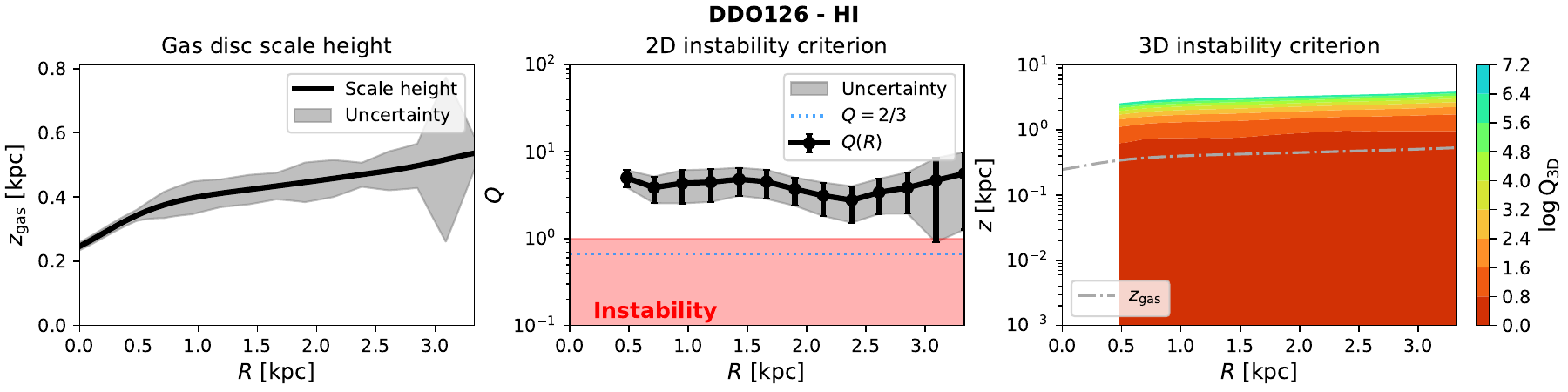} 
 \includegraphics[width=1\columnwidth]{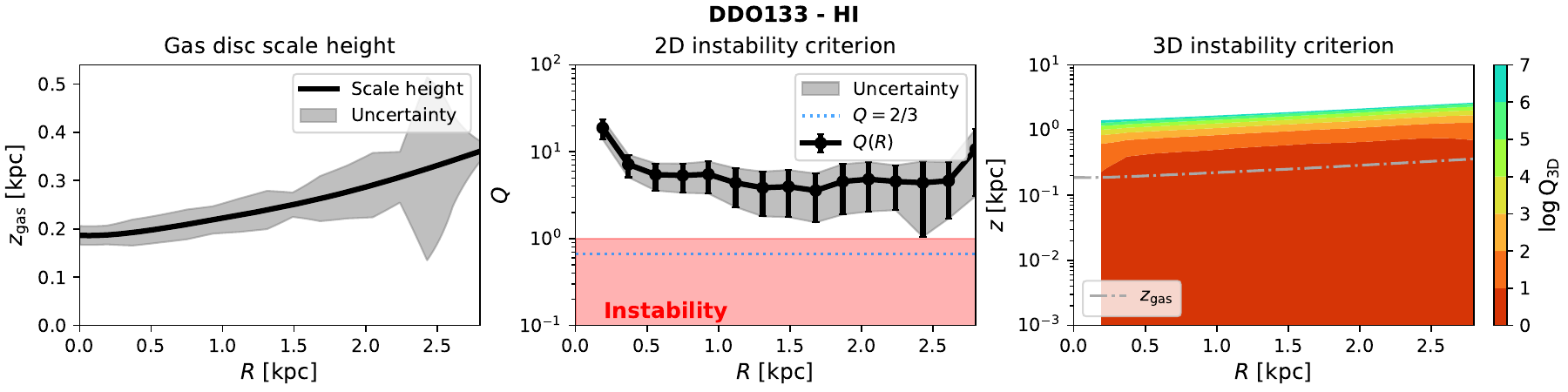} 
 \includegraphics[width=1\columnwidth]{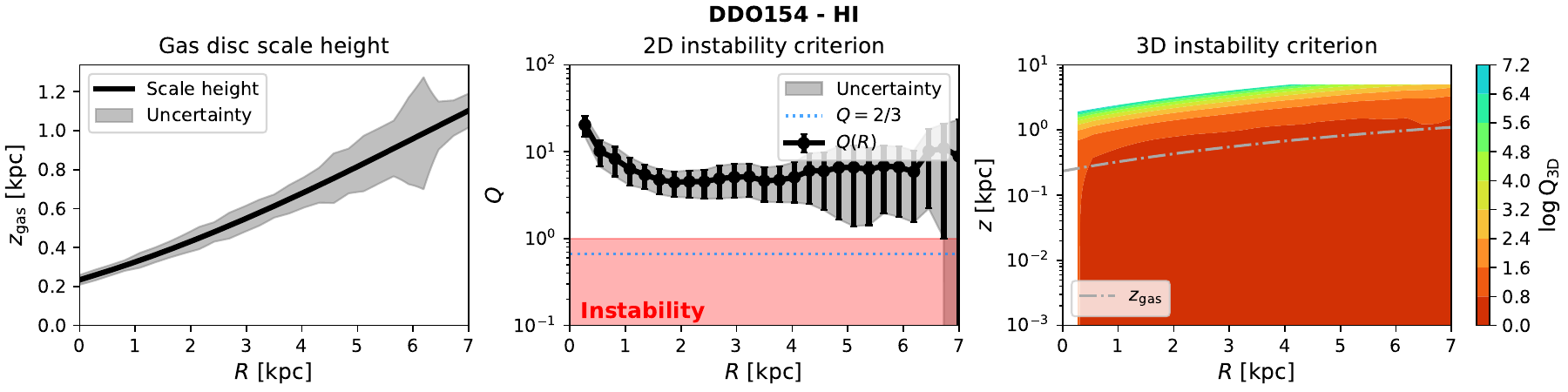} 
 \includegraphics[width=1\columnwidth]{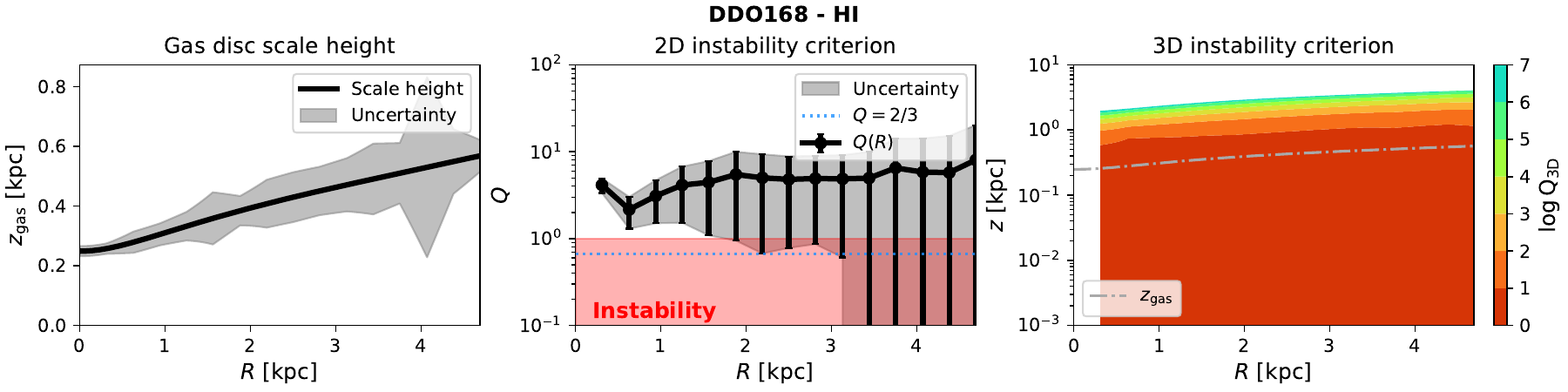} 
 \caption{continued.}
\end{figure*} 

\begin{figure*}[htp] \ContinuedFloat 
 \includegraphics[width=1\columnwidth]{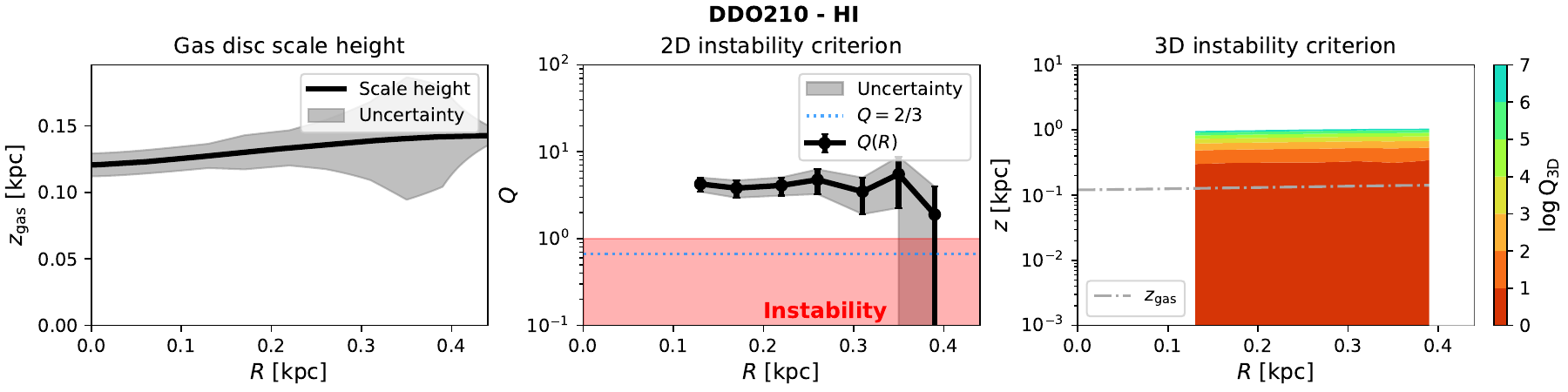} 
 \includegraphics[width=1\columnwidth]{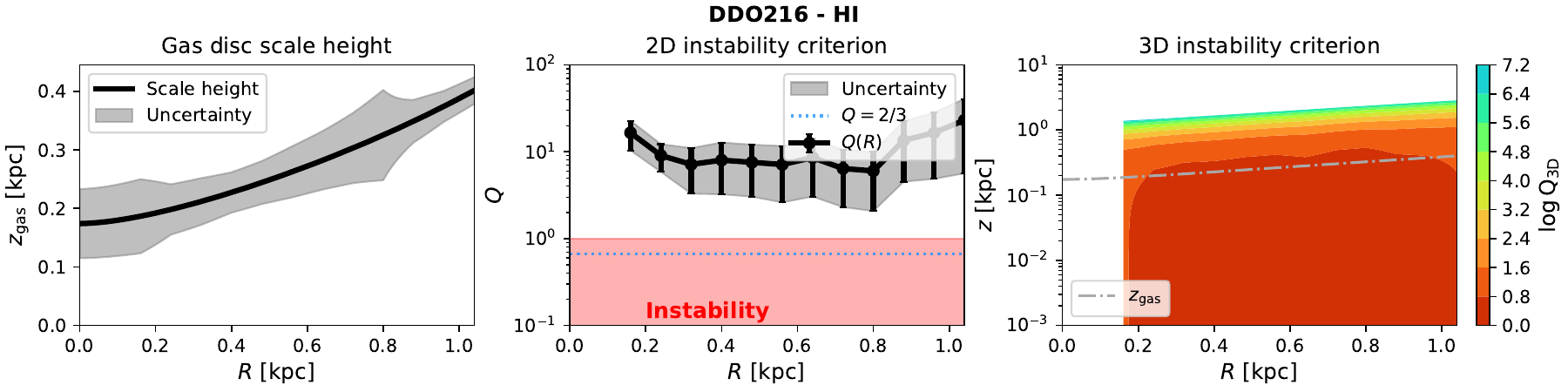} 
 \includegraphics[width=1\columnwidth]{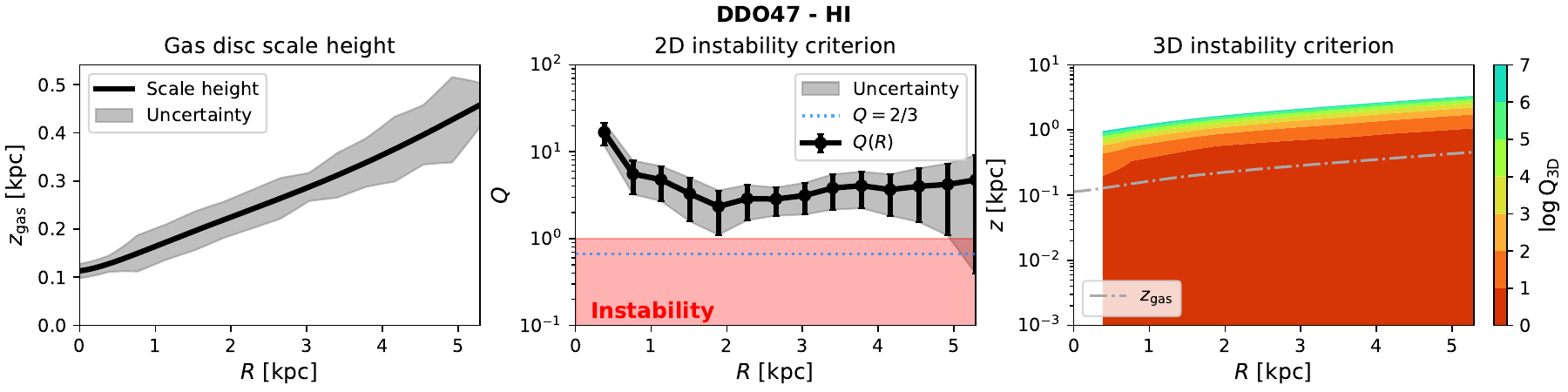} 
 \includegraphics[width=1\columnwidth]{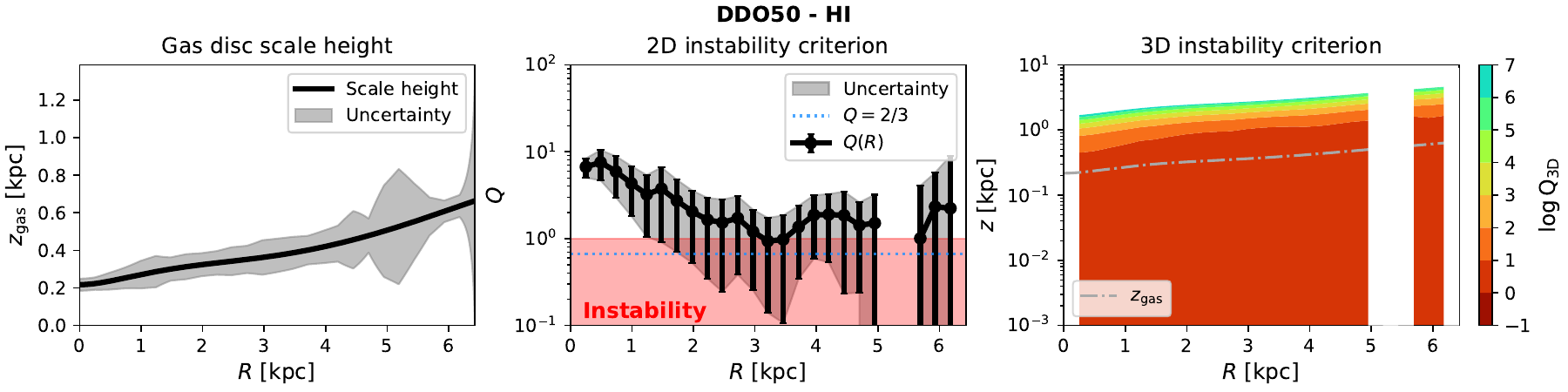} 
 \includegraphics[width=1\columnwidth]{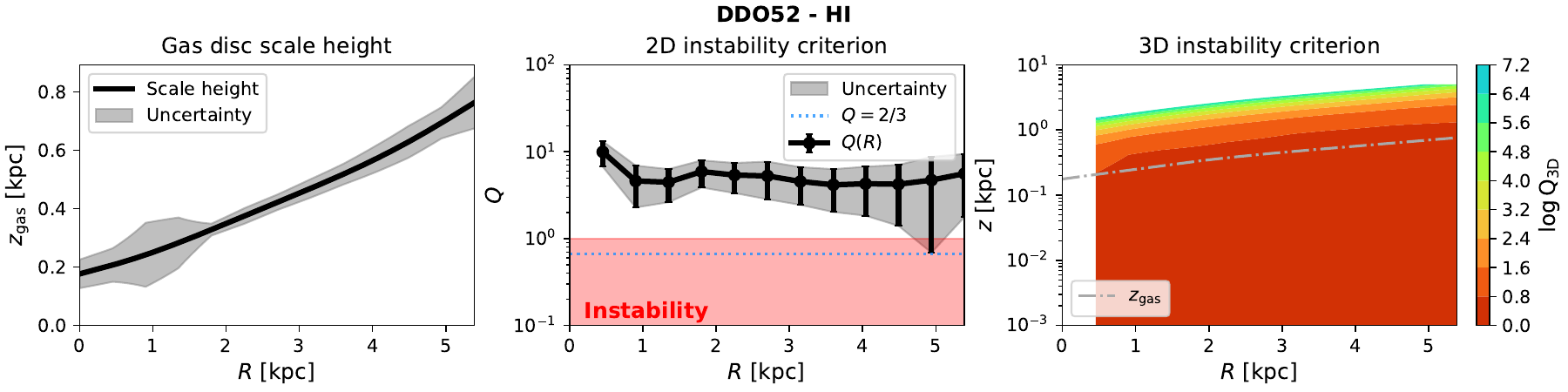} 
 \caption{continued.}
\end{figure*} 

\begin{figure*}[htp] \ContinuedFloat 
 \includegraphics[width=1\columnwidth]{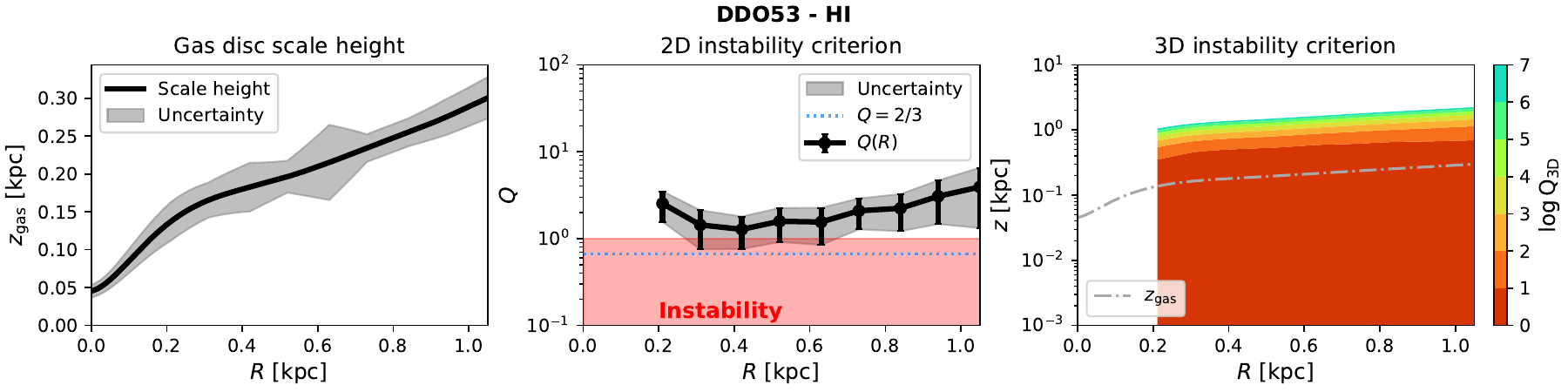} 
 \includegraphics[width=1\columnwidth]{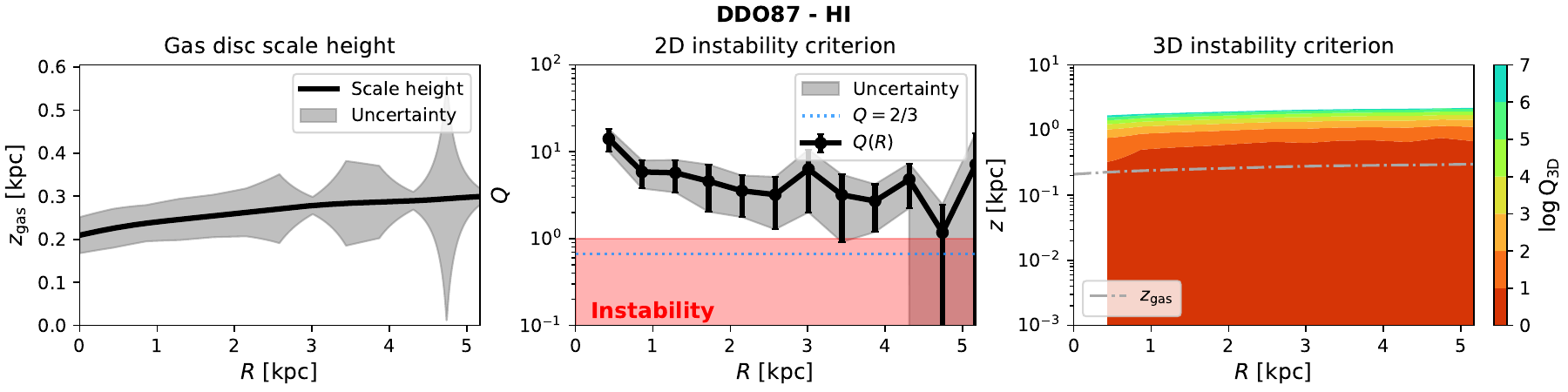} 
 \includegraphics[width=1\columnwidth]{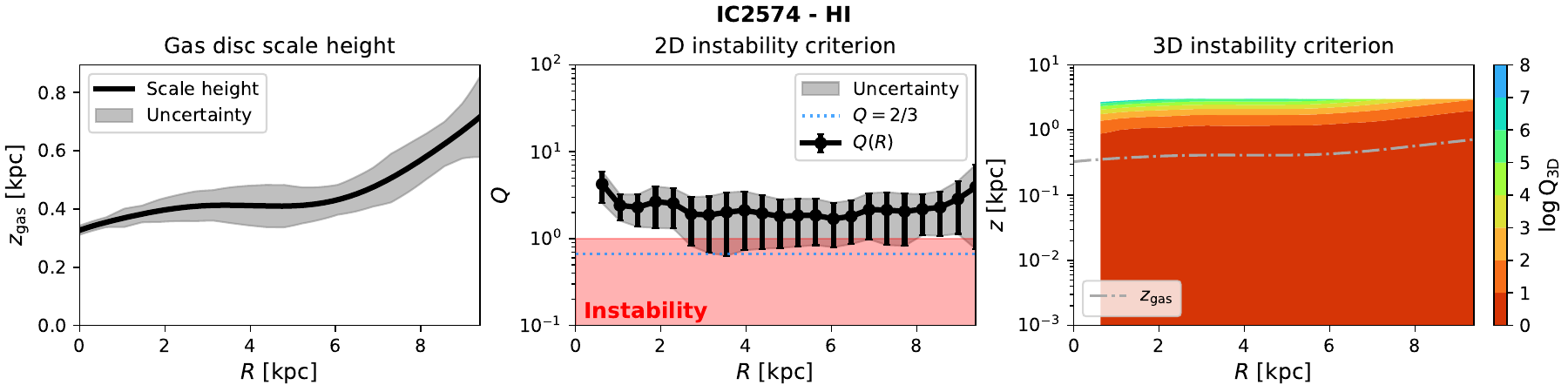} 
 \includegraphics[width=1\columnwidth]{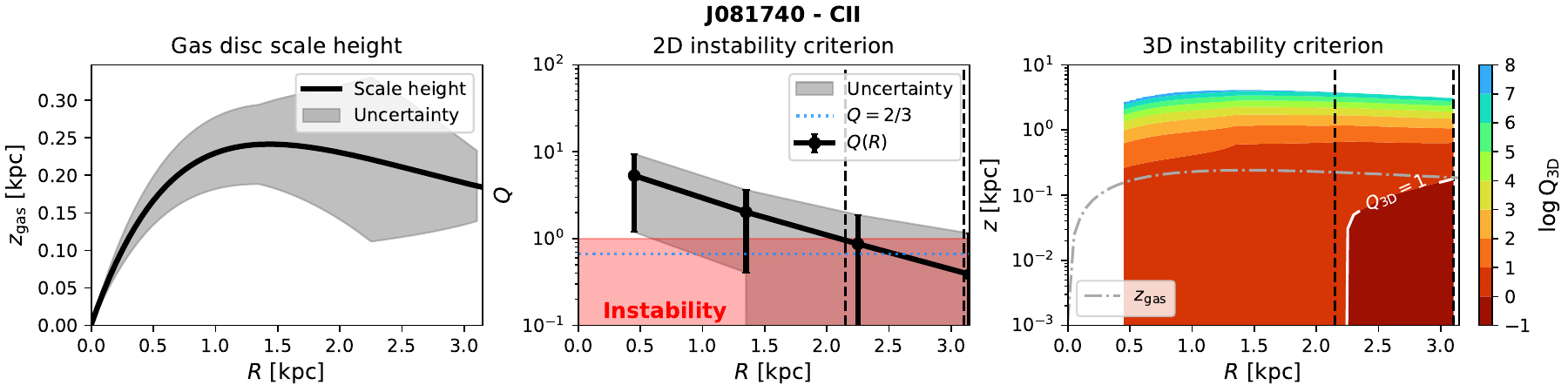} 
 \includegraphics[width=1\columnwidth]{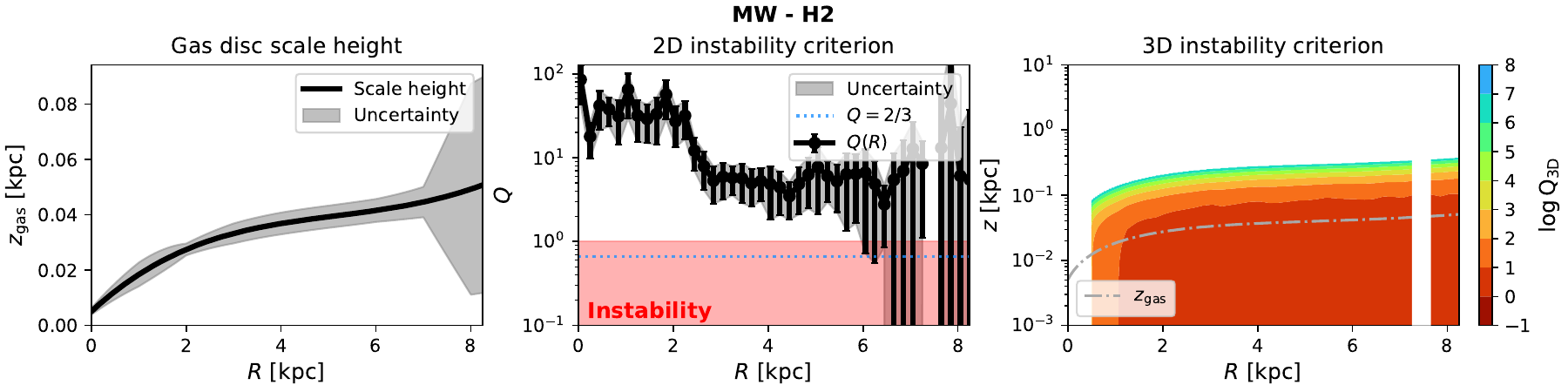} 
 \caption{continued.}
\end{figure*} 

\begin{figure*}[htp] \ContinuedFloat 
 \includegraphics[width=1\columnwidth]{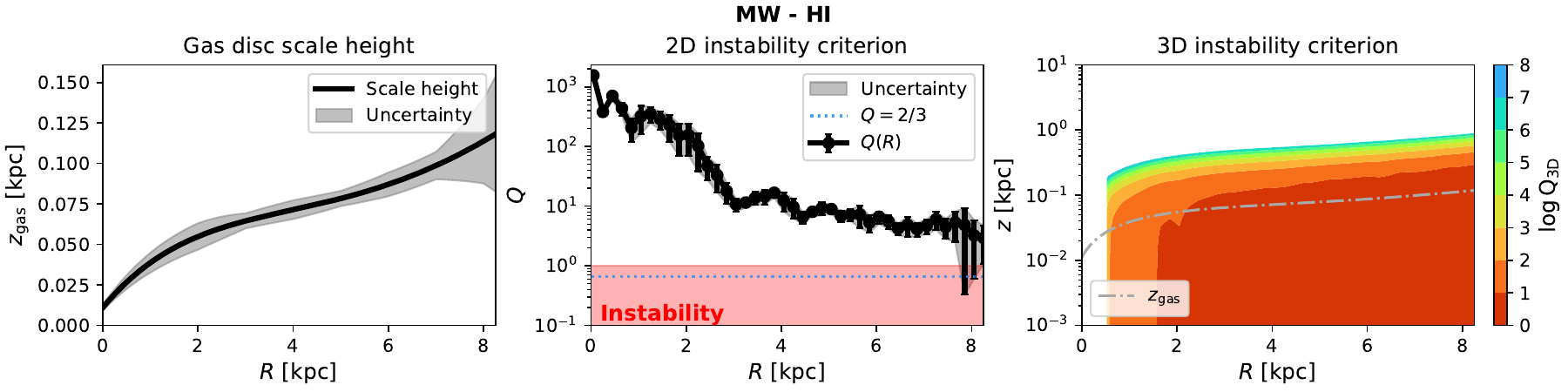} 
 \includegraphics[width=1\columnwidth]{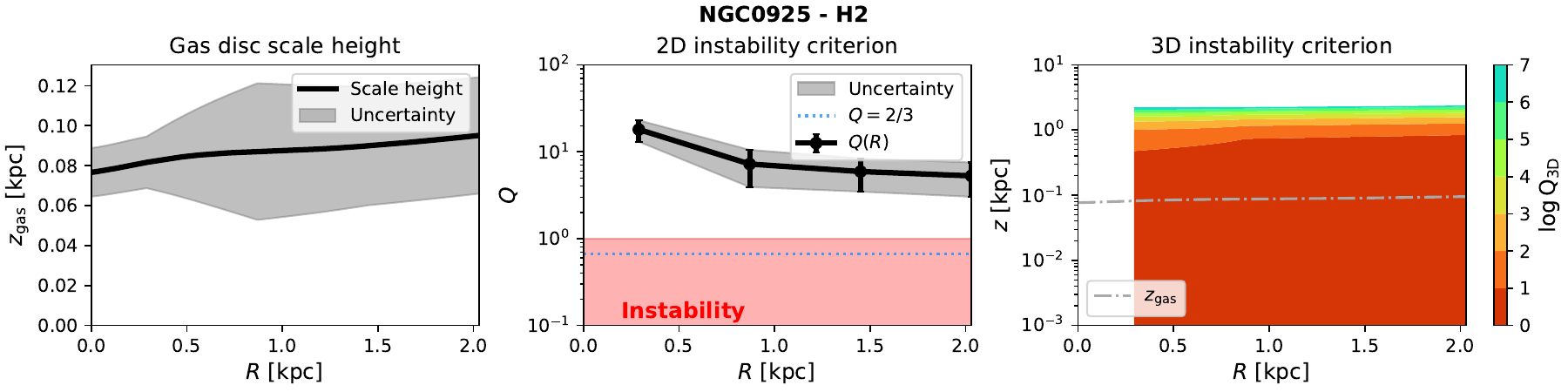}
 \includegraphics[width=1\columnwidth]{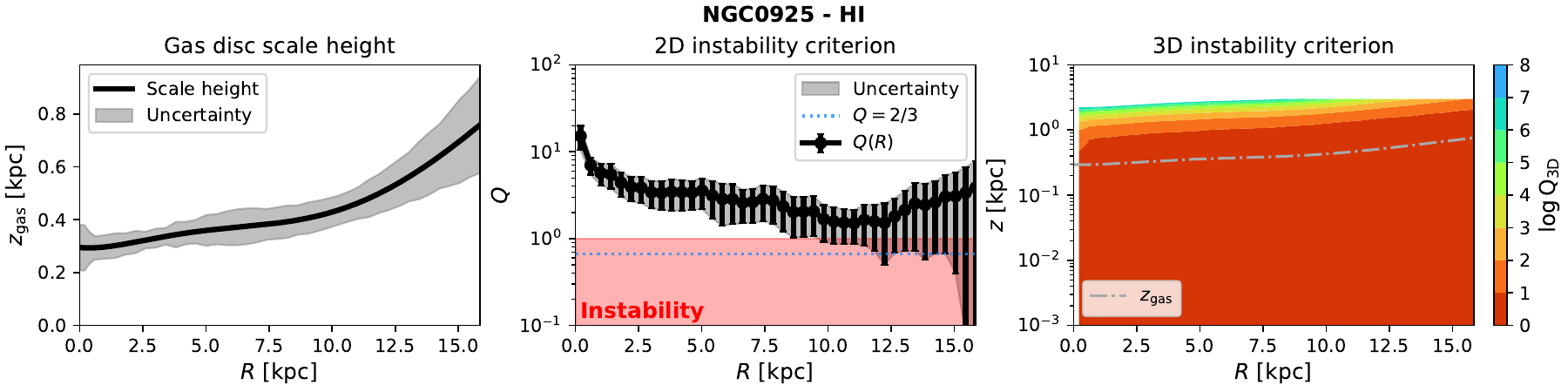} 
 \includegraphics[width=1\columnwidth]{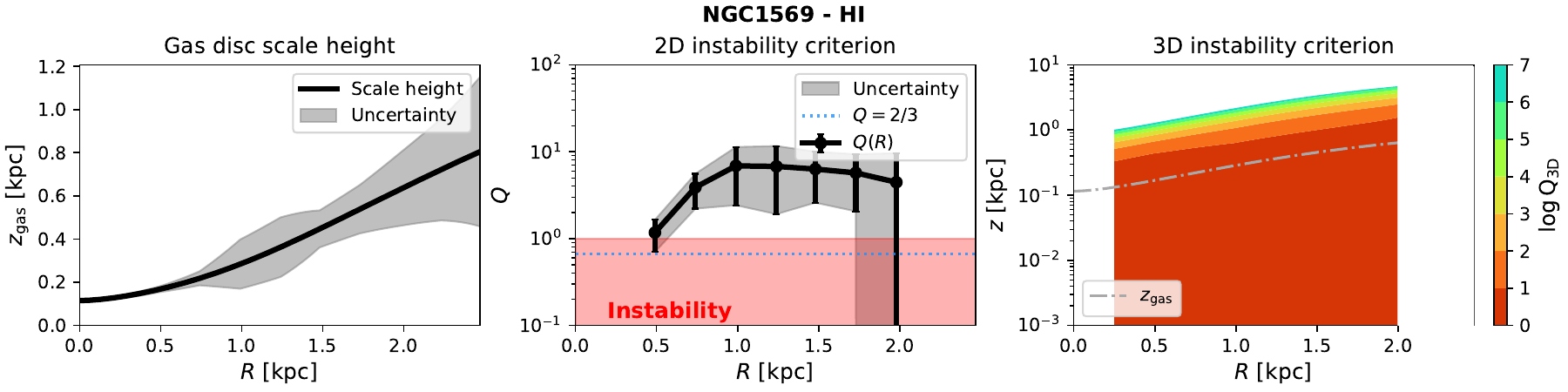} 
 \includegraphics[width=1\columnwidth]{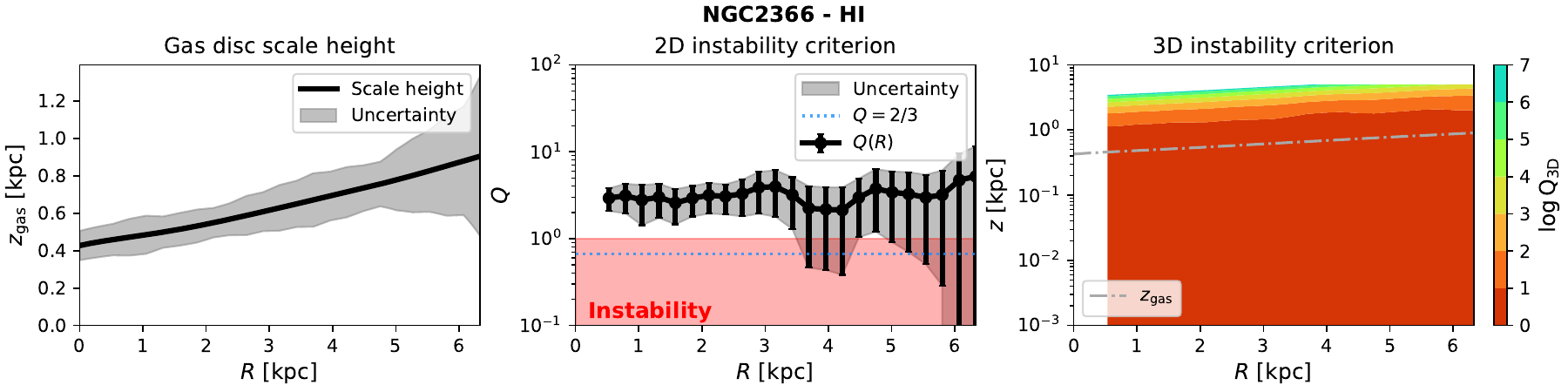} 
 \caption{continued.}
\end{figure*} 

\begin{figure*}[htp] \ContinuedFloat 
 \includegraphics[width=1\columnwidth]{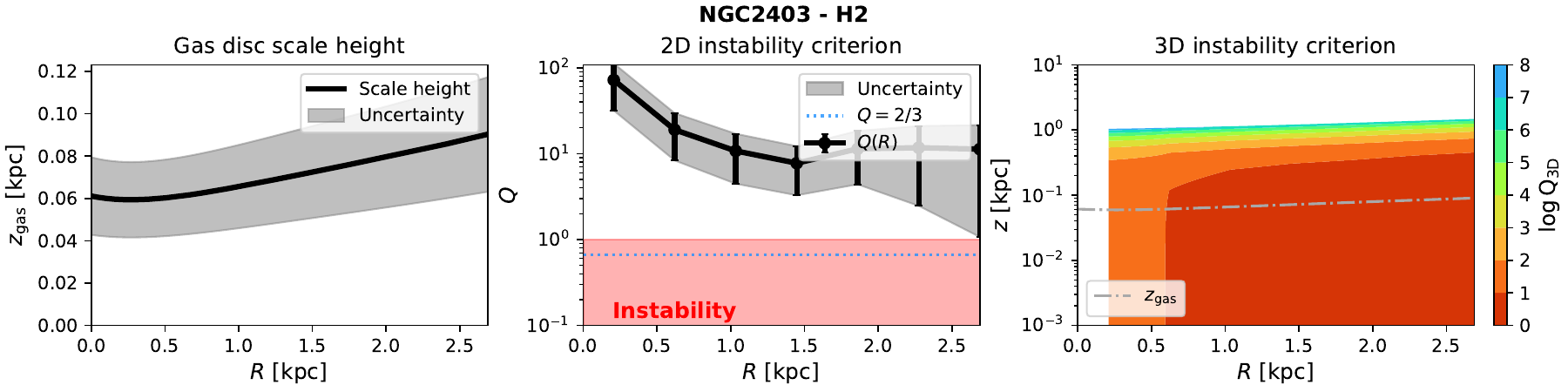} 
 \includegraphics[width=1\columnwidth]{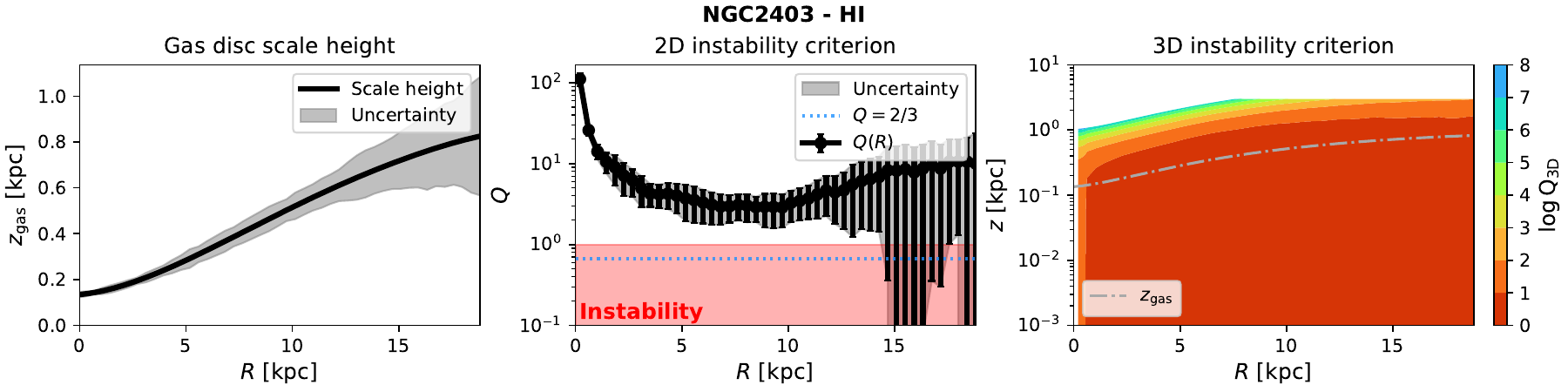}
 \includegraphics[width=1\columnwidth]{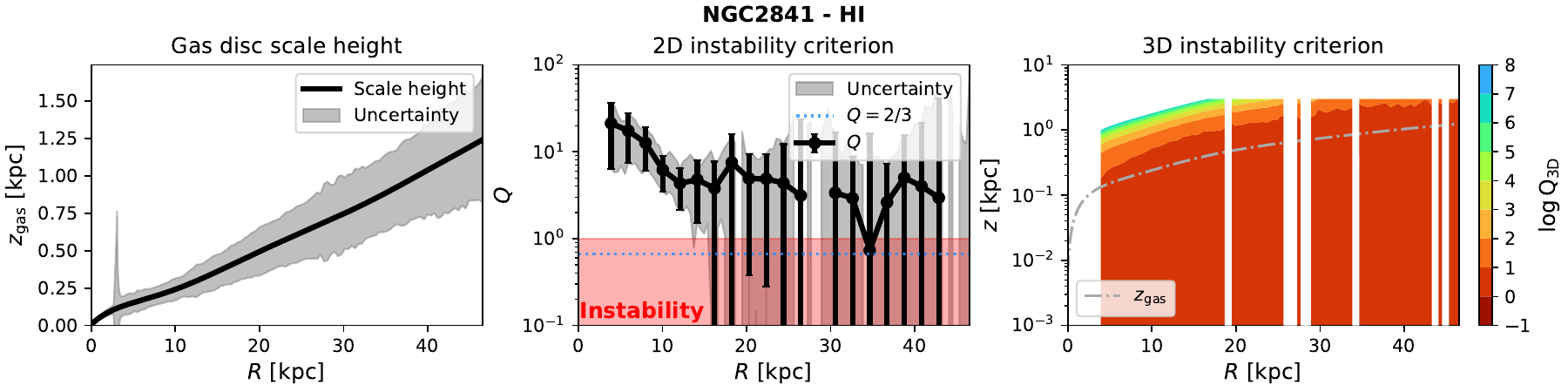}
 \includegraphics[width=1\columnwidth]{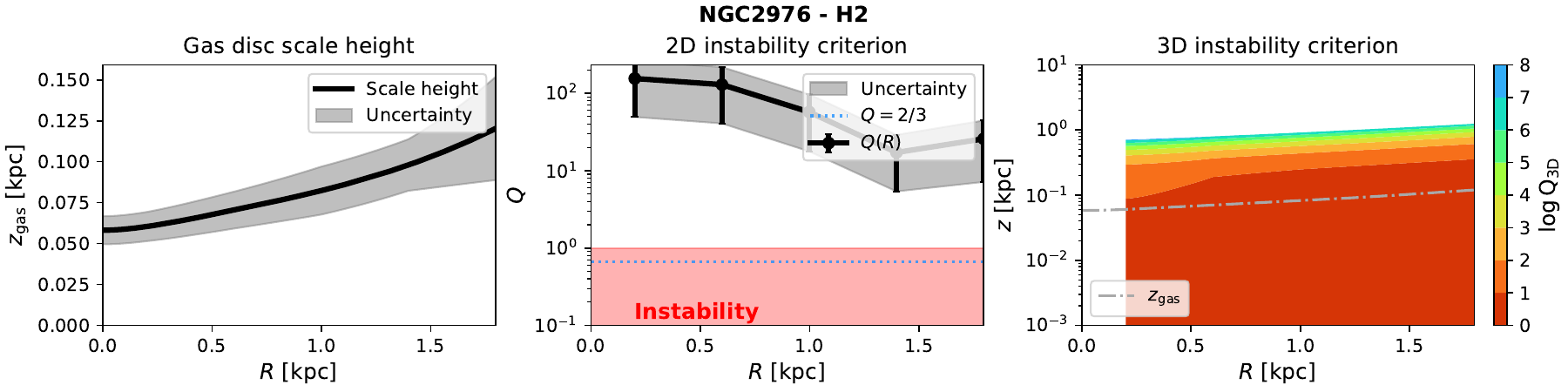}
 \includegraphics[width=1\columnwidth]{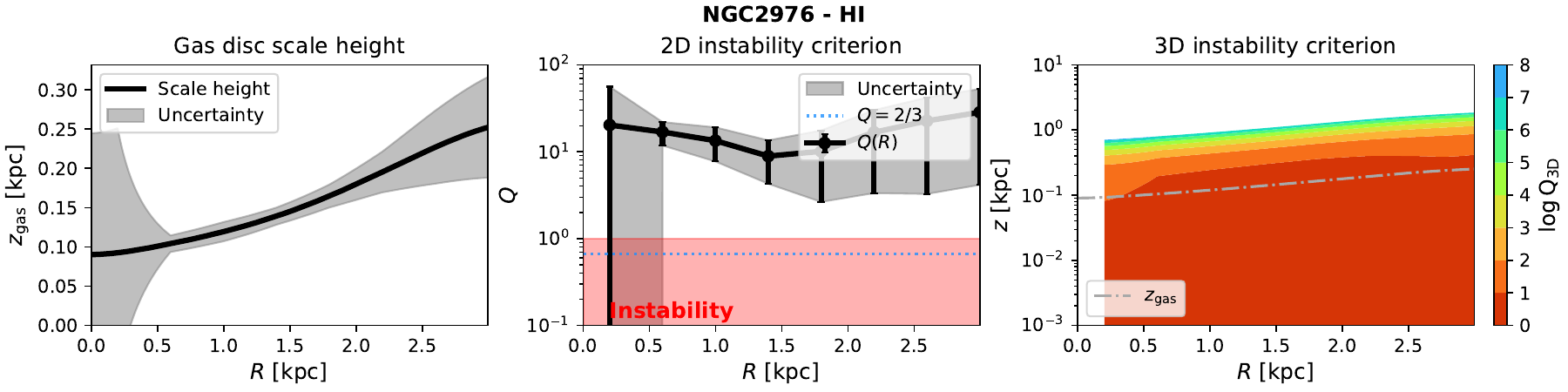}
 \caption{continued.} 
\end{figure*} 

\begin{figure*}[htp] \ContinuedFloat 
 \includegraphics[width=1\columnwidth]{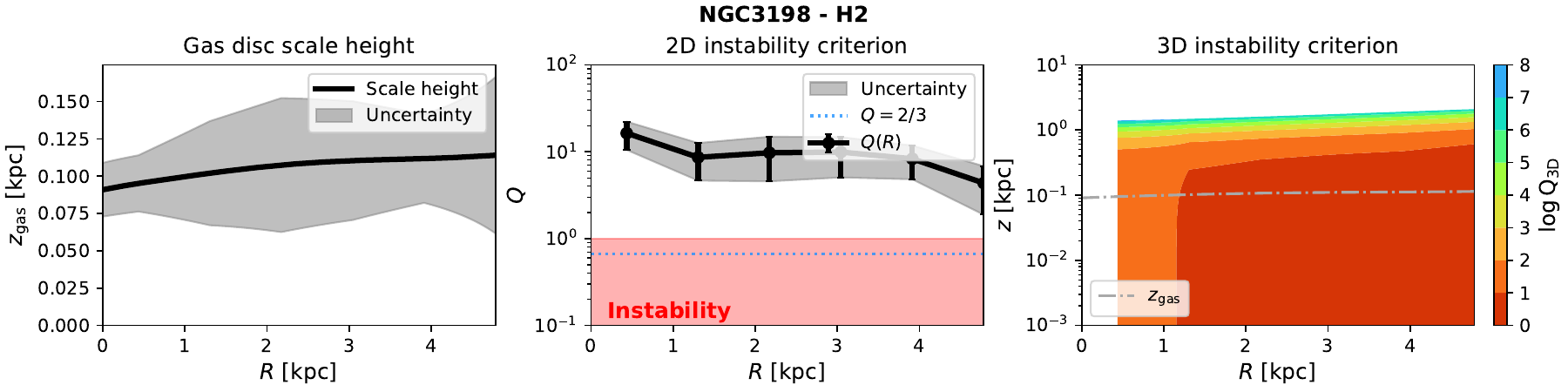} 
 \includegraphics[width=1\columnwidth]{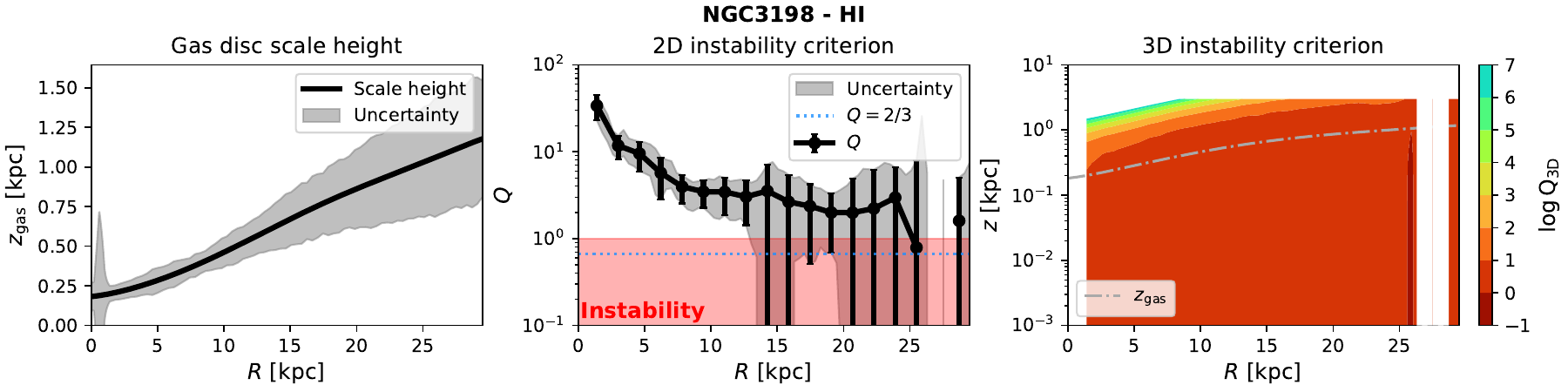} 
 \includegraphics[width=1\columnwidth]{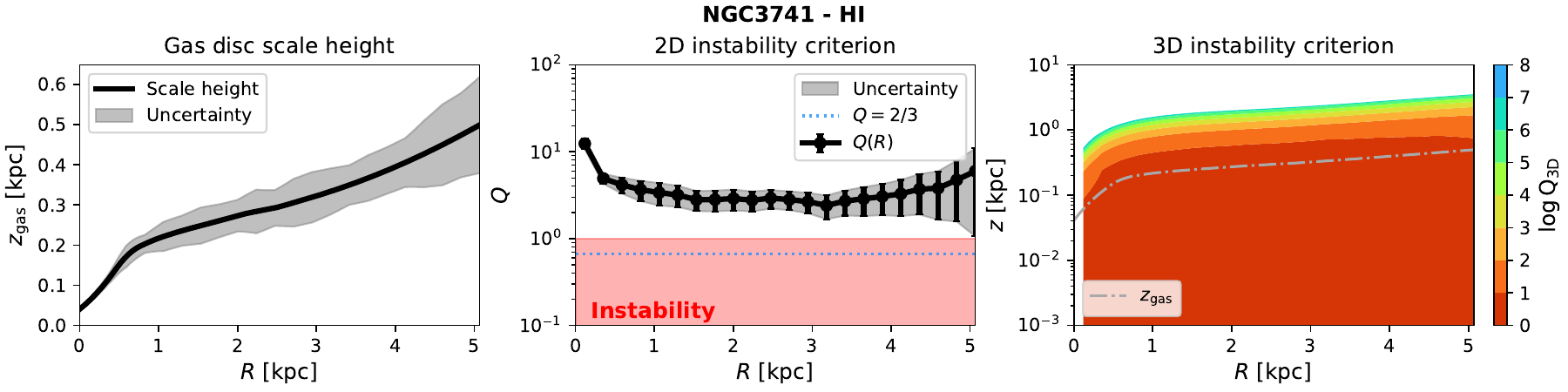} 
 \includegraphics[width=1\columnwidth]{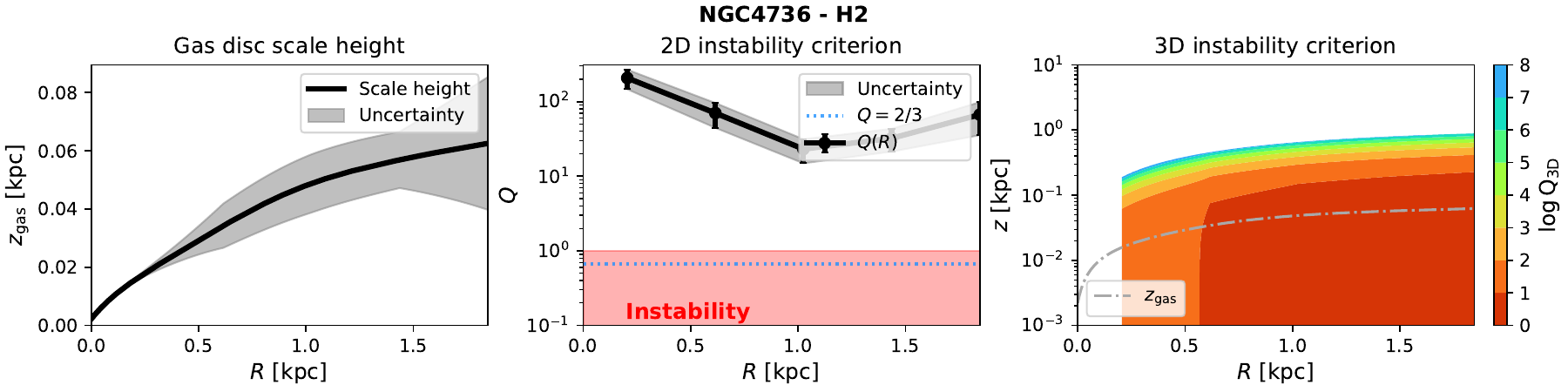} 
 \includegraphics[width=1\columnwidth]{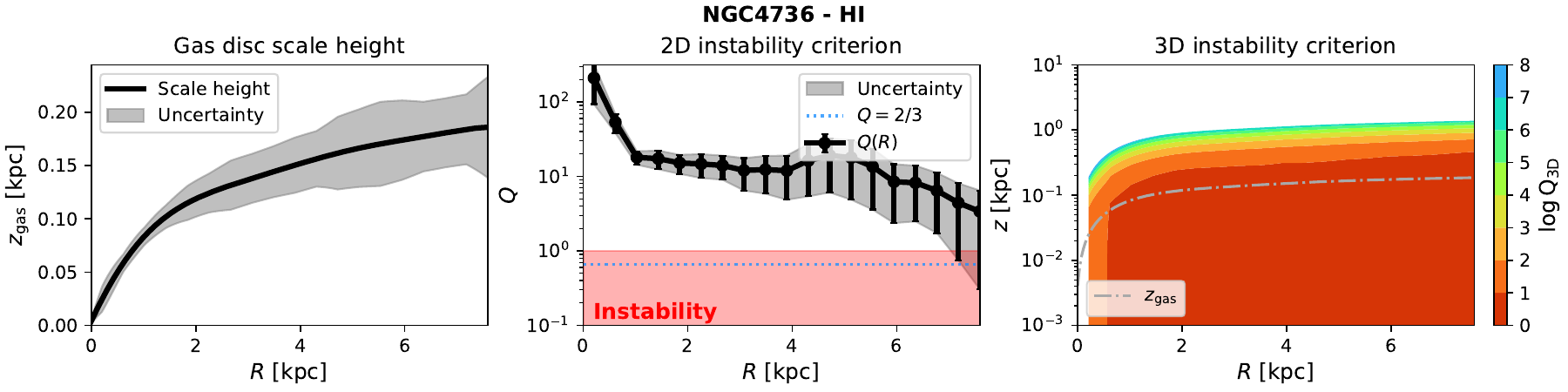} 
 \caption{continued.}
\end{figure*} 

\begin{figure*}[htp] \ContinuedFloat 
 \includegraphics[width=1\columnwidth]{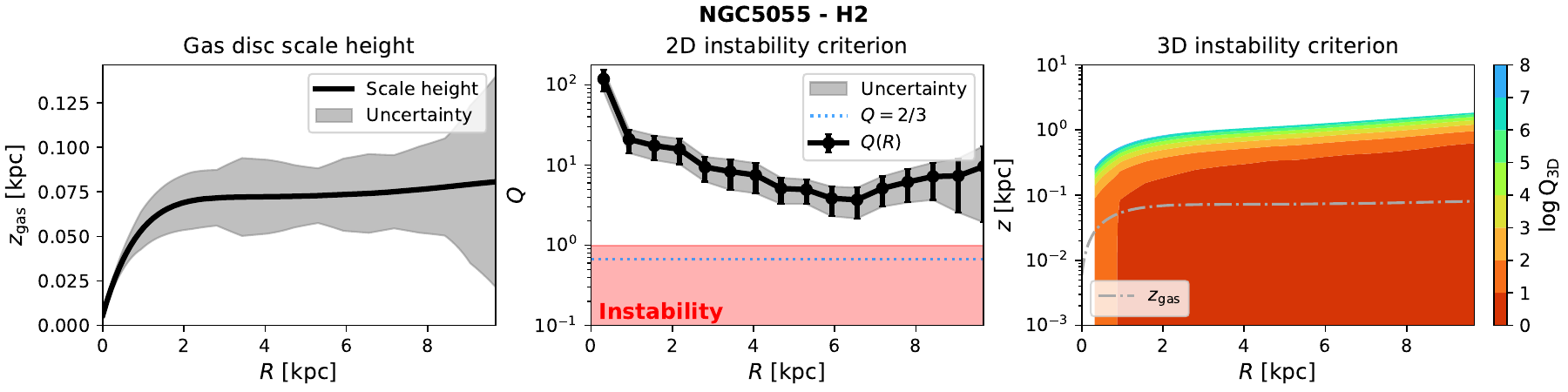} 
 \includegraphics[width=1\columnwidth]{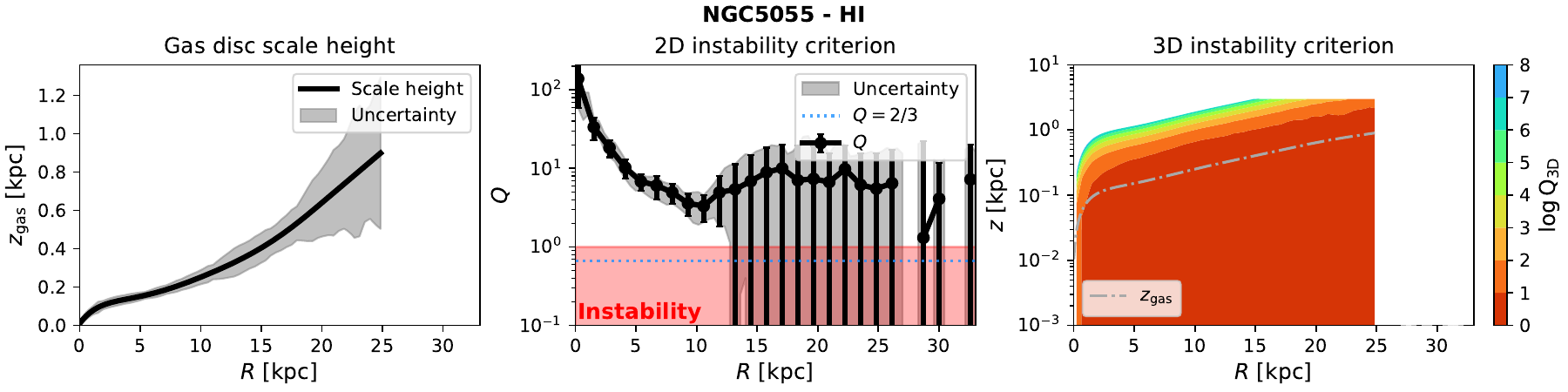} 
 \includegraphics[width=1\columnwidth]{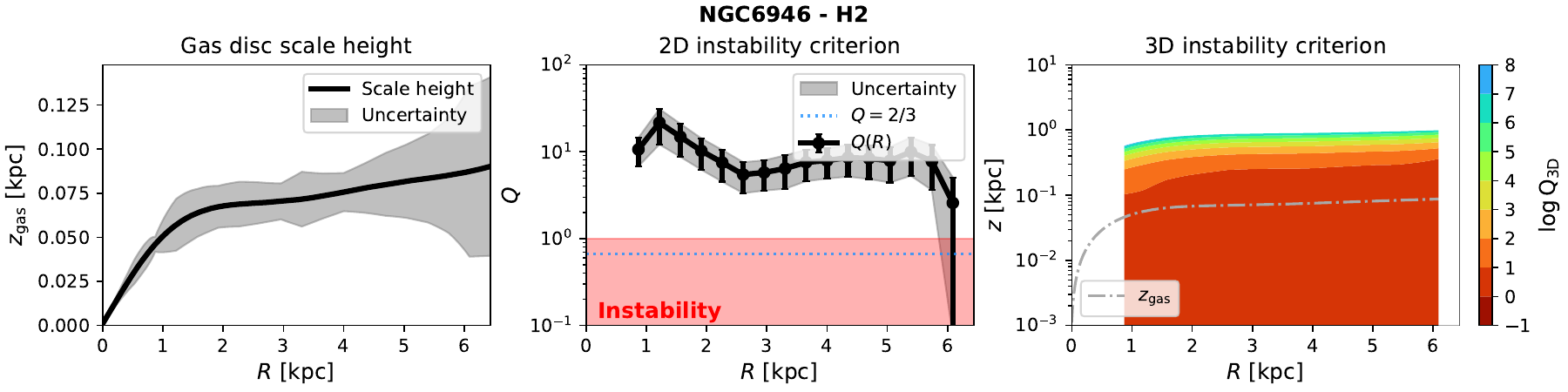} 
 \includegraphics[width=1\columnwidth]{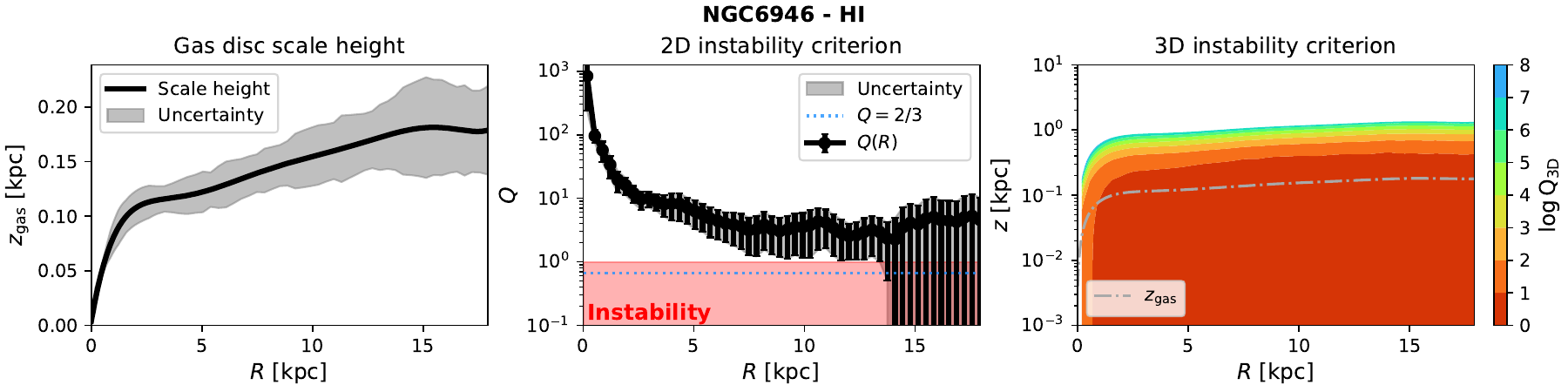} 
 \includegraphics[width=1\columnwidth]{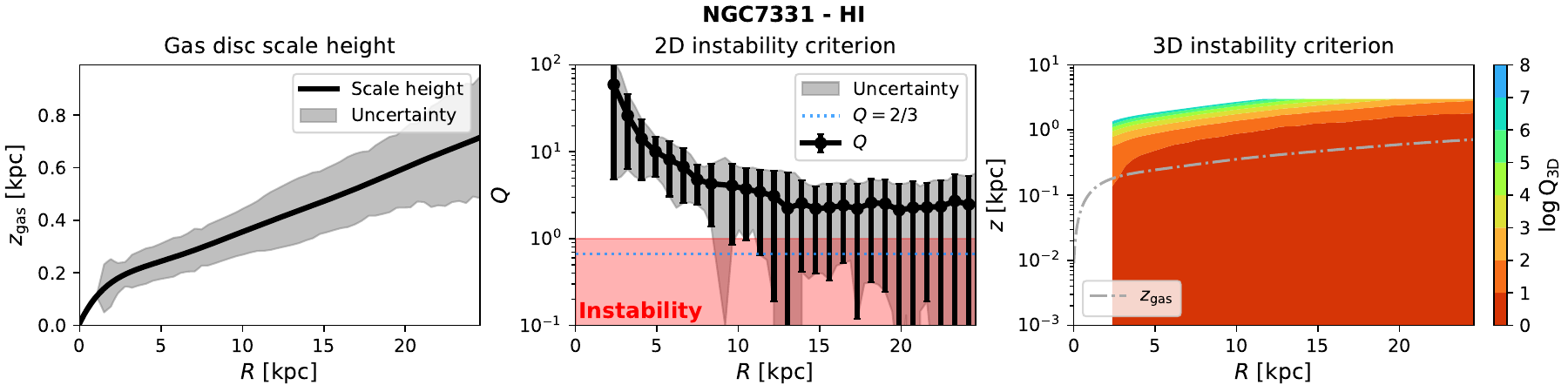} 
 \caption{continued.}
\end{figure*} 

\begin{figure*}[htp] \ContinuedFloat 
 \includegraphics[width=1\columnwidth]{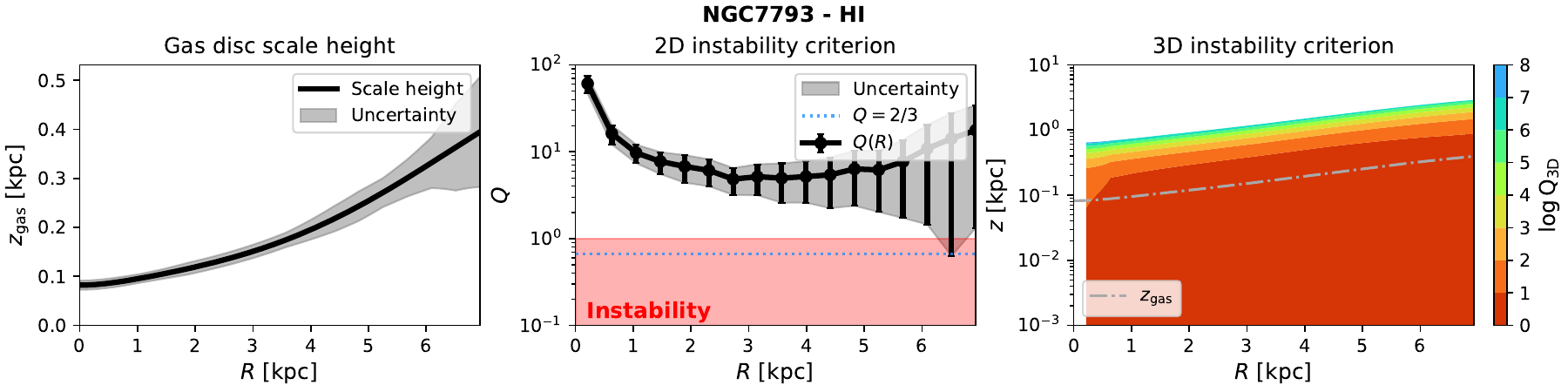} 
 \includegraphics[width=1\columnwidth]{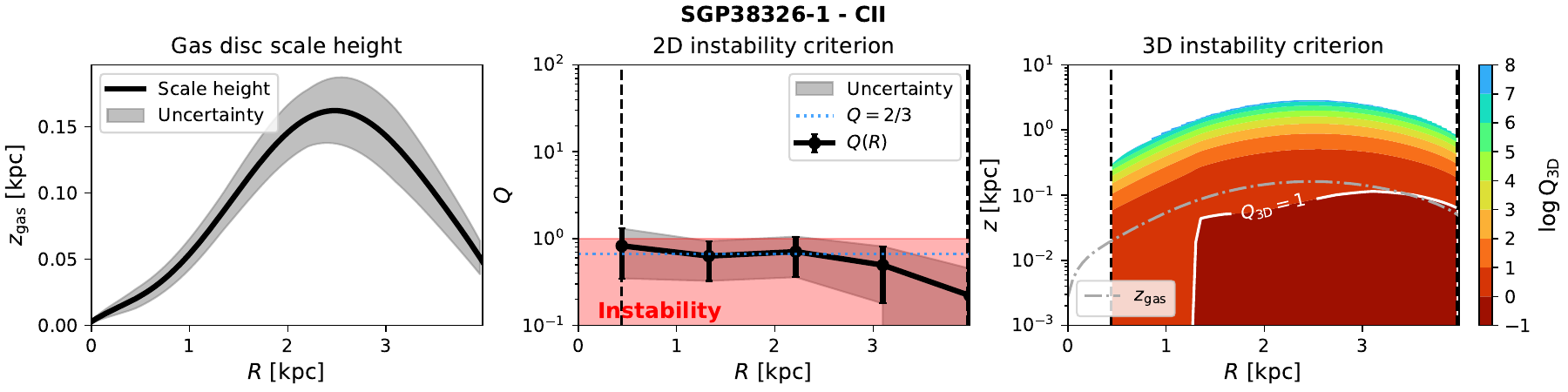} 
 \includegraphics[width=1\columnwidth]{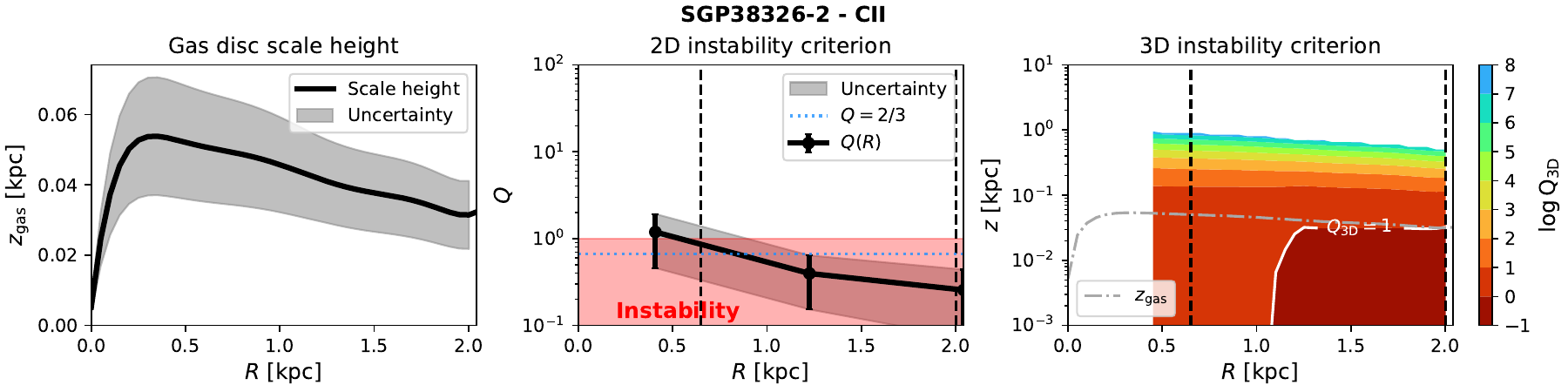} 
 \includegraphics[width=1\columnwidth]{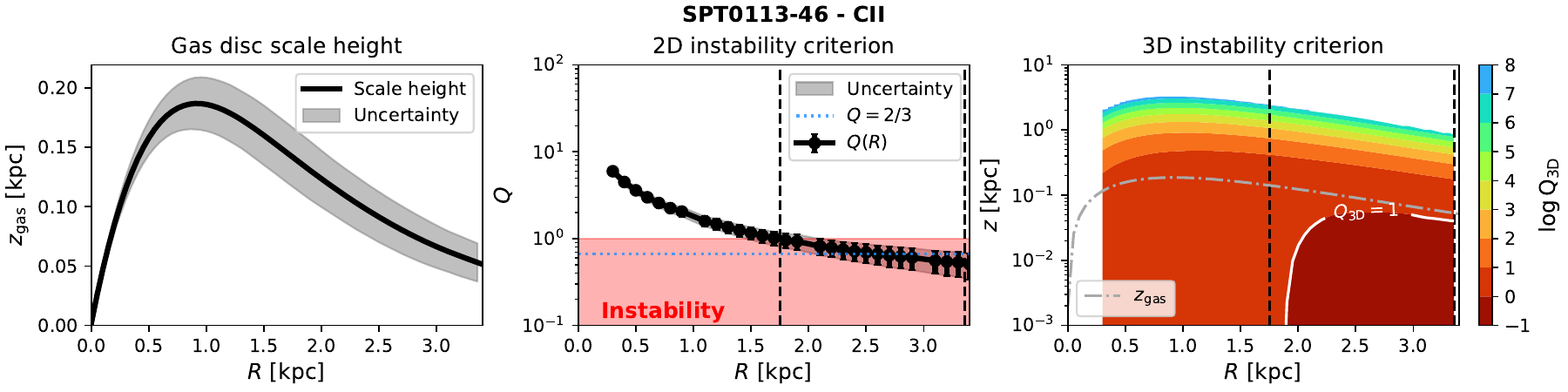} 
 \includegraphics[width=1\columnwidth]{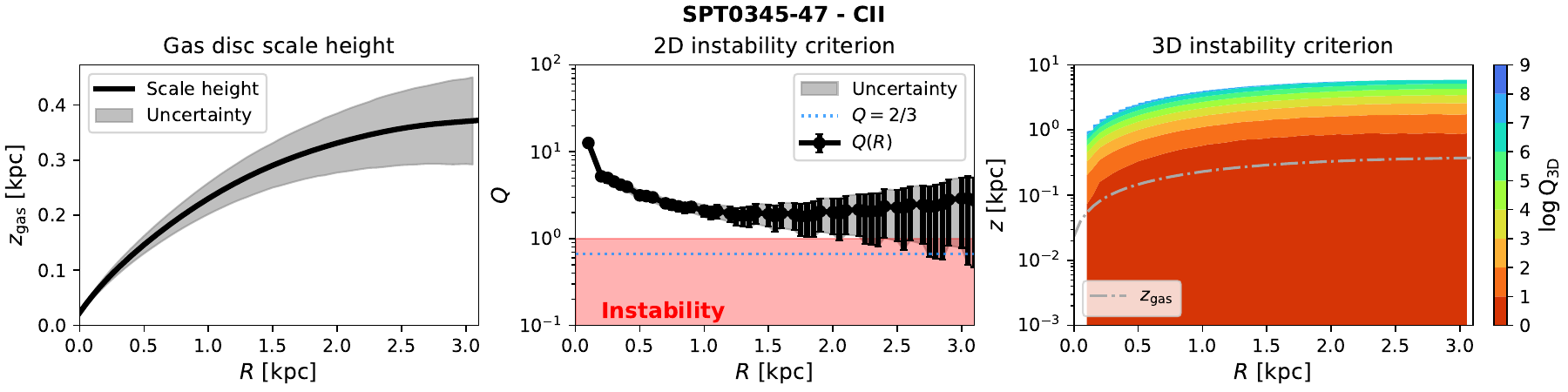} 
 \caption{continued.}
\end{figure*} 

\begin{figure*}[htp] \ContinuedFloat 
 \includegraphics[width=1\columnwidth]{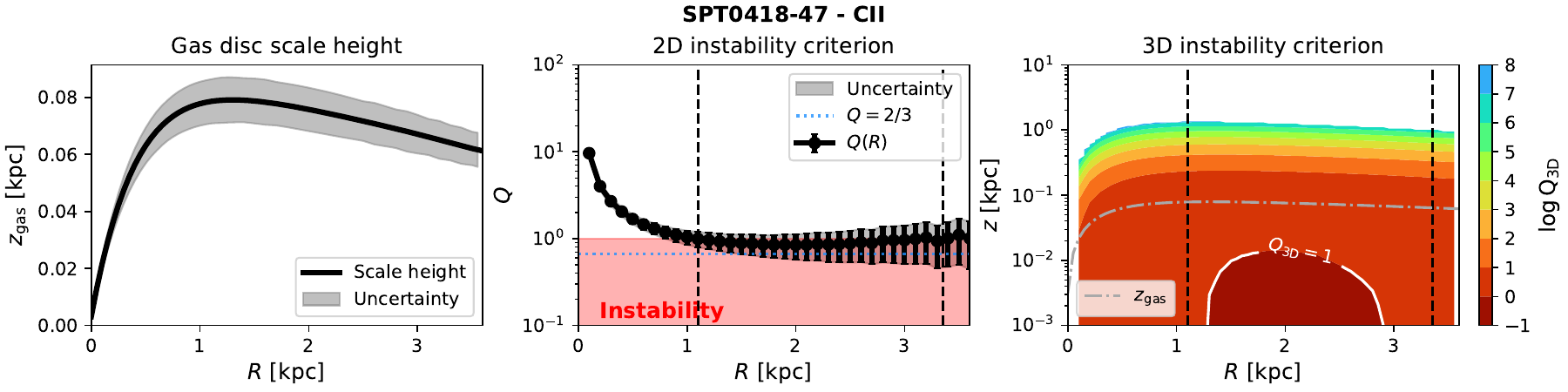} 
 \includegraphics[width=1\columnwidth]{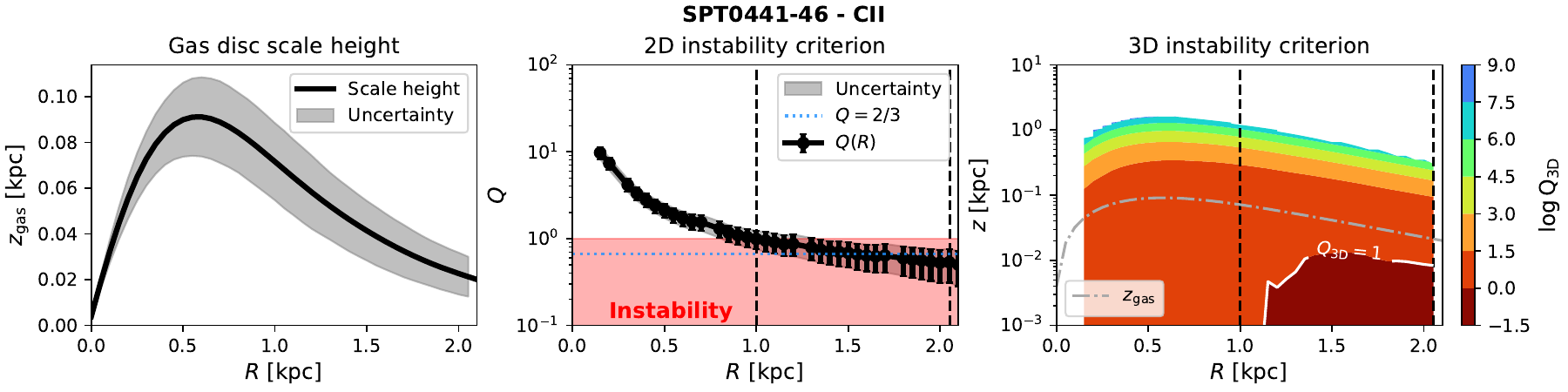} 
 \includegraphics[width=1\columnwidth]{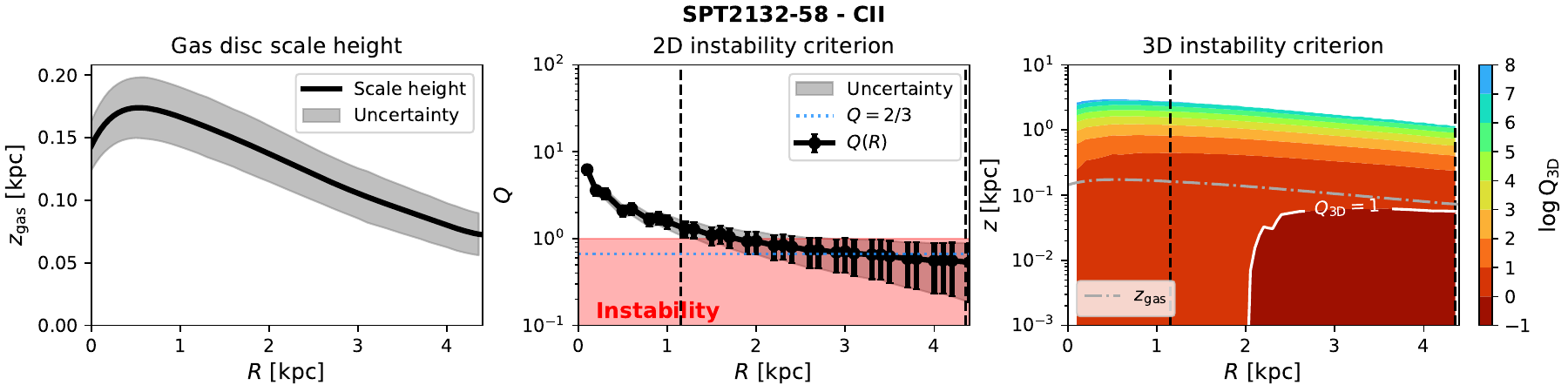} 
 \includegraphics[width=1\columnwidth]{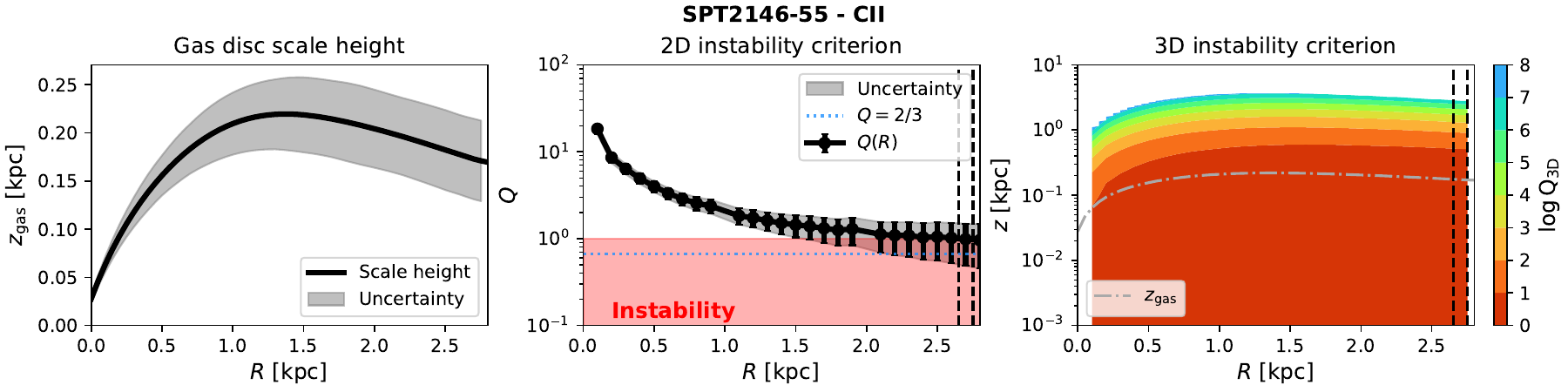} 
 \includegraphics[width=1\columnwidth]{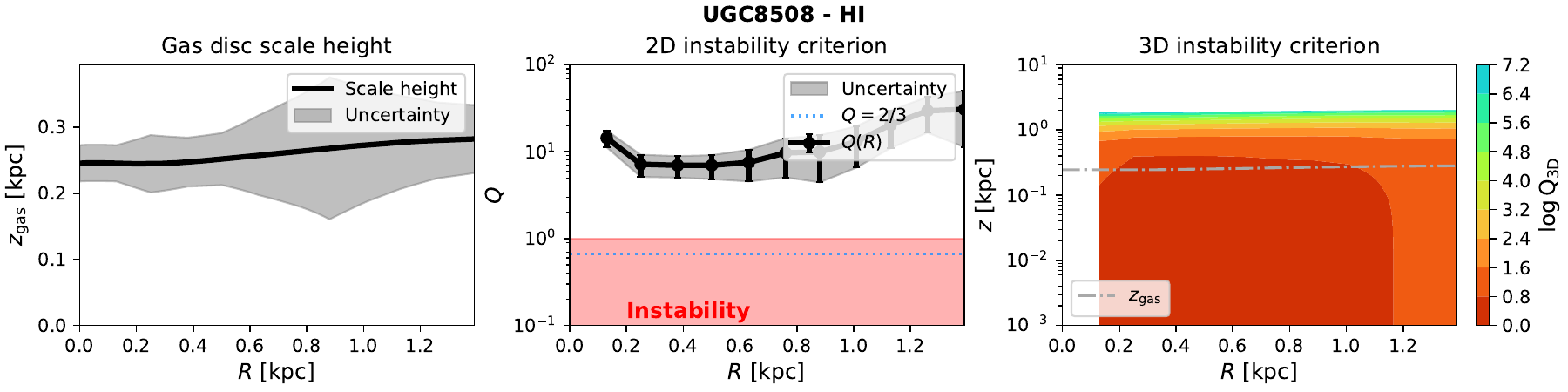} 
 \caption{continued.}
\end{figure*} 

\begin{figure*}[htp] \ContinuedFloat 
 \includegraphics[width=1\columnwidth]{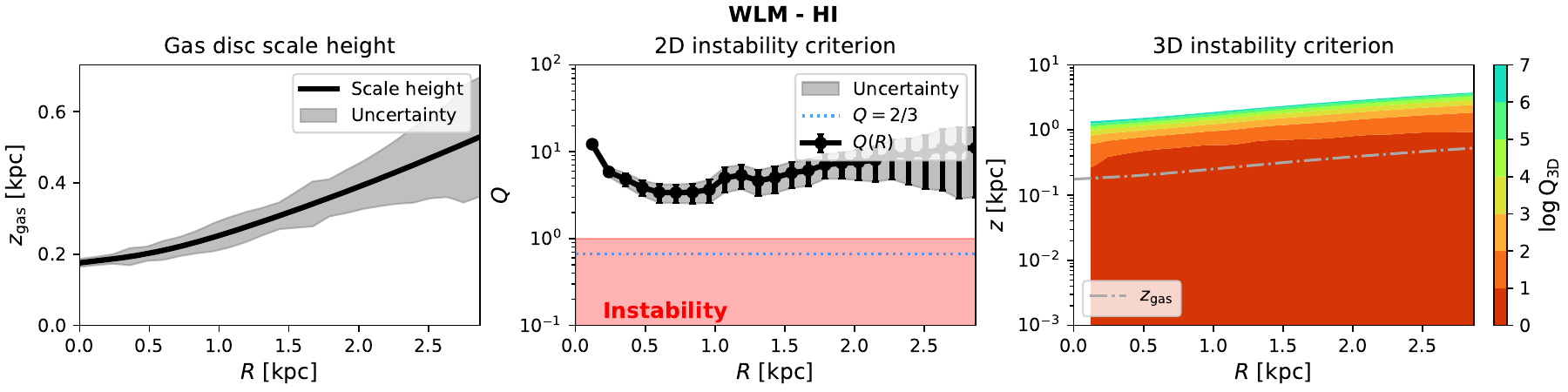} 
 \includegraphics[width=1\columnwidth]{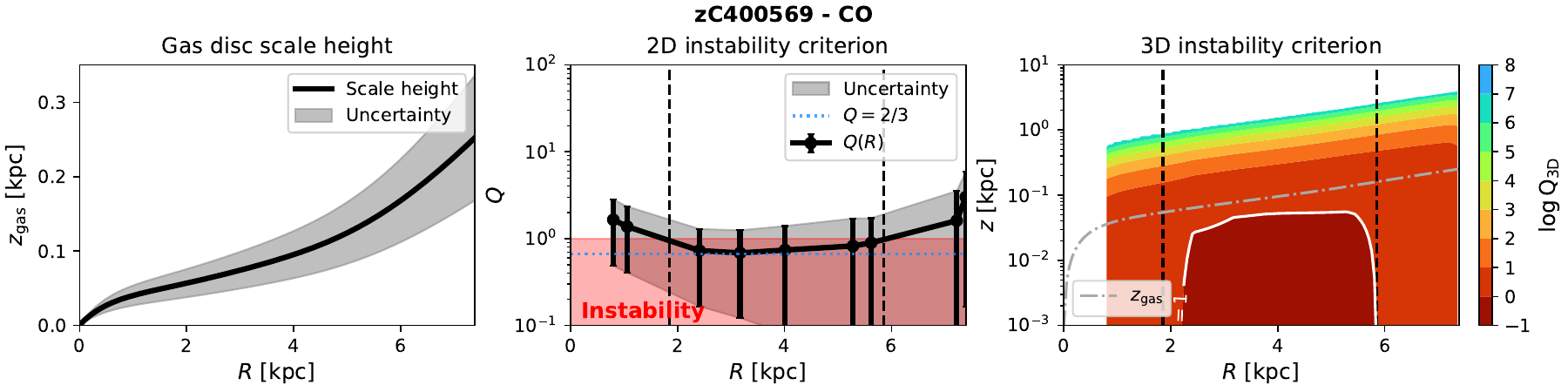}
 \includegraphics[width=1\columnwidth]{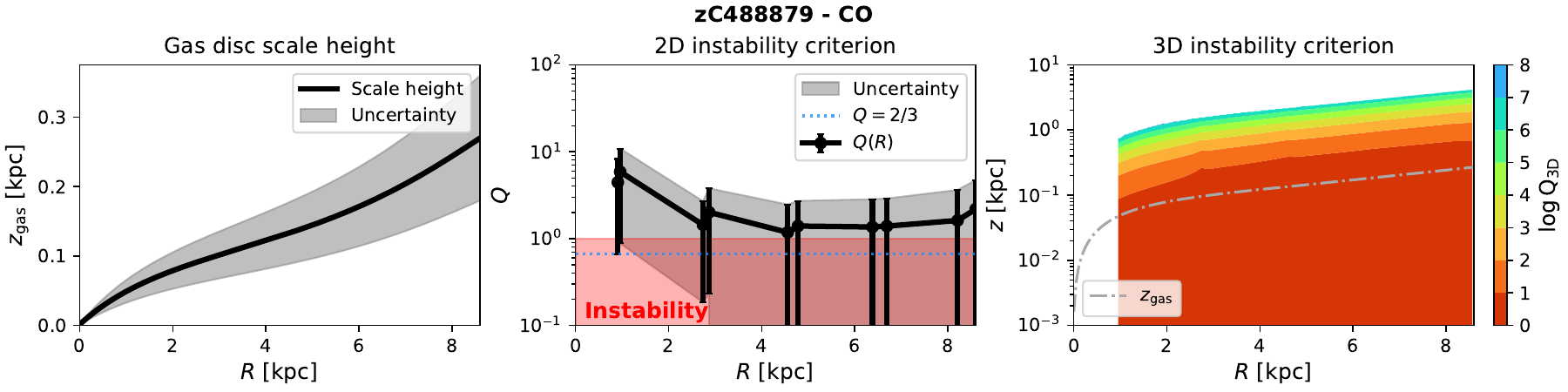}
 \caption{continued.} 
\end{figure*}

\end{appendix}

\end{document}